\documentclass[aps,pra,reprint,floatfix,superscriptaddress,nofootinbib]{revtex4-2}
\usepackage{CJK}
\usepackage[utf8]{inputenc}
\usepackage{hyperref}
\usepackage{longtable, xcolor}
\usepackage{siunitx}
\usepackage{braket,amsmath,amssymb,mathtools,amsthm, enumerate}
\usepackage[linesnumbered,ruled,vlined]{algorithm2e}

\DeclarePairedDelimiter\abs{\lvert}{\rvert}

\newtheorem{theorem}{Theorem}
\newtheorem{lemma}{Lemma}

\newtheorem{definition}[theorem]{Definition}

\begin{document}
\begin{CJK*}{GB}{} 
\title{Quantum Computer Benchmarking via Quantum Algorithms}
\author{Konstantinos Georgopoulos}
\email{k.georgopoulos2@newcastle.ac.uk}
\affiliation{School of Computing, Newcastle University, Newcastle-upon-Tyne, NE4 5TG, United Kingdom}
\author{Clive Emary}
\affiliation{Joint Quantum Centre Durham-Newcastle, School of Mathematics, Statistics and Physics, Newcastle University, Newcastle-upon-Tyne, NE1 7RU, United Kingdom}
\author{Paolo Zuliani}
\affiliation{School of Computing, Newcastle University, Newcastle-upon-Tyne, NE4 5TG, United Kingdom}
\date{\today}

\begin{abstract}
    We present a framework that utilizes quantum algorithms, an architecture aware quantum noise model and an ideal simulator to benchmark quantum computers. The benchmark metrics highlight the difference between the quantum computer evolution and the simulated noisy and ideal quantum evolutions. We utilize our framework for benchmarking three IBMQ systems. The use of multiple algorithms, including continuous-time ones, as benchmarks stresses the computers in different ways highlighting their behaviour for a diverse set of circuits. The complexity of each quantum circuit affects the efficiency of each quantum computer, with increasing circuit size resulting in more noisy behaviour. Furthermore, the use of both a continuous-time quantum algorithm and the decomposition of its Hamiltonian also allows extracting valuable comparisons regarding the efficiency of the two methods on quantum systems. The results show that our benchmarks provide sufficient and well-rounded information regarding the performance of each quantum computer.
\end{abstract}

\maketitle
\end{CJK*}

\section{Introduction}\label{sec:intro}
Benchmarking quantum computers aims to determine the performance of a quantum computing system under an appropriate set of metrics. Within the Noisy Intermediate-Scale Quantum (NISQ) era \cite{Preskill-2018}, benchmarking the capabilities and performance of quantum computers when executing quantum programs is of paramount importance, especially for assessing their scalability.

An intuitive approach to benchmarking quantum computers is establishing a set of quantum programs and measuring the performance of a quantum computer when executing each one. Such a work gains more merit as bigger quantum computers are built. There are various advantages to this approach, for example benchmarking the limits and behaviour of a quantum machine within a scaling, computationally intensive environment and particularly testing the system when performing a ``real-world" task. Various companies appear to favor such an approach. IonQ for example tested their quantum computer using the Bernstein-Vazirani \cite{BernsteinVazirani} and the Hidden Shift \cite{vanDam-2006,Rotteler-2010} algorithms. The metric for performance was the likelihood of measuring the correct output \cite{Wright-2019}. On the other hand, Google focused on the problem of quantum sampling and achieved results they claim demonstrated quantum supremacy \cite{GoogleSuprem}. While the chosen application was not particularly useful in a ``real-world" scenario, it excelled at demonstrating the computing power of the system. So a question arises, highlighting the difficulty of creating quantum program benchmarks: which benchmarks are more insightful?

The different competing quantum technologies pose a major challenge. The technologies have different topologies and thus have unique strengths and weaknesses. For example, the connectivity of an ion-trap computer provided a large advantage on some benchmarks over a superconducting quantum computer \cite{Linke-2017}.

Clearly, systems can behave differently on different benchmarks, which introduces the issue of invested interests, i.e., using benchmarks that are expected to perform well on a current system \cite{Lilja-2000}. While impressive, current quantum computers are very small compared to the computers we hope to build in the coming years. Hence, current benchmarks are also relatively simple compared to truly useful programs. While running smaller versions of real-world applications introduces error, and is accepted in classical benchmarking, this is exacerbated for quantum computers. Entirely new issues may be introduced when scaling up and it is difficult to say whether performances measured today are good indicators of future performance. For example, IonQ's computer \cite{Wright-2019} has all $11$ qubits fully-connected. This configuration is possible at this scale, but this might not be true for a system with hundred or a thousand qubits. Such a system will likely require multiple fully-connected groups of qubits and communication will need to be orchestrated between them \cite{Martonosi-2019}. This introduces additional complexity which is not found in small-scale benchmarks.

Our previous question can be refined as: which quantum algorithms would be useful for program benchmarking quantum computers in the future? Algorithms such as quantum Markov chains \cite{Gudder-2008}, Shor's \cite{Shor-1997}, Grover's \cite{Grover-1996} and quantum chemistry \cite{Linke-2017,Olson-2017,McArdle-2020} are some obvious examples. Even if, for the most part, these algorithms will remain out of reach for near-term quantum computers, there is much to be gained from analyzing their scalability and reaction to noise. Currently, classical-quantum hybrid algorithms \cite{Yudong-2019,Schuld-2020,Wecker-2015,Fahri-2014,Perruzo-2014} are popular due to their ability to make use of the limited resources of NISQ computers. Another example of an algorithm that holds great potential for benchmarking are quantum walks, due to their susceptibility to noise and clear quadratic advantage over classical random walks \cite{Szegedy}.

Within this paper we use quantum algorithms to benchmark three of the newer IBM superconducting quantum computers: the $5$-qubit Bogota and Santiago and the $7$-qubit Casablanca machines. The choice of quantum algorithms is crucial, as we want them to be (i) scalable, in order to be able to face the challenge of the growing number of qubits in quantum computers, (ii) predictable in a way that allows us to recognize its noisy behaviour, and (iii) demonstrate clear quantum advantage. With these criteria in mind, we choose five algorithms: discrete-time quantum walks \cite{Aharonov-2000}, continuous-time quantum walks \cite{Fahri-1998-ctqw}, a circuit simulating the continuous-time quantum walk by decomposing its Hamiltonian to a sequence of Pauli gates, quantum phase estimation \cite{qpe-2019} and Grover's algorithm \cite{Grover-1996}. Finally, we note that using a continuous-time quantum algorithm and its Hamiltonian decomposition for benchmarking a (digital) quantum computer has not been carried out before, to the best of our knowledge. For the experiments we make use of the IBM Qiskit development kit \cite{Qiskit,IBMQExp} to simulate and execute the quantum circuits. 

The paper is organized as follows. Section \ref{sec:prelim} introduces the preliminary methods necessary for the benchmarking process. Section \ref{sec:framework} defines the three benchmark metrics that result from the benchmarking process as well as presents a concrete framework for program benchmarking quantum computers. Moving on, the experimental process and the benchmark results for the chosen quantum computers are layed out in Section \ref{sec:results} before finally concluding the paper in Section \ref{sec:concl}.

\section{Preliminary Methods}\label{sec:prelim}
This section introduces the noise model used to simulate the behaviour of quantum computers. Additionally, it offers a brief discussion on the five quantum algorithms we use for benchmarking. Finally, it presents the quantum computers we are interested in benchmarking and the mathematical foundation of the final benchmark metrics.

\subsection{Unified Noise Model}\label{subsec:unm}
To approximate the noisy behaviour of quantum computers we use the \textit{unified noise model} (UNM) we have recently developed \cite{KGeorgo-2020-UNM}. This model combines three sources of error: (i) hardware infidelities in the form of gate, state preparation and measurement errors, (ii) decoherence in the form of thermal relaxation and (iii) dephasing of the qubits. The experiments in \cite{KGeorgo-2020-UNM} show that the UNM performs very well at approximating the behaviour of the IBMQ $15$-qubit Melbourne computer and better than other state of the art noise models.

The main characteristic of the UNM is its architecture awareness: the architectural graph that encompasses all the information regarding the connectivity of the qubits within the quantum processing unit (QPU) gets encoded within the model itself. Additionally, the model uses a number of noise parameters (see Table \ref{table:noiseparams}) calibrated from the machine itself, i.e., parameters that express the error rates of the gates, state preparations and measurements as well as the time it takes for the qubits within the QPU to decohere and dephase. Each noise parameter is unique and corresponds to each qubit individually or pair of qubits.

\bgroup
\def\arraystretch{1.2}%
\setlength\tabcolsep{0.1cm}
\begin{table}[!t]
    \centering
    \begin{tabular}{c|c|c}
        \hline
        Parameter & Error Type & No. Parameters \\ \hline \hline
        $p_{r}$ & Gate error rates & $r$ \\
        $p_{m}$ & State preparation error rates & $m$ \\ 
        $p_{s}$ & Measurement error rates & $s$ \\
        $T_{1}$ & Thermal relaxation times & $n$ \\ 
        $T_{2}$ & Dephasing times & $n$ \\ \hline
    \end{tabular}
    \caption{The noise parameters and number of noise parameters for each type of error within the UNM; $n$ is the number of qubits in the system, $m$ is the number of qubits that are measured, $s$ is the number of state preparations that occur and $r$ is the number of distinct types of gates implemented in the architecture, each considered once per qubit or pair of qubits.}
    \label{table:noiseparams}
\end{table}
\egroup

\subsection{The Quantum Algorithms}\label{subsec:qalg}
Here we give a quick overview of the quantum algorithms we use for benchmarking, i.e., (i) discrete-time quantum walks (DTQW), (ii) continuous-time quantum walks (CTQW), (iii) Pauli decomposition of the CTQW Hamiltonian (PD) (iv) quantum phase estimation (QPE) and (v) quantum search (QS). These algorithms are chosen according to three major characteristics that are interesting for benchmarking:
\begin{itemize}
    \item \textbf{Scalability.} The algorithm should be able to scale up (or down) and run on increasingly larger quantum systems.
    
    \item \textbf{Predictability.} The algorithm should produce a result that is easily predictable. An important addition to predictability is \textit{noise susceptibility}: the algorithm should provide a result whose distortion under the effects of noise is easily distinguishable from the ideal evolution.
    
    \item \textbf{Quantum advantage.} The algorithm should provide a computational speed-up over its classical counterpart or, in other words, represent a possibly relevant real-world application.
\end{itemize}

\paragraph{Discrete-time Quantum Walks.} Quantum walks (DTQW) are the quantum mechanical analogue of a classical random walk on a graph or a lattice \cite{Aharonov-2000,KGeorgo-2020-rots,Kempe-2003}. They exhibit intrinsic properties that render their evolution easily predictable and highly susceptible to noise, making them an ideal candidate for benchmarking. First of all, discrete-time quantum walks exhibit modular behaviour \cite{KGeorgo-2020-rots,Reitzner-2011-mod}. This characteristic describes the modular relationship between the parity of the number of coin-flips of the walk, the initial state and the current position of the walker, a property that gets violated in a noisy environment \cite{KGeorgo-2020-rots}. Secondly, quantum walks propagate quadratically further than classical random walks \cite{Aharonov-2000,Szegedy} thus showing clear quantum advantage over their classical counterparts.

Finally, quantum walks are a highly scalable process. The size of the state-space of a quantum walk (i.e., the number of states that the walk traverses, represented by the number of qubits in the relevant register) can easily increase to match the size of the quantum computer we are interested in benchmarking.

\paragraph{Continuous-time Quantum Walks.} Continuous-time quantum walks (CTQW) were first introduced by Fahri and Gutman in \cite{Fahri-1998-ctqw}. This algorithm, much like the DTQWs, have an easily predictable quantum evolution that is highly susceptible to quantum noise, but exhibit very different characteristics to the discrete case. First of all, the CTQW evolution is determined by a Hamiltonian, $H$, instead of a coin-flip and is driven by a unitary of the form $e^{-iHt}$. Unlike the DTQWs, continuous-time quantum walks do not exhibit modular behaviour. Although like DTQWs they feature a quadratic increase in the walker's propagation \cite{Chakra-2020,Ambainis-2020}.

Finally, CTQWs are an easily scalable process as adding qubits to circuit can scale up the size of the state-space. To the best of our knowledge, this is the first work that uses a continuous-time quantum algorithm to benchmark the performance of a digital quantum computer. There are a few ways that a continuous-time quantum walk can be implemented on a gate-based (i.e., discrete) quantum computer, for example simulating the CTQW evolution using a DTQW \cite{Childs-2009-CTQW} or approximating the CTQW through a decomposition of the unitary $e^{-iHt}$ to a sequence of universal gates. Within this paper, the latter approach is utilized.

\paragraph{Pauli Decomposition of CTQW Hamiltonian.} Decomposition of quantum Hamiltonians to a set of universal gates is a well-studied area of research \cite{Lloyd-1996-dec,Childs-2011-dec,Suzuki-1992-dec,Aharonov-2003-dec,Hedge-2016-Pdec}. For this paper, we are interested in decomposing the Hamiltonian of the CTQW using the well-known set of Pauli matrices as the universal gate set. A further detailed analysis of this process is presented in the following Section \ref{subsec:paulidec}.

This procedure, often called \textit{Hamiltonian simulation}, adheres to the criteria for a good benchmarking process via the algorithm it decomposes. In other words, since the CTQW is suitable for benchmarking, so is the circuit that implements the decomposition of the CTQW Hamiltonian. Furthermore, it provides added value to this research since one can evaluate the performance of the quantum computer when executing the Hamiltonian simulation of a quantum process (i.e., CTQW), as well as compare the decomposition with the original algorithm.

\paragraph{Quantum Phase Estimation.} The quantum phase estimation (QPE) algorithm is used to estimate the phase (or eigenvalue) of an eigenvector of a unitary operator. More precisely, given an arbitrary quantum operator $U$ and a quantum state $\ket{\psi}$ such that $U\ket{\psi}=e^{2i\pi\theta}\ket{\psi}$, the algorithm estimates the value of $\theta$, given an approximation error \cite{Kitaev-qpe,qpe-2019,Oh-2019}.

Within this paper we exploit three characteristics of QPE that make it interesting for benchmarking. First of all, it is a scalable algorithm as further accuracy of the result (i.e., the estimated phase) can be obtained by increasing the number of qubits in the system. Furthermore, the result of the QPE is predictable and highly susceptible to quantum noise. Finally, QPE offers clear quantum advantage, achieving an exponential speed-up over known classical methods, rendering the algorithm one of the most important subroutines in quantum computing and serving as the building block of major quantum algorithms, like Shor's \cite{Shor-1997} or the HHL algorithms \cite{Harrow-2009}.

\paragraph{Quantum Search.} Grover's algorithm \cite{Grover-1996} describes a process for searching for a specific item within a database. Within this paper we use quantum search (QS) to look for a specific number $s$ within a set of numbers $\mathcal{S}=\{0,\dots,2^{n-1}\}$, where $n$ is the number of qubits within the quantum system that participate in the computation.

Quantum search represents an ideal algorithm for benchmarking quantum computers. The algorithm can scale up to search for an item within a larger database simply by adding qubits to the relevant quantum register. The result is easily predictable, as it is simply the item (or number, in our case) sought, as well as highly susceptible to noise (for example, the wrong result might appear due to noise). Additionally, QS can speed up an unstructured search problem quadratically, thus making it a very alluring application for quantum computers. Finally, Grover's algorithm can serve as a general trick or subroutine to obtain quadratic runtime improvements for a variety of other algorithms through what is called amplitude amplification \cite{Brassard-2002}.

\subsection{Pauli Decomposition of CTQW Hamiltonian}\label{subsec:paulidec}
In general, an arbitrary Hamiltonian $H$ of size $N\times N$, where $N=2^{n}$ and $n$ the number of qubits in the system, can be decomposed into a sequence of Pauli operators of the set $S=\{ \sigma_{I}, \sigma_{X}, \sigma_{Y}, \sigma_{Z} \}$ with well-known matrix representation as
\begin{equation}
    H = \sum_{i_{1},\dots,i_{N}=I,x,y,z} \alpha_{i_{1},\dots,i_{N}} \left( \sigma_{i_{1}}\otimes\dots\otimes\sigma_{i_{N}} \right), \label{eq:pd}
\end{equation}
where
\begin{equation*}
     \alpha_{i_{1},\dots,i_{N}} = \frac{1}{N}\operatorname{tr}\left[ \left( \sigma_{i_{1}}\otimes\dots\otimes\sigma_{i_{N}} \right)\cdot H \right].
\end{equation*}

For this paper we implement a CTQW on an $N$-cycle with Hamiltonian defined as $H_{\text{qw}}=\gamma A=\frac{1}{d}A=\frac{1}{2}A$, where $\gamma=1/d$ is the hopping rate between the two adjacent nodes in the cycle (i.e., node degree $d=2$) and $A$ is the adjacency matrix. Note that, for the physical purposes of this evolution, the rate $\gamma_{n|i}$ has dimensions $1/t$, where $t$ is the time duration of the CTQW evolution, the Hamiltonian has the appropriate energy dimensions and, for simplicity, $\hbar=1$.

The next step is to construct the unitary evolution operator from the Hamiltonian. This can be done by exponentiating the Hamiltonian as $e^{-iHt}$ where $H$ is a sum of terms of the form of equation \eqref{eq:pd}. It is important here to consider two things. First of all, during matrix exponentiation, a decomposition of the form $e^{-i(H_{1}+H_{2})t}=e^{-iH_{1}t}e^{-iH_{2}t}$, where $H_{1}$ and $H_{2}$ are Hermitian operators, is possible if $H_{1}$ and $H_{2}$ commute, i.e., $H_{1}H_{2}-H_{2}H_{1}=0$. This rule is naturally expanded for more than two matrices on the exponent.

Secondly, in the case that not all matrices in the exponent commute, the unitary operator resulting from the Hamiltonian exponentiation needs to be decomposed using the Lie-product formula \cite{Lloyd-1996-dec} as
\begin{equation}
    e^{-i(H_{1} + H_{2} + \dots)t} \approx \left( e^{-iH_{1}t/r}e^{-iH_{2}t/r}\dots \right)^{r}
\end{equation}
where, for our case, $H=\sum_{i} H_{i}$ is the Pauli decomposition of the Hamiltonian to a sequence of Hermitian terms. This formula will create an approximation of the Hamiltonian with bounded error depending on $r$ \cite{Lloyd-1996-dec,Suzuki-1992-dec}. To ensure that the Pauli Hamiltonian decomposition exhibits error at most $\epsilon$, the bound $r$ can be taken as \cite{Lloyd-1996-dec}
\begin{equation*}
    r = \mathcal{O}\left( (||H||t)^{2}/ \epsilon \right),
\end{equation*}
where $||H||$ is the norm of the Hamiltonian $H$ and $t$ is the continuous-time duration of the quantum evolution.

Thus, using the above methodology, we can now decompose the unitary that describes the continuous-time evolution of the CTQW (i.e., $e^{-iHt}$) in a sequence of Pauli operations that \textit{approximate} said evolution. As the Pauli gates form a universal gate set, the resulting quantum circuit can be implemented on the gate-based quantum processors.

\subsection{The Quantum Computers and Circuits}\label{subsec:qcs}
Within this paper we are interested in benchmarking three computers, the $5$-qubit Bogota, Santiago and the $7$-qubit Casablanca machine. They all exhibit a quantum volume $V_{Q}=32$ \cite{IBMQExp,Moll-2018}.

\begin{figure}[!t]
    \begin{tabular}{c}
          \includegraphics[width=8.5cm]{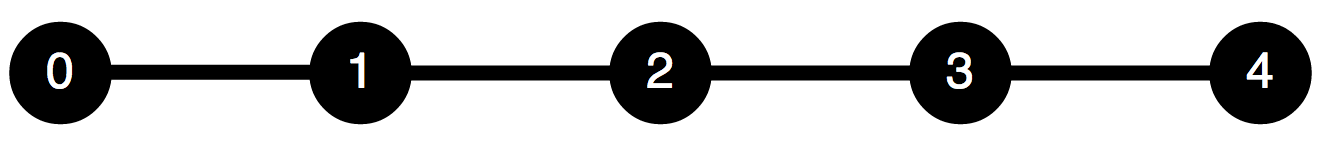} \\
          (a) \\ [6pt] 
          \includegraphics[width=5.2cm]{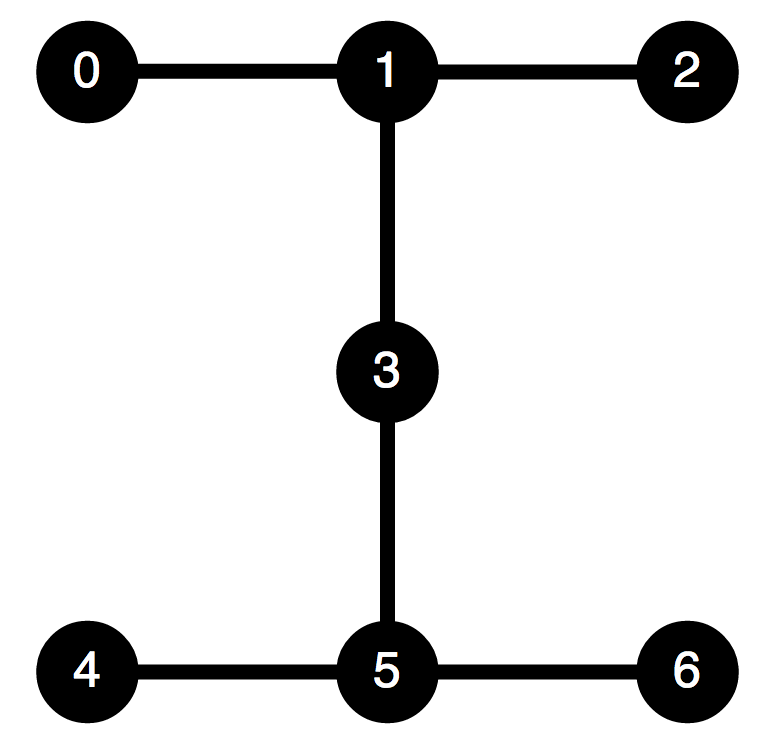} \\ 
          (b) \\[6pt]
    \end{tabular}
    \caption{Architectural graphs of the (a) IBMQ $5$-qubit Bogota and Santiago machines and (b) the IBMQ $7$-qubit Casablanca machine. A node in the graph represents a physical qubit whereas an edge represents a connection between a pair of qubits.}
    \label{fig:architectures}
\end{figure}

\paragraph{Quantum computer architectures.} Throughout our research we find that it is essential for the benchmarking procedure to be architecturally aware. This is also reflected by our choice of the UNM, an architecture aware noise model \cite{KGeorgo-2020-UNM}. Figure \ref{fig:architectures} shows the qubit connectivity of the quantum computers we benchmark in this paper.

\paragraph{Quantum circuits and characteristics.} For the implementation of quantum walks, we use a gate efficient approach that is based on inverter gates, as shown in \cite{Douglas-2009-EffWalk}. The QPE circuit is based on \cite{QPECircuit} and heavily relies on quantum Fourier transform (and its inverse) \cite{Nielsen_Chuang_2010} to estimate the relevant eigenvalue. For QS we make use of two approaches to the implementation, one with ancilla qubits (QSa) and one without (QSn) \cite{Karlsson_2018}.

For the continuous-time quantum walk, the Hamiltonian is automatically implemented by the Qiskit API when submitted for execution on the quantum computer. The Pauli decomposition (PD) of the Hamiltonian can be easily implemented on the quantum computers as it already maps the continuous-time Hamiltonian to a discrete basis gate set decomposition.

We identify four quantum circuit characteristics that are of interest for benchmarking:
\begin{enumerate}[i]
    \item the number of gates in the circuit,
    
    \item the number of active qubits, or qubits that are utilized by the quantum circuit (also called workspace),
    
    \item the depth of the circuit, i.e., the longest path between the start of the circuit and a measurement gate and
    
    \item the runtime of the circuit on the quantum computer. Table \ref{table:charactQASMcirc} shows those characteristics of the quantum circuits that implement each quantum algorithm.
\end{enumerate}

\begin{table*}[!t]
    \begin{tabular}{c}
        \bgroup
        \def\arraystretch{1.2}%
        \setlength\tabcolsep{0.36cm}
            \centering
            \begin{tabular}{c|c|c|c|c}
                \hline
                Machine & No. Gates & Size of Workspace & Depth & QC Runtime $\pm$ s.d. ($\text{ms}$) \\ \hline \hline
                $5$-qubit Bogota & $47$ & $3$ & $35$ & $2.41\pm 0.03$ \\
                $5$-qubit Santiago & $47$ & $3$ & $35$ & $2.34\pm 0.02$ \\
                $7$-qubit Casablanca & $47$ & $3$ & $35$ & $2.54\pm 0.05$ \\ \hline
            \end{tabular}
        \egroup \\ [6pt] (a) Discrete-time quantum walk circuit characteristics. \\ [6pt]
        \bgroup
        \def\arraystretch{1.2}%
        \setlength\tabcolsep{0.36cm}
            \centering
            \begin{tabular}{c|c|c|c|c}
                \hline
                Machine & No. Gates & Size of Workspace & Depth & QC Runtime $\pm$ s.d. ($\text{ms}$) \\ \hline \hline
                $5$-qubit Bogota & $19$ & $2$ & $13$ & $2.34\pm 0.04$ \\
                $5$-qubit Santiago & $19$ & $2$ & $13$ & $3.01\pm 0.07$ \\
                $7$-qubit Casablanca & $19$ & $2$ & $13$ & $2.62\pm 0.01$ \\ \hline
            \end{tabular}
        \egroup \\ (b) Continuous-time quantum walk circuit characteristics. \\ [6pt]
        \bgroup
        \def\arraystretch{1.2}%
        \setlength\tabcolsep{0.36cm}
            \centering
            \begin{tabular}{c|c|c|c|c}
                \hline
                Machine & No. Gates & Size of Workspace & Depth & QC Runtime $\pm$ s.d. ($\text{ms}$) \\ \hline \hline
                $5$-qubit Bogota & $243$ & $2$ & $183$ & $2.42\pm 0.04$ \\
                $5$-qubit Santiago & $243$ & $2$ & $183$ & $2.28\pm 0.06$ \\
                $7$-qubit Casablanca & $243$ & $2$ & $183$ & $3.76\pm 0.04$ \\ \hline
            \end{tabular}
        \egroup \\ (c) Pauli decomposition of CTQW circuit characteristics. \\ [6pt]
        \bgroup
        \def\arraystretch{1.2}%
        \setlength\tabcolsep{0.36cm}
            \centering
            \begin{tabular}{c|c|c|c|c}
                \hline
                Machine & No. Gates & Size of Workspace & Depth & QC Runtime $\pm$ s.d. ($\text{ms}$) \\ \hline \hline
                $5$-qubit Bogota & $93$ & $4$ & $66$ & $2.54\pm 0.01$ \\
                $5$-qubit Santiago & $97$ & $4$ & $72$ & $2.26\pm 0.03$ \\
                $7$-qubit Casablanca & $100$ & $4$ & $75$ & $2.68\pm 0.06$ \\ \hline
            \end{tabular}
        \egroup \\ (d) Quantum phase estimation circuit characteristics. \\ [6pt]
        \bgroup
        \def\arraystretch{1.2}%
        \setlength\tabcolsep{0.36cm}
            \centering
            \begin{tabular}{c|c|c|c|c}
                \hline
                Machine & No. Gates & Size of Workspace & Depth & QC Runtime $\pm$ s.d. ($\text{ms}$) \\ \hline \hline
                $5$-qubit Bogota & $497$ & $4$ & $358$ & $2.70\pm 0.08$ \\
                $5$-qubit Santiago & $479$ & $4$ & $343$ & $2.56\pm 0.04$ \\
                $7$-qubit Casablanca (ancilla) & $788$ & $6$ & $503$ & $2.81\pm 0.04$ \\
                $7$-qubit Casablanca (no ancilla) & $465$ & $4$ & $336$ & $2.74\pm 0.03$ \\ \hline
            \end{tabular}
        \egroup \\ (e) Quantum search circuit characteristics. \\
    \end{tabular}
    \caption{Quantum circuit characteristics for the five quantum circuits. No. gates and size of workspace are the number of gates and active qubits in the circuit respectively; depth of the circuit is the longest path between the start of the circuit and a measurement gate; QC runtime is the approximate average execution time of the circuit on the quantum computer along with standard deviation (s.d.).}
    \label{table:charactQASMcirc}
\end{table*}

\subsection{Benchmark Indicators: Hellinger Distance}\label{subsec:hd}
The end results of the benchmarking process will be in the form of a comparison of the quantum computer output distribution with the distributions resulting from the unified noise model simulations of each machine and the ideal evolution.

To compare the probability distributions and generate the benchmarks, we use the Hellinger distance (HD) \cite{Jin_2018}.
\begin{definition}[Hellinger distance]\label{def:hd}
    For probability distributions $P=\{p_{i}\}_{i\in[s]}$, $Q=\{q_{i}\}_{i\in[s]}$ supported on $[s]$, the Hellinger distance between them is defined as
    \begin{equation}
        h(P,Q) = \frac{1}{\sqrt{2}} \sqrt{ \sum_{i=1}^{k} \left( \sqrt{p_{i}} - \sqrt{q_{i}} \right)^{2} } \label{eq:hd}.
    \end{equation}
\end{definition}

The Hellinger distance is a metric satisfying the triangle inequality. It takes values between $0$ and $1$ (i.e. $h(P,Q)\in[0,1]$) with $0$ meaning that the two distributions are equal. Additionally, it is easy to compute, easy to read and it does not depend on the probability distributions having the same support. The last property is particularly useful since in many ideal output distribution of quantum circuits the probability mass is concentrated on a few states.

\section{Framework for Program Benchmarking}\label{sec:framework}
The desired benchmarks correspond to comparisons (via the Hellinger distance) between the evolution of the quantum system on the computer, the noisy simulation and the ideal case. To achieve a more memorable notation, in the following analysis the symbol $q$ denotes the quantum computer, $i$ the ideal evolution and $n$ the noisy simulation. For example, the subscript $q|n$ denotes a value that corresponds to the difference between the quantum computer evolution ($q$) and the noisy evolution ($n$). We define three distances of interest, or otherwise, three benchmark metrics, as follows. 

\begin{definition}[alpha benchmark]\label{def:alpha}
    The Hellinger distance between the probability distribution of the quantum computer evolution, $Q$, and the ideal distribution, $D$, namely $h_{\text{id}}(Q,D)$, is the alpha benchmark, with notation $\alpha_{q|i}$:
    \begin{equation}
        \alpha_{q|i} \equiv h(Q,D). \label{eq:alpha}
    \end{equation}
\end{definition}

\begin{definition}[beta benchmark]\label{def:beta}
    The Hellinger distance between the probability distribution of the quantum computer evolution, $Q$, and the distribution resulting from the noisy simulations, $N$, namely $h_{\text{nm}}(Q,N)$, is the beta benchmark, with notation $\beta_{q|n}$:
    \begin{equation}
        \beta_{q|n} \equiv h(Q,N). \label{eq:beta}
    \end{equation}
\end{definition}

\begin{definition}[gamma benchmark]\label{def:gamma}
    The Hellinger distance between the distribution resulting from the noisy simulations, $N$, and the ideal distribution, $D$, namely $h_{\text{sm}}(N,D)$, is the gamma benchmark, with notation $\gamma_{n|i}$:
    \begin{equation}
        \gamma_{n|i} \equiv h(N,D). \label{eq:gamma}
    \end{equation}
\end{definition}

As mentioned in Section \ref{subsec:hd}, the Hellinger distance is a metric that satisfies the triangle inequality. Hence, since the benchmarks established in the above definitions describe the pairwise Hellinger distances between three probability distributions (the quantum computer, $Q$, the simulated evolution, $N$, and the ideal evolution, $D$), it is possible to derive a relationship between $\alpha_{q|i}$, $\beta_{q|n}$ and $\gamma_{n|i}$ via the triangle inequality, through the following Lemma (whose proof follows simply from the triangle inequality for a metric).

\begin{lemma}[benchmarks triangle inequality]\label{lemma:triangineq}
    Given probability distributions $Q$, $N$ and $D$, as established in the benchmark definitions, the pairwise Hellinger distances between those distributions, i.e., $h(Q,D)$, $h(Q,N)$ and $h(N,D)$ follow the triangle inequality:
    \begin{equation*}
        h(Q,D) \leq h(Q,N) + h(N,D).
    \end{equation*}
    Thus, the relevant benchmarks will also follow the triangle inequality as:
    \begin{equation}
        \alpha_{q|i} \leq \beta_{q|n} + \gamma_{n|i} \label{eq:trianineq}
    \end{equation}
\end{lemma}

The above Lemma effectively means that, according to the benchmark definitions, the deviation of the quantum computer evolution from the ideal ($\alpha_{q|i}$) will never be greater than the sum of the expected (i.e., simulated) evolution derived by the levels of noise within the machine and the distance between the simulated and ideal distributions ($\beta_{q|n}+\gamma_{n|i}$). In other words, defining the benchmark metrics using the Hellinger distance allows us to quantify the confidence on the estimated level of noise during the execution of the quantum circuit.

Following the definitions and relationship between the benchmarks, we present a framework for program benchmarking quantum computers as the sequence of the six steps below:
\begin{description}
    \item[Step 1] \textbf{Select benchmark method(s).} The selection of quantum algorithms that will be used for benchmarking should adhere to the three criteria described in Section \ref{subsec:qalg}: (i) scalability, (ii) predictability and noise susceptibility and (iii) quantum advantage.
    
    \item[Step 2] \textbf{Quantum noise model and simulator.} Select or implement a noise model that approximates the noisy evolution within the quantum machine and a simulator that can execute the noise model.
    
    \item[Step 3] \textbf{Run experiments.} Design and execute a suitable number of experiments of the benchmark method(s) on the quantum computer. This step also includes calibrating the noise parameters that encapsulate the level of noise within the quantum computer at the time of the experiments.
    
    \item[Step 4] \textbf{Simulate the noisy evolution.} Use the noise model in order to simulate the noisy evolution of the quantum computer using the calibrated noise parameters from the time of the experiment, as described in Step 3.
    
    \item[Step 5] \textbf{Simulate the ideal evolution.} This can be done either through simple noise-free simulations or by calculating the probabilities through the quantum statevector.
    
    \item[Step 6] \textbf{Calculate the benchmarks.} The final benchmarks $\alpha_{q|i}$, $\beta_{q|n}$ and $\gamma_{n|i}$ are the Hellinger distances between the quantum computer, the UNM and the ideal evolution in the setting described in Definitions \ref{def:alpha}, \ref{def:beta} and \ref{def:gamma}.
\end{description}


Following the benchmarking framework we can extract meaningful results from a series of comparisons between the benchmark metrics, $\alpha_{q|i}$, $\beta_{q|n}$ and $\gamma_{n|i}$. The $\beta_{q|n}$ benchmark essentially describes how closely the noise model simulates the behaviour of the quantum computer. The $\alpha_{q|i}$ benchmark shows how far the behaviour of the quantum computer falls from the noise-free case thus giving an estimate of the overall computer performance under the effects of noise. From the comparison between the $\beta_{q|n}$ and $\alpha_{q|i}$ benchmarks we can extract valuable information: if $\beta_{q|n}<\alpha_{q|i}$ the noise levels in the quantum computer are closer to the estimated ones from the noise simulations; on the opposite case the computer operates closer to the ideal evolution. In the latter case the quantum computer behaves more efficiently with lower level of noise than expected, thus giving us more confidence regarding the computational result.

The $\gamma_{n|i}$ benchmark, even though it can be used as an estimate of the noise levels in the quantum computer, does not give any relevant information on its own. The value comes when considering the $\gamma_{n|i}$ benchmark together with the $\alpha_{q|i}$ benchmark. First of all, the closer the values of $\alpha_{q|i}$ and $\gamma_{n|i}$ are, the smaller the value of $\beta_{q|n}$ (if $\beta_{q|n}=0$ then $\alpha_{q|i}=\gamma_{n|i}$ and vice versa). Additionally, if $\alpha_{q|i}>\gamma_{n|i}$, then the noise model, and hence the noise parameters, underestimate the levels of noise during the quantum computer evolution, with the opposite being true if $\alpha_{q|i}<\gamma_{n|i}$. Moreover, the absolute difference $\abs{\alpha_{q|i}-\gamma_{n|i}}$ can quantify this noise over- or underestimation. Precisely, the smaller the absolute difference the smaller the error in estimation. This information is useful to the benchmarking process as it further highlights whether the machine is more or less noisy than estimated while also showing the efficiency of the machine calibration techniques.

Finally, using the triangle inequality from Lemma \ref{lemma:triangineq} we can make further interesting remarks. From equation \eqref{eq:trianineq} we find that $\beta_{q|n} \geq \alpha_{q|i} - \gamma_{n|i}$. Furthermore, from the above analysis we know that a comparison between $\alpha_{q|i}$ and $\gamma_{n|i}$ can tell us whether the calibrated noise parameters over- or underestimate the level of noise during the evolution on the quantum computer, and the absolute difference $\abs{\alpha_{q|i}-\gamma_{n|i}}$ can give an indication of the scale of the error in estimation. Importantly, the above inequality does not hold for absolute values when subtraction takes place (i.e., for an inequality of the form $\abs{\beta_{q|n}} \geq \abs{\alpha_{q|i} - \gamma_{n|i}}$ when $\gamma_{n|i}>\alpha_{q|i}$). Considering this, we can interpret the triangle inequality as a measure of the confidence on the calibrated parameters encapsulating a picture of the noise that is accurate enough to provide a good estimation of the quantum evolution, expressed as follows:
\begin{itemize}
    \item If $\beta_{q|n}\geq\abs{\alpha_{q|i}-\gamma_{q|i}}$, then the over- or underestimation of the noise is small enough to provide estimates of the quantum evolution with \textit{high confidence}.
    
    \item If $\beta_{q|n}<\abs{\alpha_{q|i}-\gamma_{q|i}}$, then the calibrated parameters generate \textit{low confidence} on the levels of noise.
\end{itemize}

\section{Experiments and Results}\label{sec:results}

\subsection{Experimental Setup}\label{subsec:setup}
For the benchmarking experiments we run the quantum circuits that implement the chosen quantum algorithms. In the case of the DTQW we run one step of the algorithm (i.e., one coin-flip) on a workspace (i.e., number of active qubits used by the circuit) of three qubits with initial state $\ket{0}$, as previous work shows that this configuration is satisfactory for errors to take place and the behaviour of the quantum walk to evolve in a predictable manner \cite{KGeorgo-2020-rots}. For the CTQW and its PD we use a small two-qubit state-space, as the Pauli decomposition gets excessively large for a three-qubit state space or bigger. The algorithms are implemented for a continuous time of $t=3$. Our QPE routine is tailored to estimate a phase of $\theta = 2\pi/3$ using a workspace of four qubits with theoretical probability of success estimated at $0.688$. Finally, the QS implementation performs an unstructured search for the element $\ket{s}=\ket{10}$ (decimal) within a four qubits state-space i.e., a dataset containing numbers $\ket{0}$ to $\ket{15}$. The QS algorithm is run for three iterations and shows a theoretical probability of success estimated at $0.96$. Table \ref{table:initconfig} shows more comprehensively the initial configuration for each algorithm.

\bgroup
\def\arraystretch{1.2}%
\setlength\tabcolsep{0.32cm}
\begin{table*}[!t]
    \centering
    \begin{tabular}{c|c|c|c|c}
        \hline
        Algorithm & No. Qubits & Size of Workspace & Iterations/Duration & Probability of Success \\ \hline \hline
        DTQW & $2$ & $3$ & $1$ (one coin-flip) & $p_{\text{succ}}=0.5$ in states $\ket{1}$ and $\ket{3}$ \\
        CTQW & $2$ & $2$ & $t=3$ & $p_{\ket{2}}=0.99$, $p_{\ket{1}}=p_{\ket{3}}=0.005$ \\
        PD & $2$ & $2$ & $t=3$ & $p_{\ket{2}}=0.99$, $p_{\ket{1}}=p_{\ket{3}}=0.005$ \\
        QPE & $3$ & $4$ & $1$ & $p_{\text{succ}}=0.688$ in state $\ket{3}$ \\
        QSa & $4$ & $6$ & $3$ & $p_{\text{succ}}=0.96$ in state $\ket{10}$ \\
        QSn & $4$ & $4$ & $3$ & $p_{\text{succ}}=0.96$ in state $\ket{10}$ \\ \hline
    \end{tabular}
    \caption{The initial configuration for the quantum walk (DTQW), continuous-time quantum walk (CTQW) and its Pauli decomposition (PD), quantum phase estimation (QPE) and quantum search algorithms with ancilla (QSa) and without ancilla (QSn) for the benchmarking experiments. No. qubits is the number of qubits in the state space, size of workspace is the number of active qubits utilized by the circuit, iterations is the number of repetitions of the quantum circuit and the probability of success is the theoretical probability of the algorithm to give us the correct (or expected) result.}
    \label{table:initconfig}
\end{table*}
\egroup

\begin{figure*}[!t]
    \begin{center}
          \includegraphics[width=12cm]{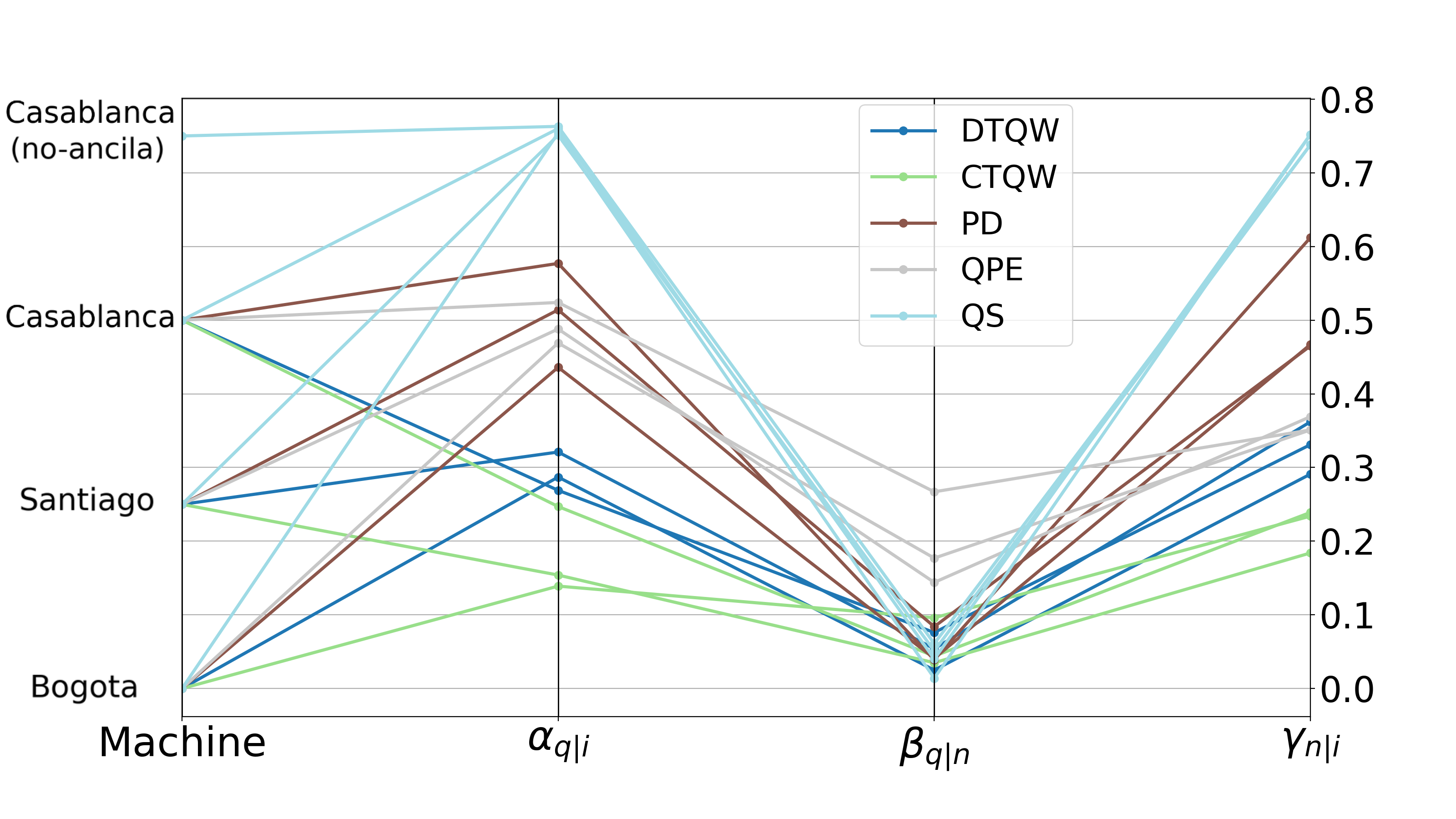} \\
    \end{center}
    \caption{Visualization of the three benchmarks, $\alpha_{q|i}$, $\beta_{q|n}$ and $\gamma_{n|i}$ for each of the three quantum computers (mapped to the left-hand side $y$-axis) when executing the five quantum algorithms.}
    \label{fig:benchmarks}
\end{figure*}

We run each algorithm independently $100{,}000$ times (see Section \ref{subsec:qalg}) on the three chosen computers (see Section \ref{subsec:qcs}) and as a simulation using the UNM (see Section \ref{subsec:unm}) and the ideal simulator \cite{Qiskit} to reproduce the noisy and the ideal quantum evolutions respectively. Finally, we compute the $\beta_{q|n}$, $\alpha_{q|i}$ and $\gamma_{n|i}$ benchmarks, as defined in Section \ref{sec:framework}. The results are shown on Table \ref{table:benchmarks} for each machine and are also visualized in Figure \ref{fig:benchmarks}.

\begin{table*}[!t]
    \begin{tabular}{c}
        \bgroup
        \def\arraystretch{1.2}%
        \setlength\tabcolsep{0.4cm}
            \centering
            \begin{tabular}{c|c|c|c|c|c|c}
                \hline
                Machine & $\alpha_{q|i}$ & $\beta_{q|n}$ & $\gamma_{n|i}$ & $\abs{\alpha_{q|i}-\gamma_{n|i}}$ & QC Runtime ($\text{sec}$) & Sim. Runtime ($\text{sec}$) \\ \hline \hline
                $5$-qubit Bogota & $0.287$ & $0.025$ & $0.291$ & $0.004$ & $240.7$ & $3{,}331.5$ \\
                $5$-qubit Santiago & $0.321$ & $0.054$ & $0.362$ & $0.041$ & $234.2$ & $3{,}119.3$ \\
                $7$-qubit Casablanca & $0.269$ & $0.076$ & $0.331$ & $0.062$ & $254.5$ & $4{,}626.6$ \\ \hline
            \end{tabular}
        \egroup \\ [6pt] (a) Quantum walk algorithm on two qubits. \\ [6pt]
        \bgroup
        \def\arraystretch{1.2}%
        \setlength\tabcolsep{0.4cm}
            \centering
            \begin{tabular}{c|c|c|c|c|c|c}
                \hline
                Machine & $\alpha_{q|i}$ & $\beta_{q|n}$ & $\gamma_{n|i}$ & $\abs{\alpha_{q|i}-\gamma_{n|i}}$ & QC Runtime ($\text{sec}$) & Sim. Runtime ($\text{sec}$) \\ \hline \hline
                $5$-qubit Bogota & $0.139$ & $0.096$ & $0.234$ & $0.095$ & $234.4$ & $3{,}026.9$ \\
                $5$-qubit Santiago & $0.154$ & $0.035$ & $0.184$ & $0.030$ & $301.7$ & $2{,}905.7$ \\
                $7$-qubit Casablanca & $0.247$ & $0.044$ & $0.239$ & $0.008$ & $262.1$ & $3{,}035.2$ \\ \hline
            \end{tabular}
        \egroup \\ (b) Continuous-time quantum walk algorithm on two qubits. \\ [6pt]
        \bgroup
        \def\arraystretch{1.2}%
        \setlength\tabcolsep{0.4cm}
            \centering
            \begin{tabular}{c|c|c|c|c|c|c}
                \hline
                Machine & $\alpha_{q|i}$ & $\beta_{q|n}$ & $\gamma_{n|i}$ & $\abs{\alpha_{q|i}-\gamma_{n|i}}$ & QC Runtime ($\text{sec}$) & Sim. Runtime ($\text{sec}$) \\ \hline \hline
                $5$-qubit Bogota & $0.436$ & $0.041$ & $0.467$ & $0.031$ & $242.7$ & $11{,}764.9$ \\
                $5$-qubit Santiago & $0.514$ & $0.084$ & $0.465$ & $0.049$ & $228.5$ & $11{,}618.6$ \\
                $7$-qubit Casablanca & $0.577$ & $0.039$ & $0.612$ & $0.035$ & $376.4$ & $12{,}861.3$ \\ \hline
            \end{tabular}
        \egroup \\ (c) Pauli decomposition of the CTQW algorithm on two qubits. \\ [6pt]
        \bgroup
        \def\arraystretch{1.2}%
        \setlength\tabcolsep{0.4cm}
            \centering
            \begin{tabular}{c|c|c|c|c|c|c}
                \hline
                Machine & $\alpha_{q|i}$ & $\beta_{q|n}$ & $\gamma_{n|i}$ & $\abs{\alpha_{q|i}-\gamma_{n|i}}$ & QC Runtime ($\text{sec}$) & Sim. Runtime ($\text{sec}$) \\ \hline \hline
                $5$-qubit Bogota & $0.469$ & $0.177$ & $0.351$ & $0.118$ & $254.2$ & $6{,}287.1$ \\
                $5$-qubit Santiago & $0.488$ & $0.144$ & $0.369$ & $0.119$ & $226.8$ & $6{,}304.2$ \\
                $7$-qubit Casablanca & $0.524$ & $0.267$ & $0.351$ & $0.173$ & $268.3$ & $7{,}741.1$ \\ \hline
            \end{tabular}
        \egroup \\ (d) Quantum phase estimation algorithm on three qubits. \\ [6pt]
        \bgroup
        \def\arraystretch{1.2}%
        \setlength\tabcolsep{0.28cm}
            \centering
            \begin{tabular}{c|c|c|c|c|c|c}
                \hline
                Machine & $\alpha_{q|i}$ & $\beta_{q|n}$ & $\gamma_{n|i}$ & $\abs{\alpha_{q|i}-\gamma_{n|i}}$ & QC Runtime ($\text{sec}$) & Sim. Runtime ($\text{sec}$) \\ \hline \hline
                $5$-qubit Bogota & $0.754$ & $0.014$ & $0.752$ & $0.002$ & $270.4$ & $22{,}834.3$ \\
                $5$-qubit Santiago & $0.751$ & $0.051$ & $0.738$ & $0.013$ & $256.7$ & $21{,}923.2$ \\
                $7$-qubit Casablanca (ancilla) & $0.760$ & $0.040$ & $0.752$ & $0.008$ & $280.1$ & $36{,}425.7$ \\
                $7$-qubit Casablanca (no ancilla) & $0.763$ & $0.061$ & $0.738$ & $0.025$ & $274.1$ & $22{,}889.9$ \\ \hline
            \end{tabular}
        \egroup \\ (e) Quantum search algorithm on four qubits. \\
    \end{tabular}
    \caption{Results from benchmarking the three machines for each of the five algorithms. The benchmark indicators $\alpha_{q|i}$, $\beta_{q|n}$ and $\gamma_{n|i}$ are the HD between the quantum computer and the UNM, the quantum computer and the ideal distribution and the UNM and the ideal distribution respectively. QC runtime is the cumulative execution time in seconds for $100{,}000$ iterations of each algorithm on the respective machine; Sim. runtime is the time it takes to simulate for $100{,}000$ iterations the behaviour of each machine when executing the algorithms.}
    \label{table:benchmarks}
\end{table*}

\subsection{Results}\label{subsec:res}
The results of the benchmark experiments are presented in Table \ref{table:benchmarks} and below for each of the algorithms used.

It is important at this point to emphasize an advantage of the structure presented for program benchmarking quantum computers. The use of the Hellinger distance to define the three benchmarks ($\alpha_{q|i}$, $\beta_{q|n}$ and $\gamma_{n|i}$) allows us to perform a comparison of the performance of different quantum algorithms on the same basis, i.e., a dimensionless quantity. In other words, through the proposed framework, apart from measuring the efficiency of a quantum processor, we can gather additional information and compare the efficiency of circuit implementations for the selected algorithms. This is done here for the two approaches on implementing the CTQW on a gate-based computer, one that is done through the IBMQ API and one that use the Pauli decomposition of the CTQW Hamiltonian (see Section \ref{subsec:paulidec}).

\subsubsection{Discrete-time Quantum Walk Benchmarking}\label{subsubsec:dtqwBench}
The DTQW implementation produces a circuit with a relatively small depth and number of gates. Our $\beta_{q|n}$ benchmark measures how closely a quantum computer operates to the expected noise levels (indicated by the noise parameters and simulated by the UNM). Thus, in the small circuit case of the DTQW, the Bogota machine operates much closer to the expected evolution than the other computers with the smallest $\beta_{q|n}$.

The $\alpha_{q|i}$ and $\gamma_{n|i}$ benchmarks lead to a coherent picture of the overall performance of the quantum computers. In the DTQW case, the $\alpha_{q|i}$ benchmark is relatively small compared to the larger algorithms. This means that the quantum computers are not as erroneous as in the larger circuit cases, an expected result. A comparison between the $\alpha_{q|i}$ and $\gamma_{n|i}$ benchmarks shows that $\alpha_{q|i} < \gamma_{n|i}$ for all the machines, indicating that the the calibrated parameters overestimate the levels of noise during the evolution of the circuit. Finally, $\beta_{q|n}\geq\abs{\alpha_{q|i}-\gamma_{q|i}}$ for all quantum computers, which implies high confidence of our noise level estimates.

\subsubsection{Continuous-time Quantum Walk Benchmarking}\label{subsubsec:ctqwBench}
The CTQW circuit is the smallest circuit we use for benchmarking, which leads us to expect that the machine evolution will be reasonably close to the ideal case for this circuit. Evidently, the $\alpha_{q|i}$ benchmark is by far the smallest for the CTQW algorithm on all three machines. As $\alpha_{q|i}<\gamma_{n|i}$ for the Bogota and Santiago machines, we can say that the noise parameters overestimate the noise for those two computers, whereas they slightly underestimate the noise for the Casablanca computer.

For the CTQW, the Santiago machine operates closer to the expected levels of noise according to its error rates, with the smallest $\beta_{q|n}$. Similarly to the DTQW, we are confident in the calibrated noise as $\beta_{q|n}\geq\abs{\alpha_{q|i}-\gamma_{q|i}}$.

\subsubsection{Pauli Decomposition Benchmarking}\label{subsubsec:pdBench}
Unlike the CTQW circuit, its Pauli decomposition is a large and quite deep circuit, which means that we should expect a very noisy evolution. The small $\beta_{q|n}$ values show that the quantum computers operate relatively close to expectations.

As shown by the large values of $\alpha_{q|i}$ for all quantum computers, the PD circuit is quite error-prone compared to the two previous cases of the quantum walk. Thus, an important conclusion we can safely draw is that, for a two-qubit continuous-time quantum walk, the Pauli decomposition of its Hamiltonian leads to a more complex and noisier circuit than the decomposition to base gates done automatically through the IBMQ API. 

The $\gamma_{n|i}$ benchmarks show that the parameters overestimate the noise of the Bogota and Casablanca machines and underestimate the noise within the Santiago computer. Furthermore, $\beta_{q|n}\geq\abs{\alpha_{q|i}-\gamma_{q|i}}$ inspires high-confidence on our noise estimates. This result is important as, through our benchmarks, we produce evidence on the performance of two different techniques for implementing the same algorithm.

\subsubsection{Quantum Phase Estimation Benchmarking}\label{subsubsec:dqpeBench}
Moving on, the QPE circuit is slightly more complex than the DTQW circuit. An overall bigger $\beta_{q|n}$ benchmark on every machine indicates that the QPE circuit execution deviates slightly more from the expected evolution compared to the other algorithms. This could be the result of random fluctuations of the calibrated parameters.

Following the $\alpha_{q|i}$ benchmark, we can extract similar results for the quantum computers as the PD circuit benchmarks, finding that the larger values indicate that the quantum computer performance is hindered by the size of the circuit. Interestingly, we see that $\alpha_{q|i}>\gamma_{n|i}$ for all the quantum computers, showcasing that the calibrated parameters underestimate the noise and we can expect noisier results from all the quantum computers for computations of similar size. We have high-confidence of our calibrated parameters as $\beta_{q|n}\geq\abs{\alpha_{q|i}-\gamma_{q|i}}$ for all experiments.

\subsubsection{Quantum Search Benchmarking}\label{subsubsec:qsBench}
Finally, the quantum search circuit represents the largest implemented algorithm and thus, we expect it to also be the noisiest. Following the above methodology, we find that the $\beta_{q|n}$ benchmark shows the smallest values for each machine, indicating that the quantum computers operated close to the predicted evolutions. The $\alpha_{q|i}$ and $\gamma_{n|i}$ benchmarks are quite close, increasing our confidence on the estimated noise levels. Nevertheless, it is safe to conclude that the machines in the case of the QS exhibit intense levels of noise, as the $\alpha_{q|i}$ benchmarks are very large.

A quick comparison between the $\alpha_{q|i}$ and $\gamma_{n|i}$ shows that the calibrated parameters offer a very good picture of the noise within the quantum computer in this experiment.

\subsubsection{Analysis}\label{subsubsec:conclBenc}
A comparison of the $\alpha_{q|i}$ and $\gamma_{n|i}$ benchmarks shows that when executing the DTQW and CTQW circuits, the machines are less error-prone than the UNM and calibrated parameters imply, with the opposite being true for the PD, QS and QPE. This informs us that the machines are better at executing small circuits, an expected result. Additionally, a larger numerical difference between $\alpha_{q|i}$ and $\gamma_{n|i}$ indicates a more efficient computer behaviour. For example, Casablanca is the most efficient in terms of noise for the DTQW case as $\alpha_{q|i}<\gamma_{n|i}$ with the biggest difference. Similarly, in the QS case the Casablanca machine is the least efficient with $\alpha_{q|i}>\gamma_{n|i}$. Here, as we use two different implementations of the QS algorithm, we can also compare them with each other. The benchmarks show that the QS circuit with ancilla is closer to the expected evolution by the UNM, shown by $\beta_{q|n}$, and to the ideal evolution, shown by $\alpha_{q|i}$, although not by much, shown by $\alpha_{q|i}$ versus $\gamma_{n|i}$. 

The results from the above analysis are crucial as they can lead towards the selection of a machine appropriate to specific circuit needs. Additionally, we provide a comparison between the continuous-time quantum walk circuit and its Pauli decomposition. This fact is clearly reflected on the benchmarks, as the values of $\alpha_{q|i}$ are much lower for the CTQW circuit, while both circuits operate within expectations (small $\beta_{q|n}$ values) and are not massively over- or underestimated by their noise parameters (small $|\alpha_{q|i}-\gamma_{n|i}|$). Thus, we can easily conclude that, for the two-qubit case, a Pauli decomposition of the CTQW Hamiltonian leads to a less efficient circuit.

Overall, we can concentrate the general benchmarking results as follows. In the smallest circuits (i.e., CTQW and DTQW case) the $\beta_{q|n}$ benchmarks show how close the computers operate to the expected levels of noise. The Casablanca machine shows the best $\alpha_{q|i}$ benchmark, i.e., it is the closest to the ideal evolution, closely followed by Bogota and with Santiago being the furthest away. The UNM and the calibrated parameters always overestimate the noise in this case as $\alpha_{q|i}<\gamma_{n|i}$, thus showing that all the quantum computers exhibit relatively low levels of noise when executing small circuits. In the slightly deeper circuit of the QPE, the Bogota machine outperforms the others, followed by Santiago and Casablanca. In this case though the $\beta_{q|n}$ benchmark indicates slightly bigger deviation from the expected level of noise which is biggest on the Casablanca machine. The UNM and calibrated parameters always underestimate the noise in the quantum computer as $\alpha_{q|i}>\gamma_{n|i}$, meaning that the quantum computers are more error-prone. Lastly, in the largest QS and PD circuits, the benchmarks indicate that all the machines operate close to the noise model with low $\beta_{q|n}$ benchmarks and exhibit very low performance as the $\alpha_{q|i}$ benchmarks are large, getting close to $1$ for the QS. The noise model slightly underestimates the noise, but in this case of very deep circuits, the benchmarks show that the machines will not produce any meaningful results, an expected outcome.

A visual representation of the comparison between the distributions resulting from the quantum computer executions and the ideal simulations for each quantum algorithm are shown in Appendix \ref{ap:benchcomp}. A further comparison between the quantum computer and the individual UNM distributions for each machine is given in Appendix \ref{ap:comp}.

\section{Discussion and Conclusions}\label{sec:concl}
In this paper we have presented an approach to benchmarking quantum computers using scaling, high-level quantum algorithms considered as attractive ``real-world" problems. We have defined three benchmark metrics, each highlighting different aspects of the machine's efficiency either as a standalone or through comparisons between them. Each benchmark metric describes the difference between two quantum evolutions and together they follow the triangle inequality.

In order to better present the main characteristics of our benchmarks, we streamline the discussion as a comparison with the vastly used metric of \textit{quantum volume} \cite{Moll-2018}, which quantifies the expected size of a circuit that can be reliably run on a quantum computer. In contrast, our program benchmarks describe the performance of the quantum computer when running a specific circuit itself. This approach has advantages and disadvantages over architecture-neutral benchmarks. First of all, our benchmark metrics show the exact performance of a QPU (and in extend, the quantum computer itself) when running the quantum circuit. Additionally, they highlight the difference with an expected (noisy) simulated evolution and an ideal (noise-free) simulated evolution, a result that better identifies the weaknesses of the machine in a more structural manner. More specifically, the metrics allow us to realise the manner and intensity that the computer deviates from the simulated evolutions. Finally, all three machines we benchmark use the same QPU technology and exhibit the same quantum volume of $32$. On the other hand, our benchmarking process exhibits different metrics and results for each quantum computer. Thus, our benchmarks capture the performance of each QPU more thoroughly and provide us with a more detailed representation of their performance.

On the con side, our benchmarking process is slower compared to the calculation of the quantum volume. This is an expected outcome as we need to run a number of experiments on the quantum computer as well as the noisy simulations. Moreover, the flip-side of the architecture-specific nature of our metrics dictates that each computer will exhibit different benchmarks when executing different algorithms.

Thus, we can conclude that our architecture-specific program benchmarks showcase the performance of a quantum computer in a ``real-world" environment, as well as highlight the performance when running a specific algorithm and carry out meaningful comparisons between related circuits (e.g., the CTQW vs the Pauli decomposition of its Hamiltonian). Additionally, we can gather information on whether the calibrated parameters over- or underestimate the levels of noise during the quantum evolution. On the other hand, architecture-neutral benchmarks like the quantum volume are more generic and excel at showcasing the limitations of QPUs when running arbitrary quantum circuits.

It is also noteworthy that the benchmarking experiments carried out during this work are not suitable for a comparative analysis regarding the efficiency of each qubit topology (see Figure \ref{fig:architectures}). One reason for this is the relatively small size and limited flexibility of the computers themselves. Nevertheless, when benchmarking larger quantum computers in the future, we believe that using our framework, one could focus, for example, on searching which architectures are more efficient for a specific task.

In conclusion, our work has shown that using quantum algorithms to benchmark quantum computers in a well-structured environment can stress different aspects and very informatively highlight the performance of a quantum computer. Additionally, it provides a way to compare the efficiency of different circuits used to implement the same task. To the best of our knowledge, this is the first work that uses a continuous-time quantum algorithm to benchmark the performance of a digital quantum machine. The results show that, for small state spaces of the continuous-time quantum walk, the Pauli decomposition does not produce an efficient circuit. This result is expected as the complexity of a Hamiltonian operating on two qubits is very small.

\section{Acknowledgements}
This work was supported by the Engineering and Physical Sciences Research Council, Centre for Doctoral Training in Cloud Computing for Big Data, United Kingdom [grant number EP/L015358/1].

\bibliography{bibphy}{}

\begin{thebibliography}{47}%
\makeatletter
\providecommand \@ifxundefined [1]{%
 \@ifx{#1\undefined}
}%
\providecommand \@ifnum [1]{%
 \ifnum #1\expandafter \@firstoftwo
 \else \expandafter \@secondoftwo
 \fi
}%
\providecommand \@ifx [1]{%
 \ifx #1\expandafter \@firstoftwo
 \else \expandafter \@secondoftwo
 \fi
}%
\providecommand \natexlab [1]{#1}%
\providecommand \enquote  [1]{``#1''}%
\providecommand \bibnamefont  [1]{#1}%
\providecommand \bibfnamefont [1]{#1}%
\providecommand \citenamefont [1]{#1}%
\providecommand \href@noop [0]{\@secondoftwo}%
\providecommand \href [0]{\begingroup \@sanitize@url \@href}%
\providecommand \@href[1]{\@@startlink{#1}\@@href}%
\providecommand \@@href[1]{\endgroup#1\@@endlink}%
\providecommand \@sanitize@url [0]{\catcode `\\12\catcode `\$12\catcode
  `\&12\catcode `\#12\catcode `\^12\catcode `\_12\catcode `\%12\relax}%
\providecommand \@@startlink[1]{}%
\providecommand \@@endlink[0]{}%
\providecommand \url  [0]{\begingroup\@sanitize@url \@url }%
\providecommand \@url [1]{\endgroup\@href {#1}{\urlprefix }}%
\providecommand \urlprefix  [0]{URL }%
\providecommand \Eprint [0]{\href }%
\providecommand \doibase [0]{https://doi.org/}%
\providecommand \selectlanguage [0]{\@gobble}%
\providecommand \bibinfo  [0]{\@secondoftwo}%
\providecommand \bibfield  [0]{\@secondoftwo}%
\providecommand \translation [1]{[#1]}%
\providecommand \BibitemOpen [0]{}%
\providecommand \bibitemStop [0]{}%
\providecommand \bibitemNoStop [0]{.\EOS\space}%
\providecommand \EOS [0]{\spacefactor3000\relax}%
\providecommand \BibitemShut  [1]{\csname bibitem#1\endcsname}%
\let\auto@bib@innerbib\@empty
\bibitem [{\citenamefont {Preskill}(2018)}]{Preskill-2018}%
  \BibitemOpen
  \bibfield  {author} {\bibinfo {author} {\bibfnamefont {J.}~\bibnamefont
  {Preskill}},\ }\bibfield  {title} {\bibinfo {title} {Quantum computing in the
  {NISQ} era and beyond},\ }\href {https://doi.org/10.22331/q-2018-08-06-79}
  {\bibfield  {journal} {\bibinfo  {journal} {Quantum}\ }\textbf {\bibinfo
  {volume} {2}},\ \bibinfo {pages} {79} (\bibinfo {year} {2018})}\BibitemShut
  {NoStop}%
\bibitem [{\citenamefont {Bernstein}\ and\ \citenamefont
  {Vazirani}(1997)}]{BernsteinVazirani}%
  \BibitemOpen
  \bibfield  {author} {\bibinfo {author} {\bibfnamefont {E.}~\bibnamefont
  {Bernstein}}\ and\ \bibinfo {author} {\bibfnamefont {U.}~\bibnamefont
  {Vazirani}},\ }\bibfield  {title} {\bibinfo {title} {Quantum complexity
  theory},\ }\href {https://doi.org/10.1137/S0097539796300921} {\bibfield
  {journal} {\bibinfo  {journal} {SIAM Journal on Computing}\ }\textbf
  {\bibinfo {volume} {26}},\ \bibinfo {pages} {1411} (\bibinfo {year}
  {1997})}\BibitemShut {NoStop}%
\bibitem [{\citenamefont {van Dam}\ \emph {et~al.}(2006)\citenamefont {van
  Dam}, \citenamefont {Hallgren},\ and\ \citenamefont {Ip}}]{vanDam-2006}%
  \BibitemOpen
  \bibfield  {author} {\bibinfo {author} {\bibfnamefont {W.}~\bibnamefont {van
  Dam}}, \bibinfo {author} {\bibfnamefont {S.}~\bibnamefont {Hallgren}},\ and\
  \bibinfo {author} {\bibfnamefont {L.}~\bibnamefont {Ip}},\ }\bibfield
  {title} {\bibinfo {title} {Quantum algorithms for some hidden shift
  problems},\ }\href {https://doi.org/10.1137/S009753970343141X} {\bibfield
  {journal} {\bibinfo  {journal} {SIAM Journal on Computing}\ }\textbf
  {\bibinfo {volume} {36}},\ \bibinfo {pages} {763} (\bibinfo {year}
  {2006})}\BibitemShut {NoStop}%
\bibitem [{\citenamefont {R{\"o}tteler}(2010)}]{Rotteler-2010}%
  \BibitemOpen
  \bibfield  {author} {\bibinfo {author} {\bibfnamefont {M.}~\bibnamefont
  {R{\"o}tteler}},\ }\bibinfo {title} {Quantum algorithms for highly non-linear
  boolean functions},\ in\ \href {https://doi.org/10.1137/1.9781611973075.37}
  {\emph {\bibinfo {booktitle} {Proceedings of the 2010 Annual ACM-SIAM
  Symposium on Discrete Algorithms}}}\ (\bibinfo {year} {2010})\ pp.\ \bibinfo
  {pages} {448--457}\BibitemShut {NoStop}%
\bibitem [{\citenamefont {Wright}\ \emph {et~al.}(2019)\citenamefont {Wright},
  \citenamefont {Beck}, \citenamefont {Debnath},\ and\ \citenamefont {et.
  al.}}]{Wright-2019}%
  \BibitemOpen
  \bibfield  {author} {\bibinfo {author} {\bibfnamefont {K.}~\bibnamefont
  {Wright}}, \bibinfo {author} {\bibfnamefont {K.~M.}\ \bibnamefont {Beck}},
  \bibinfo {author} {\bibfnamefont {S.}~\bibnamefont {Debnath}},\ and\ \bibinfo
  {author} {\bibnamefont {et. al.}},\ }\bibfield  {title} {\bibinfo {title}
  {Benchmarking an 11-qubit quantum computer},\ }\href
  {https://doi.org/10.1038/s41467-019-13534-2} {\bibfield  {journal} {\bibinfo
  {journal} {Nature Communications}\ }\textbf {\bibinfo {volume} {10}},\
  \bibinfo {pages} {5464} (\bibinfo {year} {2019})}\BibitemShut {NoStop}%
\bibitem [{\citenamefont {Arute}\ \emph {et~al.}(2019)\citenamefont {Arute},
  \citenamefont {Arya}, \citenamefont {Babbush},\ and\ \citenamefont {et.
  al.}}]{GoogleSuprem}%
  \BibitemOpen
  \bibfield  {author} {\bibinfo {author} {\bibfnamefont {F.}~\bibnamefont
  {Arute}}, \bibinfo {author} {\bibfnamefont {K.}~\bibnamefont {Arya}},
  \bibinfo {author} {\bibfnamefont {R.}~\bibnamefont {Babbush}},\ and\ \bibinfo
  {author} {\bibnamefont {et. al.}},\ }\bibfield  {title} {\bibinfo {title}
  {Quantum supremacy using a programmable superconducting processor},\ }\href
  {https://doi.org/10.1038/s41586-019-1666-5} {\bibfield  {journal} {\bibinfo
  {journal} {Nature}\ }\textbf {\bibinfo {volume} {574}},\ \bibinfo {pages}
  {505} (\bibinfo {year} {2019})}\BibitemShut {NoStop}%
\bibitem [{\citenamefont {Linke}\ \emph {et~al.}(2017)\citenamefont {Linke},
  \citenamefont {Maslov}, \citenamefont {Roetteler}, \citenamefont {Debnath},
  \citenamefont {Figgatt}, \citenamefont {Landsman}, \citenamefont {Wright},\
  and\ \citenamefont {Monroe}}]{Linke-2017}%
  \BibitemOpen
  \bibfield  {author} {\bibinfo {author} {\bibfnamefont {N.~M.}\ \bibnamefont
  {Linke}}, \bibinfo {author} {\bibfnamefont {D.}~\bibnamefont {Maslov}},
  \bibinfo {author} {\bibfnamefont {M.}~\bibnamefont {Roetteler}}, \bibinfo
  {author} {\bibfnamefont {S.}~\bibnamefont {Debnath}}, \bibinfo {author}
  {\bibfnamefont {C.}~\bibnamefont {Figgatt}}, \bibinfo {author} {\bibfnamefont
  {K.~A.}\ \bibnamefont {Landsman}}, \bibinfo {author} {\bibfnamefont
  {K.}~\bibnamefont {Wright}},\ and\ \bibinfo {author} {\bibfnamefont
  {C.}~\bibnamefont {Monroe}},\ }\bibfield  {title} {\bibinfo {title}
  {Experimental comparison of two quantum computing architectures},\ }\href
  {https://doi.org/10.1073/pnas.1618020114} {\bibfield  {journal} {\bibinfo
  {journal} {Proceedings of the National Academy of Sciences of the United
  States of America}\ }\textbf {\bibinfo {volume} {114}},\ \bibinfo {pages}
  {3305} (\bibinfo {year} {2017})}\BibitemShut {NoStop}%
\bibitem [{\citenamefont {Lilja}(2000)}]{Lilja-2000}%
  \BibitemOpen
  \bibfield  {author} {\bibinfo {author} {\bibfnamefont {D.~J.}\ \bibnamefont
  {Lilja}},\ }\href {https://doi.org/10.1017/CBO9780511612398} {\emph {\bibinfo
  {title} {Measuring Computer Performance: A Practitioner's Guide}}}\ (\bibinfo
   {publisher} {Cambridge University Press},\ \bibinfo {year}
  {2000})\BibitemShut {NoStop}%
\bibitem [{\citenamefont {Martonosi}\ and\ \citenamefont
  {Roetteler}(2019)}]{Martonosi-2019}%
  \BibitemOpen
  \bibfield  {author} {\bibinfo {author} {\bibfnamefont {M.}~\bibnamefont
  {Martonosi}}\ and\ \bibinfo {author} {\bibfnamefont {M.}~\bibnamefont
  {Roetteler}},\ }\href@noop {} {\bibinfo {title} {Next steps in quantum
  computing: Computer science's role}} (\bibinfo {year} {2019}),\ \Eprint
  {https://arxiv.org/abs/1903.10541} {arXiv:1903.10541 [cs.ET]} \BibitemShut
  {NoStop}%
\bibitem [{\citenamefont {{Gudder}}(2008)}]{Gudder-2008}%
  \BibitemOpen
  \bibfield  {author} {\bibinfo {author} {\bibfnamefont {S.}~\bibnamefont
  {{Gudder}}},\ }\bibfield  {title} {\bibinfo {title} {{Quantum Markov
  chains}},\ }\href {https://doi.org/10.1063/1.2953952} {\bibfield  {journal}
  {\bibinfo  {journal} {Journal of Mathematical Physics}\ }\textbf {\bibinfo
  {volume} {49}},\ \bibinfo {eid} {072105} (\bibinfo {year}
  {2008})}\BibitemShut {NoStop}%
\bibitem [{\citenamefont {Shor}(1997)}]{Shor-1997}%
  \BibitemOpen
  \bibfield  {author} {\bibinfo {author} {\bibfnamefont {P.~W.}\ \bibnamefont
  {Shor}},\ }\bibfield  {title} {\bibinfo {title} {Polynomial-time algorithms
  for prime factorization and discrete logarithms on a quantum computer},\
  }\href {https://doi.org/10.1137/s0097539795293172} {\bibfield  {journal}
  {\bibinfo  {journal} {SIAM Journal on Computing}\ }\textbf {\bibinfo {volume}
  {26}},\ \bibinfo {pages} {1484} (\bibinfo {year} {1997})}\BibitemShut
  {NoStop}%
\bibitem [{\citenamefont {Grover}(1996)}]{Grover-1996}%
  \BibitemOpen
  \bibfield  {author} {\bibinfo {author} {\bibfnamefont {L.~K.}\ \bibnamefont
  {Grover}},\ }\bibfield  {title} {\bibinfo {title} {A fast quantum mechanical
  algorithm for database search},\ }in\ \href
  {https://doi.org/10.1145/237814.237866} {\emph {\bibinfo {booktitle}
  {Proceedings of the Twenty-Eighth Annual ACM Symposium on Theory of
  Computing}}},\ \bibinfo {series and number} {STOC '96}\ (\bibinfo
  {publisher} {Association for Computing Machinery},\ \bibinfo {address} {New
  York, NY, USA},\ \bibinfo {year} {1996})\ pp.\ \bibinfo {pages}
  {212--219}\BibitemShut {NoStop}%
\bibitem [{\citenamefont {Olson}\ \emph {et~al.}(2017)\citenamefont {Olson},
  \citenamefont {Cao}, \citenamefont {Romero}, \citenamefont {Johnson},
  \citenamefont {Dallaire-Demers}, \citenamefont {Sawaya}, \citenamefont
  {Narang}, \citenamefont {Kivlichan}, \citenamefont {Wasielewski},\ and\
  \citenamefont {Aspuru-Guzik}}]{Olson-2017}%
  \BibitemOpen
  \bibfield  {author} {\bibinfo {author} {\bibfnamefont {J.}~\bibnamefont
  {Olson}}, \bibinfo {author} {\bibfnamefont {Y.}~\bibnamefont {Cao}}, \bibinfo
  {author} {\bibfnamefont {J.}~\bibnamefont {Romero}}, \bibinfo {author}
  {\bibfnamefont {P.}~\bibnamefont {Johnson}}, \bibinfo {author} {\bibfnamefont
  {P.-L.}\ \bibnamefont {Dallaire-Demers}}, \bibinfo {author} {\bibfnamefont
  {N.}~\bibnamefont {Sawaya}}, \bibinfo {author} {\bibfnamefont
  {P.}~\bibnamefont {Narang}}, \bibinfo {author} {\bibfnamefont
  {I.}~\bibnamefont {Kivlichan}}, \bibinfo {author} {\bibfnamefont
  {M.}~\bibnamefont {Wasielewski}},\ and\ \bibinfo {author} {\bibfnamefont
  {A.}~\bibnamefont {Aspuru-Guzik}},\ }\href@noop {} {\bibinfo {title} {Quantum
  information and computation for chemistry}} (\bibinfo {year} {2017}),\
  \Eprint {https://arxiv.org/abs/1706.05413} {arXiv:1706.05413 [quant-ph]}
  \BibitemShut {NoStop}%
\bibitem [{\citenamefont {McArdle}\ \emph {et~al.}(2020)\citenamefont
  {McArdle}, \citenamefont {Endo}, \citenamefont {Aspuru-Guzik}, \citenamefont
  {Benjamin},\ and\ \citenamefont {Yuan}}]{McArdle-2020}%
  \BibitemOpen
  \bibfield  {author} {\bibinfo {author} {\bibfnamefont {S.}~\bibnamefont
  {McArdle}}, \bibinfo {author} {\bibfnamefont {S.}~\bibnamefont {Endo}},
  \bibinfo {author} {\bibfnamefont {A.}~\bibnamefont {Aspuru-Guzik}}, \bibinfo
  {author} {\bibfnamefont {S.~C.}\ \bibnamefont {Benjamin}},\ and\ \bibinfo
  {author} {\bibfnamefont {X.}~\bibnamefont {Yuan}},\ }\bibfield  {title}
  {\bibinfo {title} {Quantum computational chemistry},\ }\href
  {https://doi.org/10.1103/RevModPhys.92.015003} {\bibfield  {journal}
  {\bibinfo  {journal} {Rev. Mod. Phys.}\ }\textbf {\bibinfo {volume} {92}},\
  \bibinfo {pages} {015003} (\bibinfo {year} {2020})}\BibitemShut {NoStop}%
\bibitem [{\citenamefont {Cao}\ \emph {et~al.}(2019)\citenamefont {Cao},
  \citenamefont {Romero}, \citenamefont {Olson}, \citenamefont {Degroote},
  \citenamefont {Johnson}, \citenamefont {Kieferov{\'a}}, \citenamefont
  {Kivlichan}, \citenamefont {Menke}, \citenamefont {Peropadre}, \citenamefont
  {Sawaya}, \citenamefont {Sim}, \citenamefont {Veis},\ and\ \citenamefont
  {Aspuru-Guzik}}]{Yudong-2019}%
  \BibitemOpen
  \bibfield  {author} {\bibinfo {author} {\bibfnamefont {Y.}~\bibnamefont
  {Cao}}, \bibinfo {author} {\bibfnamefont {J.}~\bibnamefont {Romero}},
  \bibinfo {author} {\bibfnamefont {J.~P.}\ \bibnamefont {Olson}}, \bibinfo
  {author} {\bibfnamefont {M.}~\bibnamefont {Degroote}}, \bibinfo {author}
  {\bibfnamefont {P.~D.}\ \bibnamefont {Johnson}}, \bibinfo {author}
  {\bibfnamefont {M.}~\bibnamefont {Kieferov{\'a}}}, \bibinfo {author}
  {\bibfnamefont {I.~D.}\ \bibnamefont {Kivlichan}}, \bibinfo {author}
  {\bibfnamefont {T.}~\bibnamefont {Menke}}, \bibinfo {author} {\bibfnamefont
  {B.}~\bibnamefont {Peropadre}}, \bibinfo {author} {\bibfnamefont {N.~P.~D.}\
  \bibnamefont {Sawaya}}, \bibinfo {author} {\bibfnamefont {S.}~\bibnamefont
  {Sim}}, \bibinfo {author} {\bibfnamefont {L.}~\bibnamefont {Veis}},\ and\
  \bibinfo {author} {\bibfnamefont {A.}~\bibnamefont {Aspuru-Guzik}},\
  }\bibfield  {title} {\bibinfo {title} {Quantum chemistry in the age of
  quantum computing},\ }\bibfield  {booktitle} {\emph {\bibinfo {booktitle}
  {Chemical Reviews}},\ }\href {https://doi.org/10.1021/acs.chemrev.8b00803}
  {\bibfield  {journal} {\bibinfo  {journal} {Chemical Reviews}\ }\textbf
  {\bibinfo {volume} {119}},\ \bibinfo {pages} {10856} (\bibinfo {year}
  {2019})}\BibitemShut {NoStop}%
\bibitem [{\citenamefont {Schuld}\ \emph {et~al.}(2020)\citenamefont {Schuld},
  \citenamefont {Bocharov}, \citenamefont {Svore},\ and\ \citenamefont
  {Wiebe}}]{Schuld-2020}%
  \BibitemOpen
  \bibfield  {author} {\bibinfo {author} {\bibfnamefont {M.}~\bibnamefont
  {Schuld}}, \bibinfo {author} {\bibfnamefont {A.}~\bibnamefont {Bocharov}},
  \bibinfo {author} {\bibfnamefont {K.~M.}\ \bibnamefont {Svore}},\ and\
  \bibinfo {author} {\bibfnamefont {N.}~\bibnamefont {Wiebe}},\ }\bibfield
  {title} {\bibinfo {title} {Circuit-centric quantum classifiers},\ }\href
  {https://doi.org/10.1103/PhysRevA.101.032308} {\bibfield  {journal} {\bibinfo
   {journal} {Phys. Rev. A}\ }\textbf {\bibinfo {volume} {101}},\ \bibinfo
  {pages} {032308} (\bibinfo {year} {2020})}\BibitemShut {NoStop}%
\bibitem [{\citenamefont {Wecker}\ \emph {et~al.}(2015)\citenamefont {Wecker},
  \citenamefont {Hastings},\ and\ \citenamefont {Troyer}}]{Wecker-2015}%
  \BibitemOpen
  \bibfield  {author} {\bibinfo {author} {\bibfnamefont {D.}~\bibnamefont
  {Wecker}}, \bibinfo {author} {\bibfnamefont {M.~B.}\ \bibnamefont
  {Hastings}},\ and\ \bibinfo {author} {\bibfnamefont {M.}~\bibnamefont
  {Troyer}},\ }\bibfield  {title} {\bibinfo {title} {Progress towards practical
  quantum variational algorithms},\ }\href
  {https://doi.org/10.1103/PhysRevA.92.042303} {\bibfield  {journal} {\bibinfo
  {journal} {Phys. Rev. A}\ }\textbf {\bibinfo {volume} {92}},\ \bibinfo
  {pages} {042303} (\bibinfo {year} {2015})}\BibitemShut {NoStop}%
\bibitem [{\citenamefont {Farhi}\ \emph {et~al.}(2014)\citenamefont {Farhi},
  \citenamefont {Goldstone},\ and\ \citenamefont {Gutmann}}]{Fahri-2014}%
  \BibitemOpen
  \bibfield  {author} {\bibinfo {author} {\bibfnamefont {E.}~\bibnamefont
  {Farhi}}, \bibinfo {author} {\bibfnamefont {J.}~\bibnamefont {Goldstone}},\
  and\ \bibinfo {author} {\bibfnamefont {S.}~\bibnamefont {Gutmann}},\
  }\href@noop {} {\bibinfo {title} {A quantum approximate optimization
  algorithm}} (\bibinfo {year} {2014}),\ \Eprint
  {https://arxiv.org/abs/1411.4028} {arXiv:1411.4028 [quant-ph]} \BibitemShut
  {NoStop}%
\bibitem [{\citenamefont {Peruzzo}\ \emph {et~al.}(2014)\citenamefont
  {Peruzzo}, \citenamefont {McClean}, \citenamefont {Shadbolt}, \citenamefont
  {Yung}, \citenamefont {Zhou}, \citenamefont {Love}, \citenamefont
  {Aspuru-Guzik},\ and\ \citenamefont {O'Brien}}]{Perruzo-2014}%
  \BibitemOpen
  \bibfield  {author} {\bibinfo {author} {\bibfnamefont {A.}~\bibnamefont
  {Peruzzo}}, \bibinfo {author} {\bibfnamefont {J.}~\bibnamefont {McClean}},
  \bibinfo {author} {\bibfnamefont {P.}~\bibnamefont {Shadbolt}}, \bibinfo
  {author} {\bibfnamefont {M.-H.}\ \bibnamefont {Yung}}, \bibinfo {author}
  {\bibfnamefont {X.-Q.}\ \bibnamefont {Zhou}}, \bibinfo {author}
  {\bibfnamefont {P.~J.}\ \bibnamefont {Love}}, \bibinfo {author}
  {\bibfnamefont {A.}~\bibnamefont {Aspuru-Guzik}},\ and\ \bibinfo {author}
  {\bibfnamefont {J.~L.}\ \bibnamefont {O'Brien}},\ }\bibfield  {title}
  {\bibinfo {title} {A variational eigenvalue solver on a photonic quantum
  processor},\ }\href {https://doi.org/10.1038/ncomms5213} {\bibfield
  {journal} {\bibinfo  {journal} {Nature Communications}\ }\textbf {\bibinfo
  {volume} {5}},\ \bibinfo {pages} {4213} (\bibinfo {year} {2014})}\BibitemShut
  {NoStop}%
\bibitem [{\citenamefont {Szegedy}(2004)}]{Szegedy}%
  \BibitemOpen
  \bibfield  {author} {\bibinfo {author} {\bibfnamefont {M.}~\bibnamefont
  {Szegedy}},\ }\bibfield  {title} {\bibinfo {title} {Quantum speed-up of
  {M}arkov chain based algorithms},\ }in\ \href
  {https://doi.org/10.1109/FOCS.2004.53} {\emph {\bibinfo {booktitle}
  {Proceedings of the 45th Annual IEEE Symposium on Foundations of Computer
  Science}}},\ \bibinfo {series and number} {FOCS '04}\ (\bibinfo  {publisher}
  {IEEE Computer Society},\ \bibinfo {address} {Washington, DC, USA},\ \bibinfo
  {year} {2004})\ pp.\ \bibinfo {pages} {32--41}\BibitemShut {NoStop}%
\bibitem [{\citenamefont {Aharonov}\ \emph {et~al.}(2001)\citenamefont
  {Aharonov}, \citenamefont {Ambainis}, \citenamefont {Kempe},\ and\
  \citenamefont {Vazirani}}]{Aharonov-2000}%
  \BibitemOpen
  \bibfield  {author} {\bibinfo {author} {\bibfnamefont {D.}~\bibnamefont
  {Aharonov}}, \bibinfo {author} {\bibfnamefont {A.}~\bibnamefont {Ambainis}},
  \bibinfo {author} {\bibfnamefont {J.}~\bibnamefont {Kempe}},\ and\ \bibinfo
  {author} {\bibfnamefont {U.}~\bibnamefont {Vazirani}},\ }\bibfield  {title}
  {\bibinfo {title} {Quantum walks on graphs},\ }in\ \href
  {https://doi.org/10.1145/380752.380758} {\emph {\bibinfo {booktitle}
  {Proceedings of the Thirty-Third Annual ACM Symposium on Theory of
  Computing}}},\ \bibinfo {series and number} {STOC '01}\ (\bibinfo
  {publisher} {Association for Computing Machinery},\ \bibinfo {address} {New
  York, NY, USA},\ \bibinfo {year} {2001})\ pp.\ \bibinfo {pages}
  {50--59}\BibitemShut {NoStop}%
\bibitem [{\citenamefont {Farhi}\ and\ \citenamefont
  {Gutmann}(1998)}]{Fahri-1998-ctqw}%
  \BibitemOpen
  \bibfield  {author} {\bibinfo {author} {\bibfnamefont {E.}~\bibnamefont
  {Farhi}}\ and\ \bibinfo {author} {\bibfnamefont {S.}~\bibnamefont
  {Gutmann}},\ }\bibfield  {title} {\bibinfo {title} {Quantum computation and
  decision trees},\ }\href {https://doi.org/10.1103/PhysRevA.58.915} {\bibfield
   {journal} {\bibinfo  {journal} {Phys. Rev. A}\ }\textbf {\bibinfo {volume}
  {58}},\ \bibinfo {pages} {915} (\bibinfo {year} {1998})}\BibitemShut
  {NoStop}%
\bibitem [{\citenamefont {Tansuwannont}\ \emph {et~al.}(2019)\citenamefont
  {Tansuwannont}, \citenamefont {Limkumnerd}, \citenamefont {Suwanna},\ and\
  \citenamefont {Kalasuwan}}]{qpe-2019}%
  \BibitemOpen
  \bibfield  {author} {\bibinfo {author} {\bibfnamefont {T.}~\bibnamefont
  {Tansuwannont}}, \bibinfo {author} {\bibfnamefont {S.}~\bibnamefont
  {Limkumnerd}}, \bibinfo {author} {\bibfnamefont {S.}~\bibnamefont
  {Suwanna}},\ and\ \bibinfo {author} {\bibfnamefont {P.}~\bibnamefont
  {Kalasuwan}},\ }\bibfield  {title} {\bibinfo {title} {Quantum phase
  estimation algorithm for finding polynomial roots},\ }\href
  {https://doi.org/doi:10.1515/phys-2019-0087} {\bibfield  {journal} {\bibinfo
  {journal} {Open Physics}\ }\textbf {\bibinfo {volume} {17}},\ \bibinfo
  {pages} {839} (\bibinfo {year} {2019})}\BibitemShut {NoStop}%
\bibitem [{Qis(2021)}]{Qiskit}%
  \BibitemOpen
  \href@noop {} {\bibinfo {title} {Ibm qiskit}},\ \bibinfo {howpublished}
  {\url{https://qiskit.org/}} (\bibinfo {year} {Last accessed April
  2021})\BibitemShut {NoStop}%
\bibitem [{IBM(2021)}]{IBMQExp}%
  \BibitemOpen
  \href@noop {} {\bibinfo {title} {Ibm quantum experience}},\ \bibinfo
  {howpublished}
  {\url{https://www.ibm.com/quantum-computing/technology/experience}} (\bibinfo
  {year} {Last accessed April 2021})\BibitemShut {NoStop}%
\bibitem [{\citenamefont {Georgopoulos}\ \emph
  {et~al.}(2021{\natexlab{a}})\citenamefont {Georgopoulos}, \citenamefont
  {Emary},\ and\ \citenamefont {Zuliani}}]{KGeorgo-2020-UNM}%
  \BibitemOpen
  \bibfield  {author} {\bibinfo {author} {\bibfnamefont {K.}~\bibnamefont
  {Georgopoulos}}, \bibinfo {author} {\bibfnamefont {C.}~\bibnamefont
  {Emary}},\ and\ \bibinfo {author} {\bibfnamefont {P.}~\bibnamefont
  {Zuliani}},\ }\href@noop {} {\bibinfo {title} {Modelling and simulating the
  noisy behaviour of near-term quantum computers}} (\bibinfo {year}
  {2021}{\natexlab{a}}),\ \Eprint {https://arxiv.org/abs/2101.02109}
  {arXiv:2101.02109 [quant-ph]} \BibitemShut {NoStop}%
\bibitem [{\citenamefont {Georgopoulos}\ \emph
  {et~al.}(2021{\natexlab{b}})\citenamefont {Georgopoulos}, \citenamefont
  {Emary},\ and\ \citenamefont {Zuliani}}]{KGeorgo-2020-rots}%
  \BibitemOpen
  \bibfield  {author} {\bibinfo {author} {\bibfnamefont {K.}~\bibnamefont
  {Georgopoulos}}, \bibinfo {author} {\bibfnamefont {C.}~\bibnamefont
  {Emary}},\ and\ \bibinfo {author} {\bibfnamefont {P.}~\bibnamefont
  {Zuliani}},\ }\bibfield  {title} {\bibinfo {title} {Comparison of
  quantum-walk implementations on noisy intermediate-scale quantum computers},\
  }\href {http://dx.doi.org/10.1103/PhysRevA.103.022408} {\bibfield  {journal}
  {\bibinfo  {journal} {Physical Review A}\ }\textbf {\bibinfo {volume} {103}}
  (\bibinfo {year} {2021}{\natexlab{b}})}\BibitemShut {NoStop}%
\bibitem [{\citenamefont {Kempe}(2003)}]{Kempe-2003}%
  \BibitemOpen
  \bibfield  {author} {\bibinfo {author} {\bibfnamefont {J.}~\bibnamefont
  {Kempe}},\ }\bibfield  {title} {\bibinfo {title} {Quantum random walks: An
  introductory overview},\ }\href
  {https://doi.org/10.1080/00107151031000110776} {\bibfield  {journal}
  {\bibinfo  {journal} {Contemporary Physics}\ }\textbf {\bibinfo {volume}
  {44}},\ \bibinfo {pages} {307} (\bibinfo {year} {2003})}\BibitemShut
  {NoStop}%
\bibitem [{\citenamefont {Reitzner}\ \emph {et~al.}(2011)\citenamefont
  {Reitzner}, \citenamefont {Nagaj},\ and\ \citenamefont
  {Bu\v{z}ek}}]{Reitzner-2011-mod}%
  \BibitemOpen
  \bibfield  {author} {\bibinfo {author} {\bibfnamefont {D.}~\bibnamefont
  {Reitzner}}, \bibinfo {author} {\bibfnamefont {D.}~\bibnamefont {Nagaj}},\
  and\ \bibinfo {author} {\bibfnamefont {V.}~\bibnamefont {Bu\v{z}ek}},\
  }\bibfield  {title} {\bibinfo {title} {Quantum walks},\ }\href
  {https://doi.org/10.2478/v10155-011-0006-6} {\bibfield  {journal} {\bibinfo
  {journal} {Acta Physica Slovaca Reviews and Tutorials}\ }\textbf {\bibinfo
  {volume} {61}},\ \bibinfo {pages} {603} (\bibinfo {year} {2011})}\BibitemShut
  {NoStop}%
\bibitem [{\citenamefont {Chakraborty}\ \emph {et~al.}(2020)\citenamefont
  {Chakraborty}, \citenamefont {Novo},\ and\ \citenamefont
  {Roland}}]{Chakra-2020}%
  \BibitemOpen
  \bibfield  {author} {\bibinfo {author} {\bibfnamefont {S.}~\bibnamefont
  {Chakraborty}}, \bibinfo {author} {\bibfnamefont {L.}~\bibnamefont {Novo}},\
  and\ \bibinfo {author} {\bibfnamefont {J.}~\bibnamefont {Roland}},\
  }\bibfield  {title} {\bibinfo {title} {Finding a marked node on any graph via
  continuous-time quantum walks},\ }\href
  {https://doi.org/10.1103/PhysRevA.102.022227} {\bibfield  {journal} {\bibinfo
   {journal} {Phys. Rev. A}\ }\textbf {\bibinfo {volume} {102}},\ \bibinfo
  {pages} {022227} (\bibinfo {year} {2020})}\BibitemShut {NoStop}%
\bibitem [{\citenamefont {Ambainis}\ \emph {et~al.}(2020)\citenamefont
  {Ambainis}, \citenamefont {Gily\'{e}n}, \citenamefont {Jeffery},\ and\
  \citenamefont {Kokainis}}]{Ambainis-2020}%
  \BibitemOpen
  \bibfield  {author} {\bibinfo {author} {\bibfnamefont {A.}~\bibnamefont
  {Ambainis}}, \bibinfo {author} {\bibfnamefont {A.}~\bibnamefont
  {Gily\'{e}n}}, \bibinfo {author} {\bibfnamefont {S.}~\bibnamefont
  {Jeffery}},\ and\ \bibinfo {author} {\bibfnamefont {M.}~\bibnamefont
  {Kokainis}},\ }\bibfield  {title} {\bibinfo {title} {Quadratic speedup for
  finding marked vertices by quantum walks},\ }in\ \href
  {https://doi.org/10.1145/3357713.3384252} {\emph {\bibinfo {booktitle}
  {Proceedings of the 52nd Annual ACM SIGACT Symposium on Theory of
  Computing}}},\ \bibinfo {series and number} {STOC 2020}\ (\bibinfo
  {publisher} {Association for Computing Machinery},\ \bibinfo {address} {New
  York, NY, USA},\ \bibinfo {year} {2020})\ pp.\ \bibinfo {pages}
  {412--424}\BibitemShut {NoStop}%
\bibitem [{\citenamefont {Childs}(2009)}]{Childs-2009-CTQW}%
  \BibitemOpen
  \bibfield  {author} {\bibinfo {author} {\bibfnamefont {A.~M.}\ \bibnamefont
  {Childs}},\ }\bibfield  {title} {\bibinfo {title} {On the relationship
  between continuous- and discrete-time quantum walk},\ }\href
  {https://doi.org/10.1007/s00220-009-0930-1} {\bibfield  {journal} {\bibinfo
  {journal} {Communications in Mathematical Physics}\ }\textbf {\bibinfo
  {volume} {294}},\ \bibinfo {pages} {581} (\bibinfo {year}
  {2009})}\BibitemShut {NoStop}%
\bibitem [{\citenamefont {Lloyd}(1996)}]{Lloyd-1996-dec}%
  \BibitemOpen
  \bibfield  {author} {\bibinfo {author} {\bibfnamefont {S.}~\bibnamefont
  {Lloyd}},\ }\bibfield  {title} {\bibinfo {title} {Universal quantum
  simulators},\ }\href {https://doi.org/10.1126/science.273.5278.1073}
  {\bibfield  {journal} {\bibinfo  {journal} {Science}\ }\textbf {\bibinfo
  {volume} {273}},\ \bibinfo {pages} {1073} (\bibinfo {year}
  {1996})}\BibitemShut {NoStop}%
\bibitem [{\citenamefont {Childs}\ and\ \citenamefont
  {Kothari}(2011)}]{Childs-2011-dec}%
  \BibitemOpen
  \bibfield  {author} {\bibinfo {author} {\bibfnamefont {A.~M.}\ \bibnamefont
  {Childs}}\ and\ \bibinfo {author} {\bibfnamefont {R.}~\bibnamefont
  {Kothari}},\ }\bibfield  {title} {\bibinfo {title} {Simulating sparse
  hamiltonians with star decompositions},\ }in\ \href@noop {} {\emph {\bibinfo
  {booktitle} {Theory of Quantum Computation, Communication, and
  Cryptography}}},\ \bibinfo {editor} {edited by\ \bibinfo {editor}
  {\bibfnamefont {W.}~\bibnamefont {van Dam}}, \bibinfo {editor} {\bibfnamefont
  {V.~M.}\ \bibnamefont {Kendon}},\ and\ \bibinfo {editor} {\bibfnamefont
  {S.}~\bibnamefont {Severini}}}\ (\bibinfo  {publisher} {Springer Berlin
  Heidelberg},\ \bibinfo {address} {Berlin, Heidelberg},\ \bibinfo {year}
  {2011})\ pp.\ \bibinfo {pages} {94--103}\BibitemShut {NoStop}%
\bibitem [{\citenamefont {Suzuki}(1992)}]{Suzuki-1992-dec}%
  \BibitemOpen
  \bibfield  {author} {\bibinfo {author} {\bibfnamefont {M.}~\bibnamefont
  {Suzuki}},\ }\bibfield  {title} {\bibinfo {title} {General nonsymmetric
  higher-order decomposition of exponential operators and symplectic
  integrators},\ }\href {https://doi.org/10.1143/JPSJ.61.3015} {\bibfield
  {journal} {\bibinfo  {journal} {Journal of the Physical Society of Japan}\
  }\textbf {\bibinfo {volume} {61}},\ \bibinfo {pages} {3015} (\bibinfo {year}
  {1992})},\ \Eprint
  {https://arxiv.org/abs/https://doi.org/10.1143/JPSJ.61.3015}
  {https://doi.org/10.1143/JPSJ.61.3015} \BibitemShut {NoStop}%
\bibitem [{\citenamefont {Aharonov}\ and\ \citenamefont
  {Ta-Shma}(2003)}]{Aharonov-2003-dec}%
  \BibitemOpen
  \bibfield  {author} {\bibinfo {author} {\bibfnamefont {D.}~\bibnamefont
  {Aharonov}}\ and\ \bibinfo {author} {\bibfnamefont {A.}~\bibnamefont
  {Ta-Shma}},\ }\bibfield  {title} {\bibinfo {title} {Adiabatic quantum state
  generation and statistical zero knowledge},\ }in\ \href
  {https://doi.org/10.1145/780542.780546} {\emph {\bibinfo {booktitle}
  {Proceedings of the Thirty-Fifth Annual ACM Symposium on Theory of
  Computing}}},\ \bibinfo {series and number} {STOC '03}\ (\bibinfo
  {publisher} {Association for Computing Machinery},\ \bibinfo {address} {New
  York, NY, USA},\ \bibinfo {year} {2003})\ pp.\ \bibinfo {pages}
  {20--29}\BibitemShut {NoStop}%
\bibitem [{\citenamefont {Hegde}\ \emph {et~al.}(2016)\citenamefont {Hegde},
  \citenamefont {Rao},\ and\ \citenamefont {Mahesh}}]{Hedge-2016-Pdec}%
  \BibitemOpen
  \bibfield  {author} {\bibinfo {author} {\bibfnamefont {S.~S.}\ \bibnamefont
  {Hegde}}, \bibinfo {author} {\bibfnamefont {K.~R.~K.}\ \bibnamefont {Rao}},\
  and\ \bibinfo {author} {\bibfnamefont {T.~S.}\ \bibnamefont {Mahesh}},\
  }\href@noop {} {\bibinfo {title} {Pauli decomposition over commuting subsets:
  Applications in gate synthesis, state preparation, and quantum simulations}}
  (\bibinfo {year} {2016}),\ \Eprint {https://arxiv.org/abs/1603.06867}
  {arXiv:1603.06867 [quant-ph]} \BibitemShut {NoStop}%
\bibitem [{\citenamefont {A.Yu.Kitaev}(1995)}]{Kitaev-qpe}%
  \BibitemOpen
  \bibfield  {author} {\bibinfo {author} {\bibnamefont {A.Yu.Kitaev}},\
  }\bibfield  {title} {\bibinfo {title} {Quantum measurements and the abelian
  stabilizer problem}\ }(\bibinfo {year} {1995})\BibitemShut {NoStop}%
\bibitem [{\citenamefont {{Mohammadbagherpoor}}\ \emph
  {et~al.}(2019)\citenamefont {{Mohammadbagherpoor}}, \citenamefont {{Oh}},
  \citenamefont {{Dreher}}, \citenamefont {{Singh}}, \citenamefont {{Yu}},\
  and\ \citenamefont {{Rindos}}}]{Oh-2019}%
  \BibitemOpen
  \bibfield  {author} {\bibinfo {author} {\bibfnamefont {H.}~\bibnamefont
  {{Mohammadbagherpoor}}}, \bibinfo {author} {\bibfnamefont {Y.}~\bibnamefont
  {{Oh}}}, \bibinfo {author} {\bibfnamefont {P.}~\bibnamefont {{Dreher}}},
  \bibinfo {author} {\bibfnamefont {A.}~\bibnamefont {{Singh}}}, \bibinfo
  {author} {\bibfnamefont {X.}~\bibnamefont {{Yu}}},\ and\ \bibinfo {author}
  {\bibfnamefont {A.~J.}\ \bibnamefont {{Rindos}}},\ }\bibfield  {title}
  {\bibinfo {title} {An improved implementation approach for quantum phase
  estimation on quantum computers},\ }in\ \href
  {https://doi.org/10.1109/ICRC.2019.8914702} {\emph {\bibinfo {booktitle}
  {2019 IEEE International Conference on Rebooting Computing (ICRC)}}}\
  (\bibinfo {year} {2019})\ pp.\ \bibinfo {pages} {1--9}\BibitemShut {NoStop}%
\bibitem [{\citenamefont {Harrow}\ \emph {et~al.}(2009)\citenamefont {Harrow},
  \citenamefont {Hassidim},\ and\ \citenamefont {Lloyd}}]{Harrow-2009}%
  \BibitemOpen
  \bibfield  {author} {\bibinfo {author} {\bibfnamefont {A.~W.}\ \bibnamefont
  {Harrow}}, \bibinfo {author} {\bibfnamefont {A.}~\bibnamefont {Hassidim}},\
  and\ \bibinfo {author} {\bibfnamefont {S.}~\bibnamefont {Lloyd}},\ }\bibfield
   {title} {\bibinfo {title} {Quantum algorithm for linear systems of
  equations},\ }\href {https://doi.org/10.1103/PhysRevLett.103.150502}
  {\bibfield  {journal} {\bibinfo  {journal} {Phys. Rev. Lett.}\ }\textbf
  {\bibinfo {volume} {103}},\ \bibinfo {pages} {150502} (\bibinfo {year}
  {2009})}\BibitemShut {NoStop}%
\bibitem [{\citenamefont {Brassard}\ \emph {et~al.}(2002)\citenamefont
  {Brassard}, \citenamefont {Hoyer}, \citenamefont {Mosca},\ and\ \citenamefont
  {Tapp}}]{Brassard-2002}%
  \BibitemOpen
  \bibfield  {author} {\bibinfo {author} {\bibfnamefont {G.}~\bibnamefont
  {Brassard}}, \bibinfo {author} {\bibfnamefont {P.}~\bibnamefont {Hoyer}},
  \bibinfo {author} {\bibfnamefont {M.}~\bibnamefont {Mosca}},\ and\ \bibinfo
  {author} {\bibfnamefont {A.}~\bibnamefont {Tapp}},\ }\bibfield  {title}
  {\bibinfo {title} {Quantum amplitude amplification and estimation},\ }\href
  {https://doi.org/10.1090/conm/305/05215} {\bibfield  {journal} {\bibinfo
  {journal} {Quantum Computation and Information}\ ,\ \bibinfo {pages} {53}}
  (\bibinfo {year} {2002})}\BibitemShut {NoStop}%
\bibitem [{\citenamefont {Moll}\ \emph {et~al.}(2018)\citenamefont {Moll},
  \citenamefont {Barkoutsos}, \citenamefont {Bishop}, \citenamefont {Chow},
  \citenamefont {Cross}, \citenamefont {Egger}, \citenamefont {Filipp},
  \citenamefont {Fuhrer}, \citenamefont {Gambetta}, \citenamefont {Ganzhorn},\
  and\ \citenamefont {et~al.}}]{Moll-2018}%
  \BibitemOpen
  \bibfield  {author} {\bibinfo {author} {\bibfnamefont {N.}~\bibnamefont
  {Moll}}, \bibinfo {author} {\bibfnamefont {P.}~\bibnamefont {Barkoutsos}},
  \bibinfo {author} {\bibfnamefont {L.~S.}\ \bibnamefont {Bishop}}, \bibinfo
  {author} {\bibfnamefont {J.~M.}\ \bibnamefont {Chow}}, \bibinfo {author}
  {\bibfnamefont {A.}~\bibnamefont {Cross}}, \bibinfo {author} {\bibfnamefont
  {D.~J.}\ \bibnamefont {Egger}}, \bibinfo {author} {\bibfnamefont
  {S.}~\bibnamefont {Filipp}}, \bibinfo {author} {\bibfnamefont
  {A.}~\bibnamefont {Fuhrer}}, \bibinfo {author} {\bibfnamefont {J.~M.}\
  \bibnamefont {Gambetta}}, \bibinfo {author} {\bibfnamefont {M.}~\bibnamefont
  {Ganzhorn}},\ and\ \bibinfo {author} {\bibnamefont {et~al.}},\ }\bibfield
  {title} {\bibinfo {title} {Quantum optimization using variational algorithms
  on near-term quantum devices},\ }\href
  {https://doi.org/10.1088/2058-9565/aab822} {\bibfield  {journal} {\bibinfo
  {journal} {Quantum Science and Technology}\ }\textbf {\bibinfo {volume}
  {3}},\ \bibinfo {pages} {030503} (\bibinfo {year} {2018})}\BibitemShut
  {NoStop}%
\bibitem [{\citenamefont {Douglas}\ and\ \citenamefont
  {Wang}(2009)}]{Douglas-2009-EffWalk}%
  \BibitemOpen
  \bibfield  {author} {\bibinfo {author} {\bibfnamefont {B.~L.}\ \bibnamefont
  {Douglas}}\ and\ \bibinfo {author} {\bibfnamefont {J.~B.}\ \bibnamefont
  {Wang}},\ }\bibfield  {title} {\bibinfo {title} {Efficient quantum circuit
  implementation of quantum walks},\ }\href
  {https://doi.org/10.1103/PhysRevA.79.052335} {\bibfield  {journal} {\bibinfo
  {journal} {Phys. Rev. A}\ }\textbf {\bibinfo {volume} {79}},\ \bibinfo
  {pages} {052335} (\bibinfo {year} {2009})}\BibitemShut {NoStop}%
\bibitem [{QPE(2021)}]{QPECircuit}%
  \BibitemOpen
  \href@noop {} {\bibinfo {title} {Quantum phase estimation: Circuit
  implementation in qiskit}},\ \bibinfo {howpublished}
  {\url{https://qiskit.org/textbook/ch-algorithms/quantum-phase-estimation.html}}
  (\bibinfo {year} {Last accessed April 2021})\BibitemShut {NoStop}%
\bibitem [{\citenamefont {Nielsen}\ and\ \citenamefont
  {Chuang}(2010)}]{Nielsen_Chuang_2010}%
  \BibitemOpen
  \bibfield  {author} {\bibinfo {author} {\bibfnamefont {M.~A.}\ \bibnamefont
  {Nielsen}}\ and\ \bibinfo {author} {\bibfnamefont {I.~L.}\ \bibnamefont
  {Chuang}},\ }\bibinfo {title} {The quantum fourier transform and its
  applications},\ in\ \href {https://doi.org/10.1017/CBO9780511976667.009}
  {\emph {\bibinfo {booktitle} {Quantum Computation and Quantum Information:
  10th Anniversary Edition}}}\ (\bibinfo  {publisher} {Cambridge University
  Press},\ \bibinfo {year} {2010})\ pp.\ \bibinfo {pages}
  {216--247}\BibitemShut {NoStop}%
\bibitem [{\citenamefont {Str{\"o}mberg}\ and\ \citenamefont
  {Blomkvist-Karlsson}(2018)}]{Karlsson_2018}%
  \BibitemOpen
  \bibfield  {author} {\bibinfo {author} {\bibfnamefont {P.}~\bibnamefont
  {Str{\"o}mberg}}\ and\ \bibinfo {author} {\bibfnamefont {V.}~\bibnamefont
  {Blomkvist-Karlsson}},\ }\emph {\bibinfo {title} {4-qubit {G}rover's
  algorithm implemented for the IBMQx5 architecture}},\ \href
  {http://urn.kb.se/resolve?urn=urn:nbn:se:kth:diva-229797} {Ph.D. thesis}
  (\bibinfo {year} {2018})\BibitemShut {NoStop}%
\bibitem [{\citenamefont {Jin}\ and\ \citenamefont {Fei}(2018)}]{Jin_2018}%
  \BibitemOpen
  \bibfield  {author} {\bibinfo {author} {\bibfnamefont {Z.-X.}\ \bibnamefont
  {Jin}}\ and\ \bibinfo {author} {\bibfnamefont {S.-M.}\ \bibnamefont {Fei}},\
  }\bibfield  {title} {\bibinfo {title} {Quantifying quantum coherence and
  nonclassical correlation based on {H}ellinger distance},\ }\bibfield
  {journal} {\bibinfo  {journal} {Physical Review A}\ }\textbf {\bibinfo
  {volume} {97}},\ \href {https://doi.org/10.1103/physreva.97.062342}
  {10.1103/physreva.97.062342} (\bibinfo {year} {2018})\BibitemShut {NoStop}%
\end{thebibliography}%


%

\appendix

\section{Visual Comparison of the Quantum Computer Distributions}\label{ap:benchcomp}
\begin{figure*}[!pt]
    \begin{tabular}{cc}
          \includegraphics[width=7.5cm]{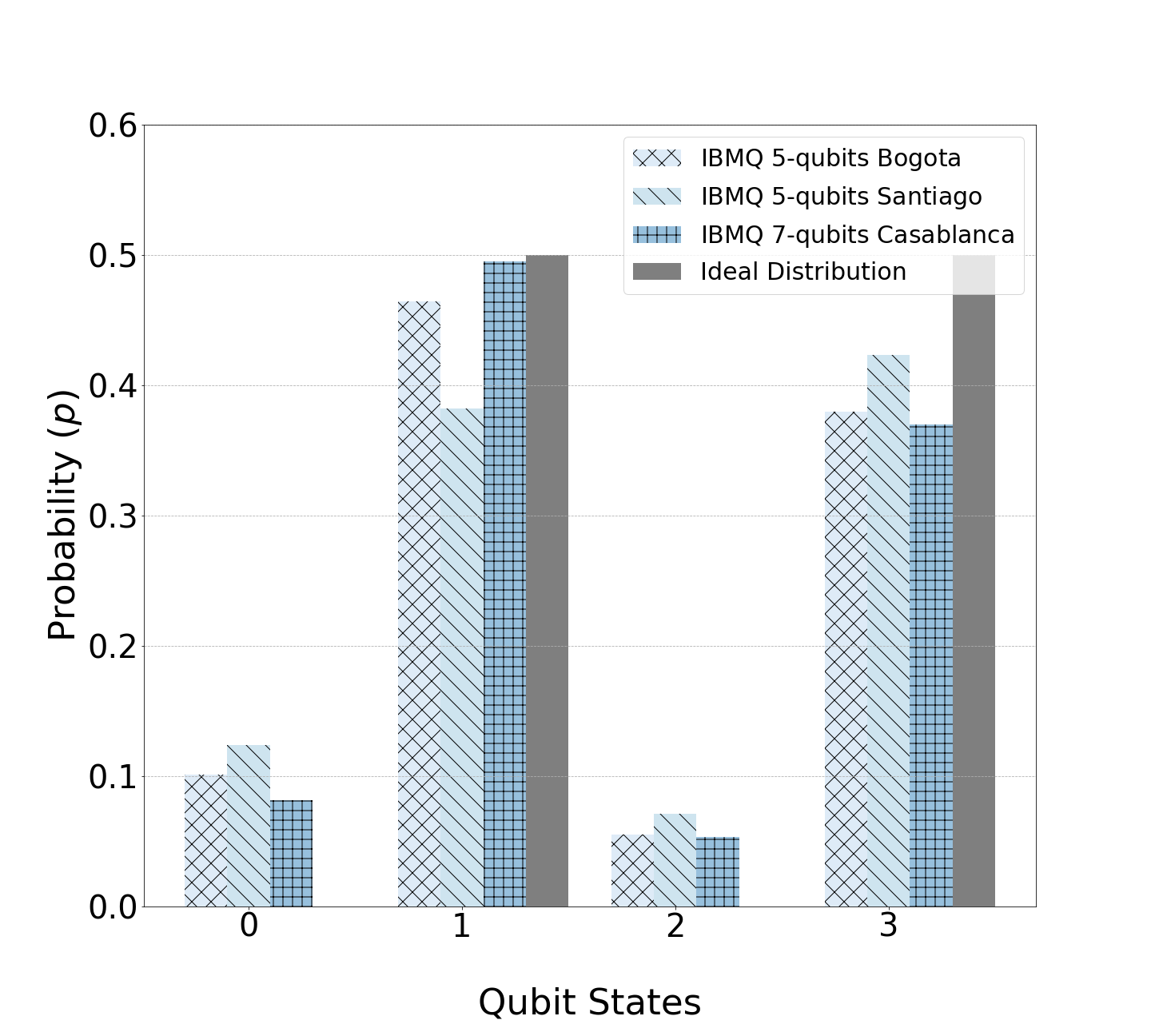} & \hspace{3em} \includegraphics[width=7.5cm]{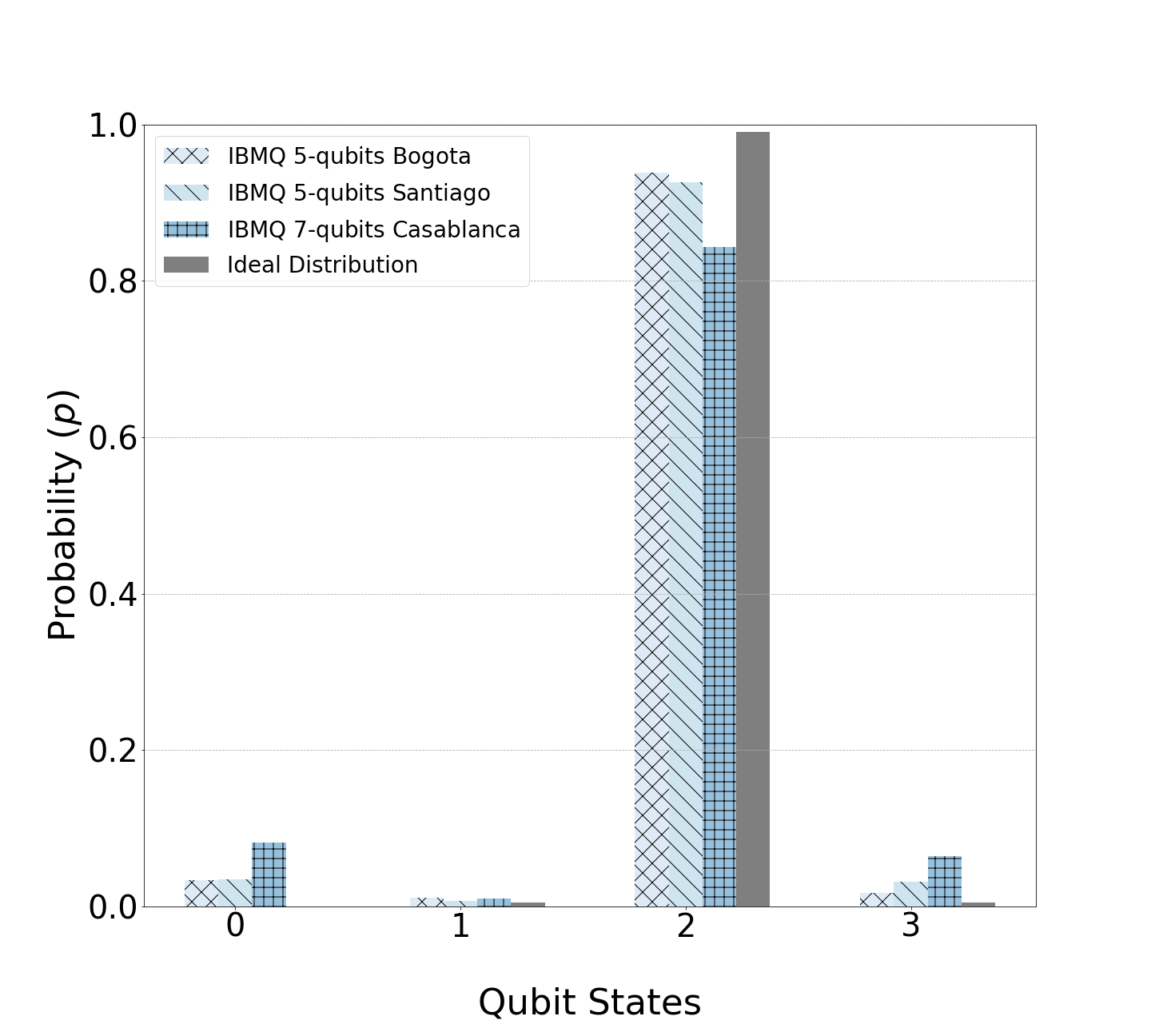} \\
        (a) Quantum walk distributions. & \hspace{3em} (b) Continuous-time quantum walk distributions. \\[6pt]
    \end{tabular}
    \begin{tabular}{cc}
          \includegraphics[width=7.5cm]{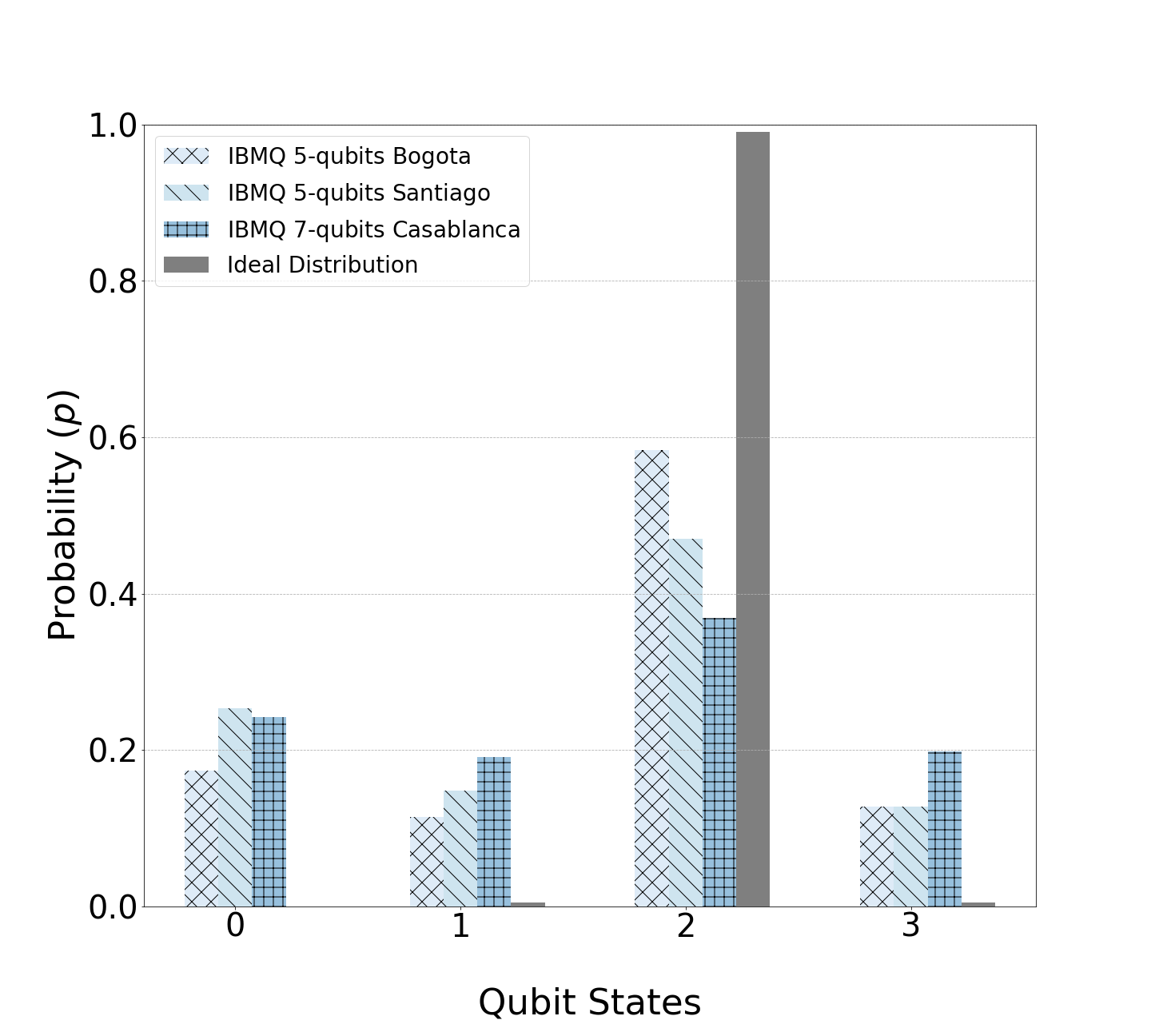} & \hspace{3em} \includegraphics[width=7.5cm]{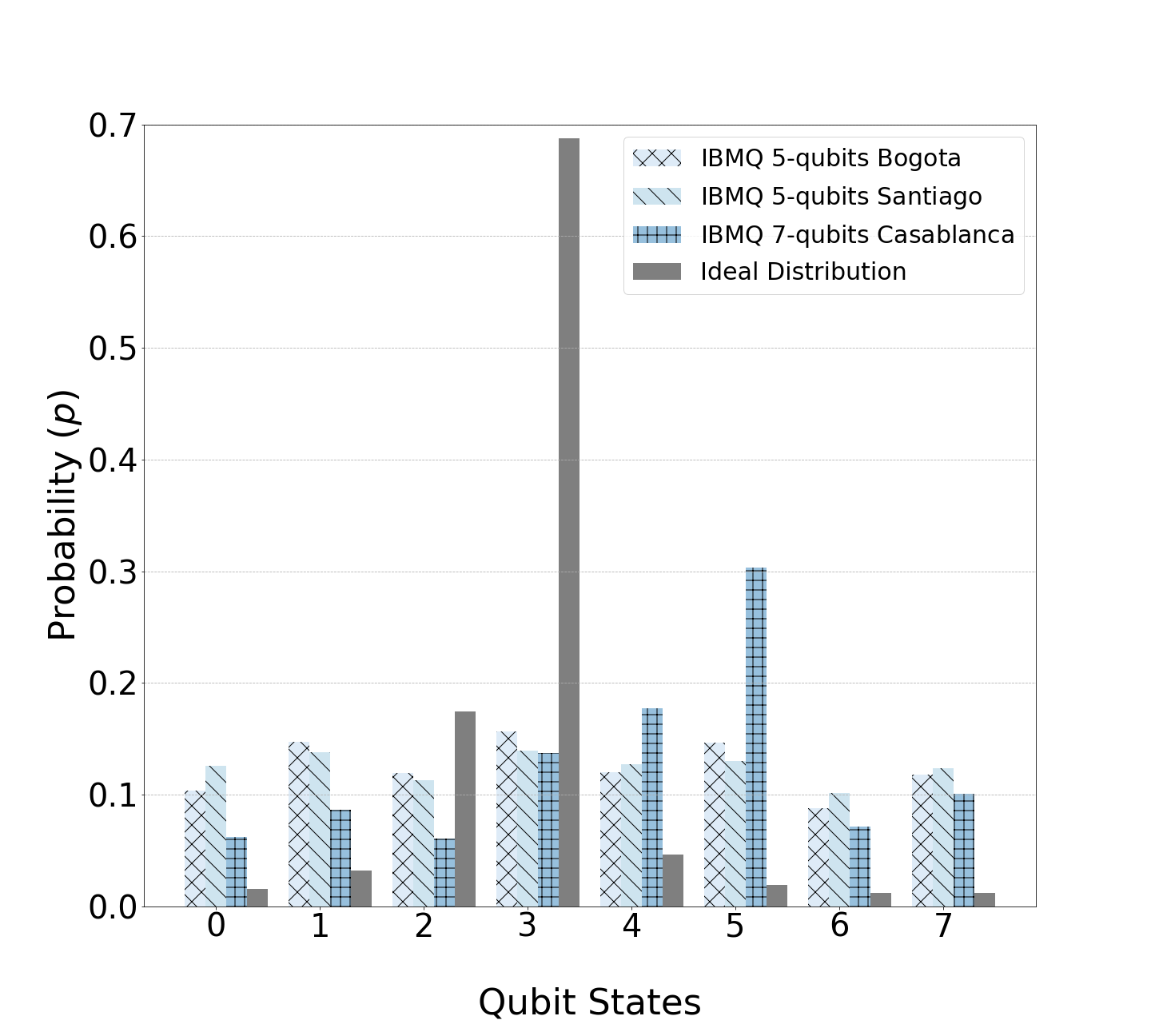} \\
        (c) Pauli decomposition of CTQW distributions. & \hspace{3em} (d) Quantum phase estimation distributions. \\[6pt]
    \end{tabular}
    \begin{tabular}{c}
          \includegraphics[width=7.5cm]{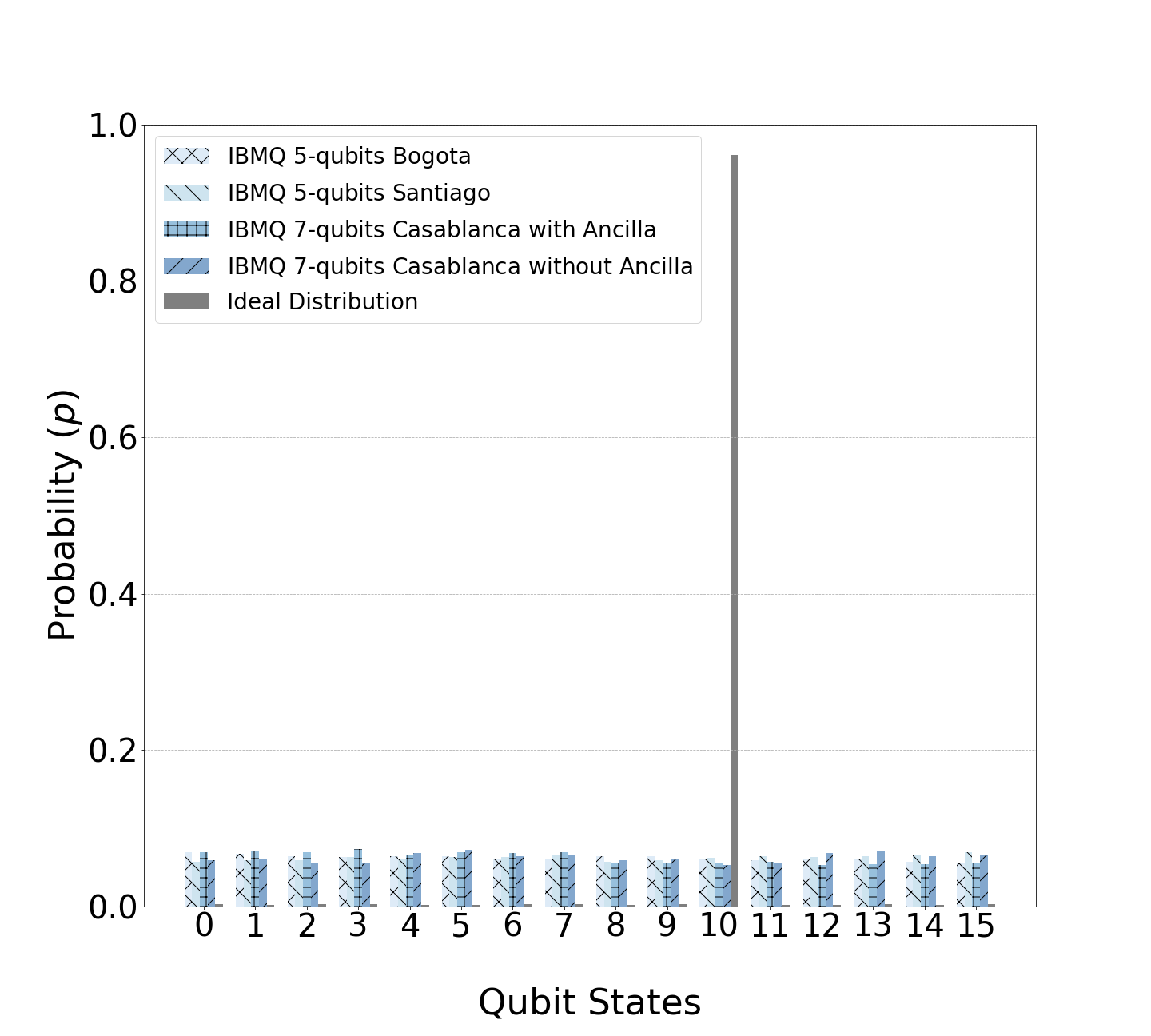} \\
        (e) Quantum search distributions. \\
    \end{tabular}
    \caption{Comparison of the probability distributions for each of the algorithms when executed on the quantum machines and when simulated in a noise-free environment.}
    \label{fig:qcvsideal}
\end{figure*}

In order to enhance the results from the benchmarking process we offer in Figure \ref{fig:qcvsideal} a visual comparison of the probability distributions that produce the Hellinger distances of Table \ref{table:benchmarks}. Each individual graph in the Figure portrays the probability distributions for the execution of one of the quantum algorithms used for benchmarking (see Section \ref{subsec:qalg}). It also includes the probability distribution of the ideal evolution of the respective quantum algorithm.

\section{Comparison of Probability Distributions}\label{ap:comp}
\begin{figure*}[!tp]
    \begin{tabular}{cc}
          \includegraphics[width=6.5cm]{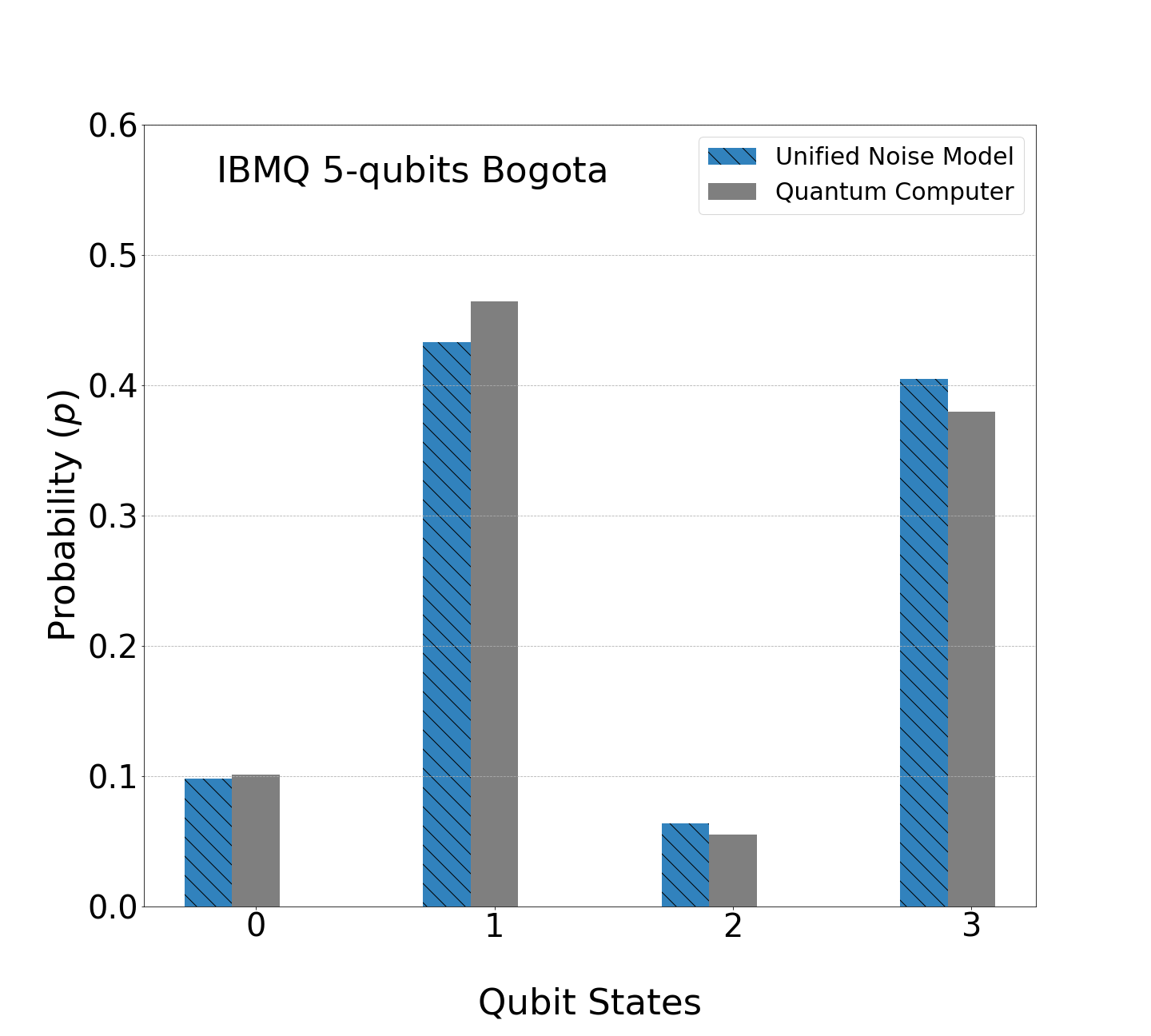} & \hspace{3em} \includegraphics[width=6.5cm]{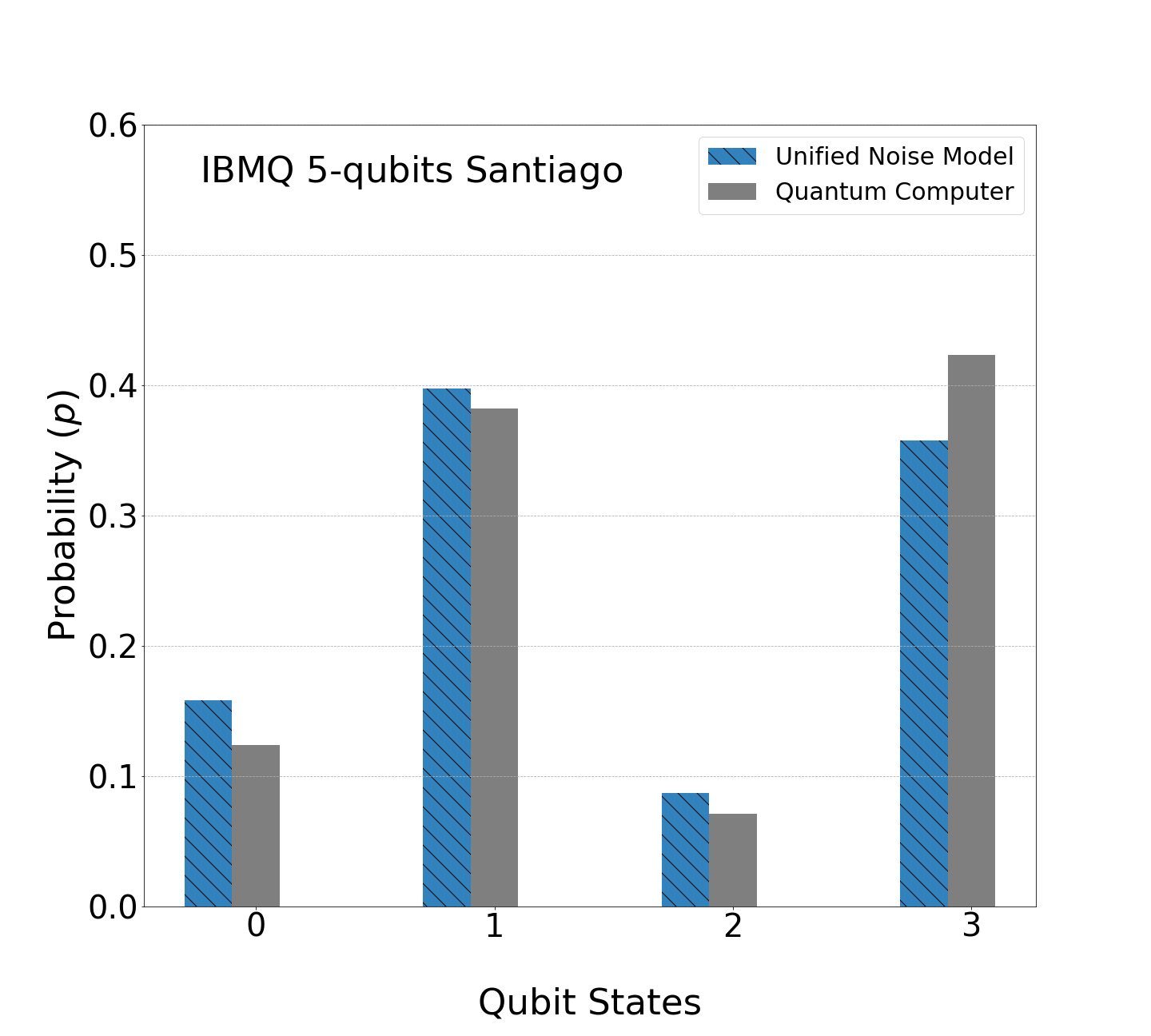} \\
        (a) & \hspace{3em} (b) \\[6pt]
          \includegraphics[width=6.5cm]{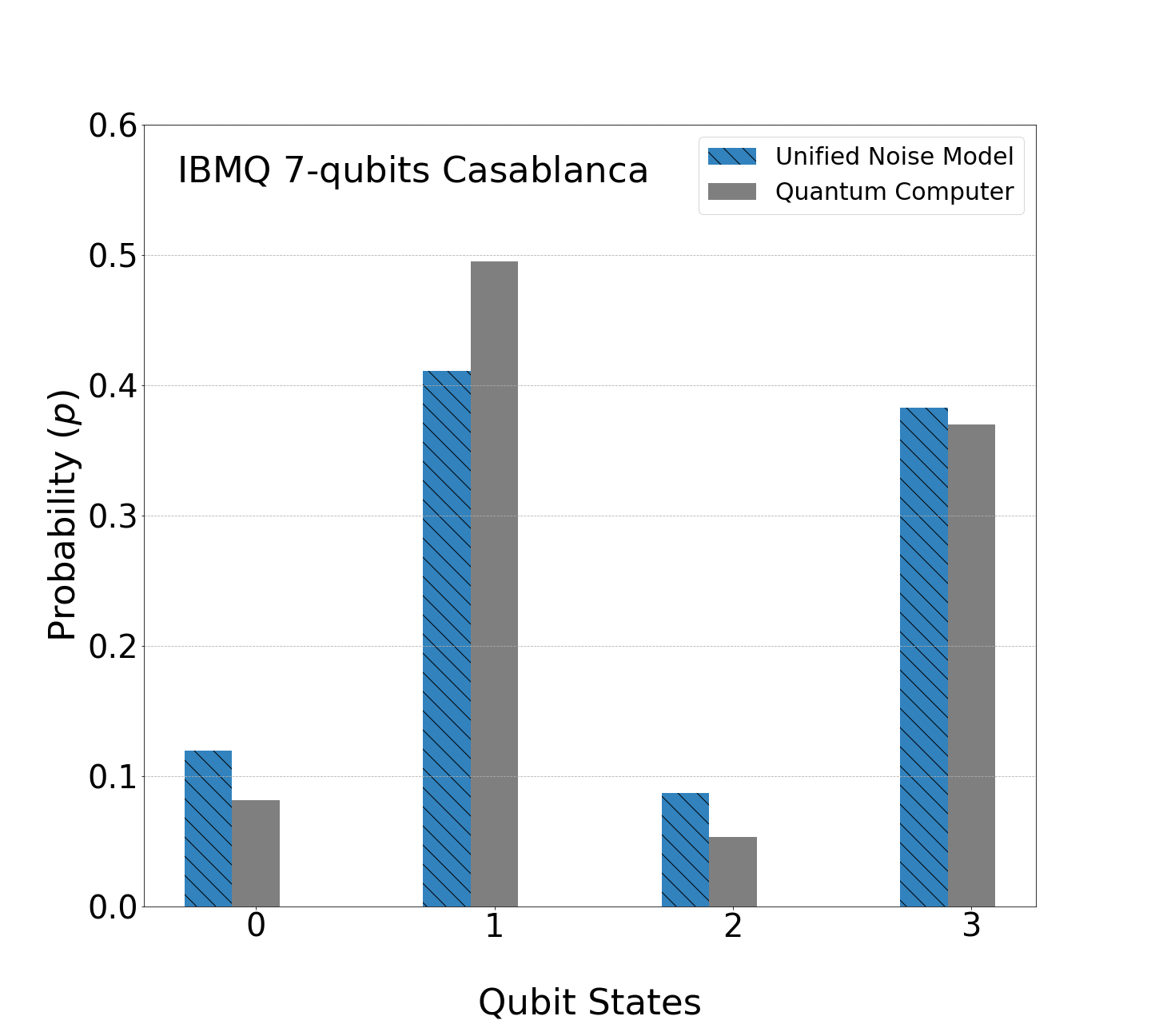} & \hspace{3em}
          \includegraphics[width=6.5cm]{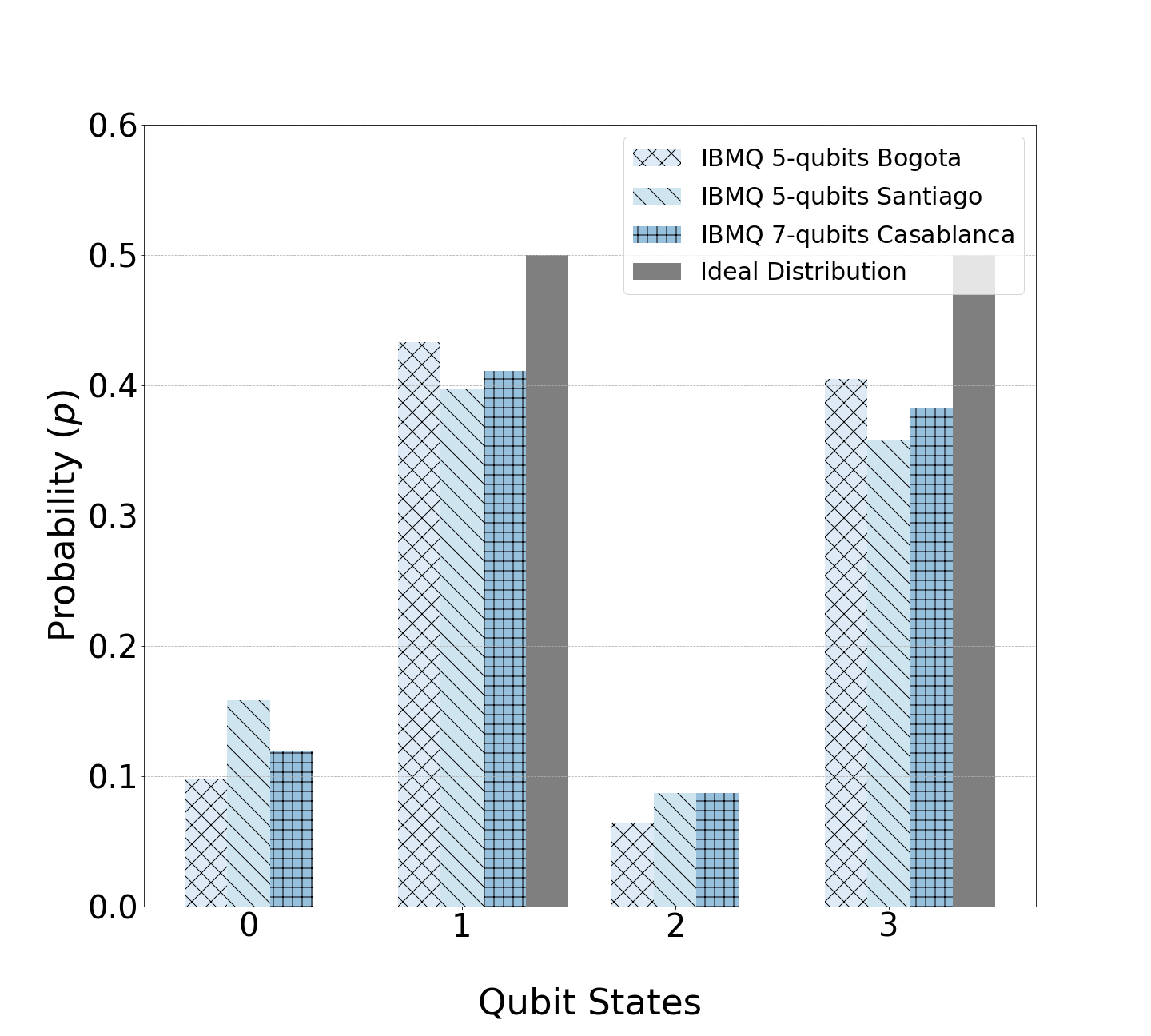} \\
        (c) & \hspace{3em} (d) \\[6pt]
    \end{tabular}
    \caption{Comparison between the probability distributions of the discrete-time quantum walk execution on each quantum machine: (a) the IBMQ $5$-qubit Bogota, (b) IBMQ $5$-qubit Santiago and (c) IBMQ $7$-qubit Casablanca machines; (d) comparison between the UNM simulations for each machine and the ideal distribution.}
    \label{fig:2qQW}
\end{figure*}

\begin{figure*}[!tp]
    \begin{tabular}{cc}
          \includegraphics[width=6.5cm]{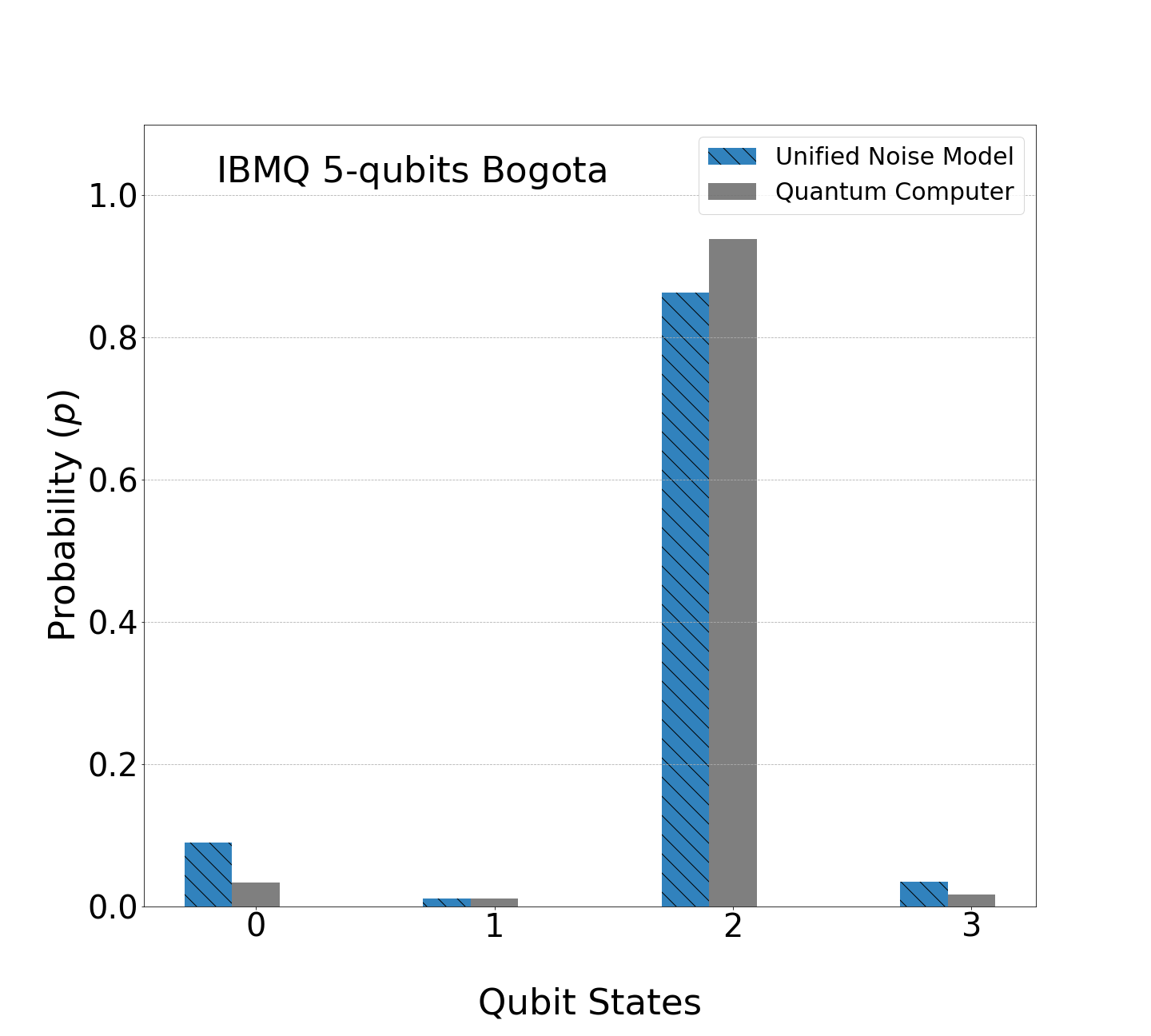} & \hspace{3em} \includegraphics[width=6.5cm]{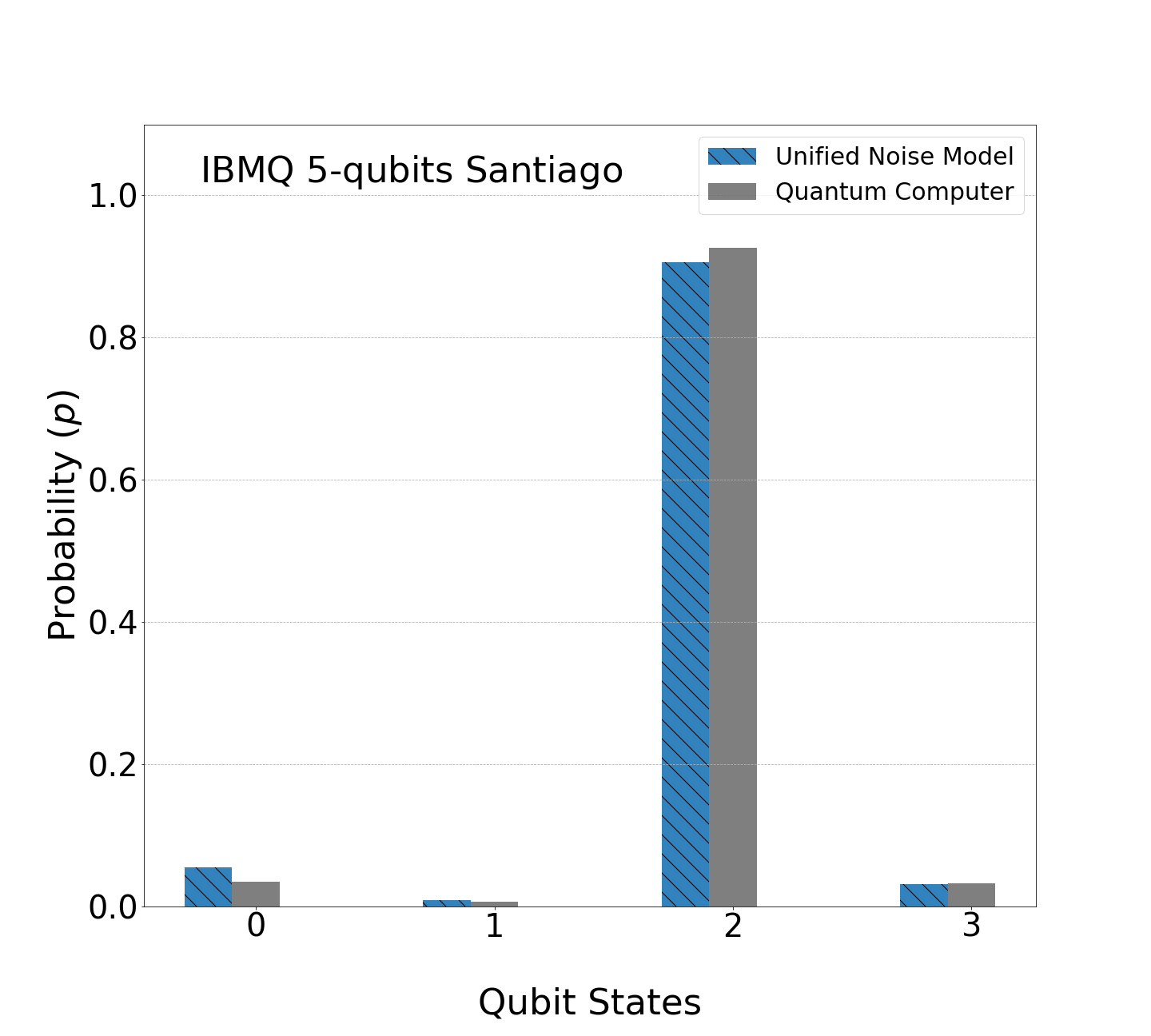} \\
        (a) & \hspace{3em} (b) \\[6pt]
          \includegraphics[width=6.5cm]{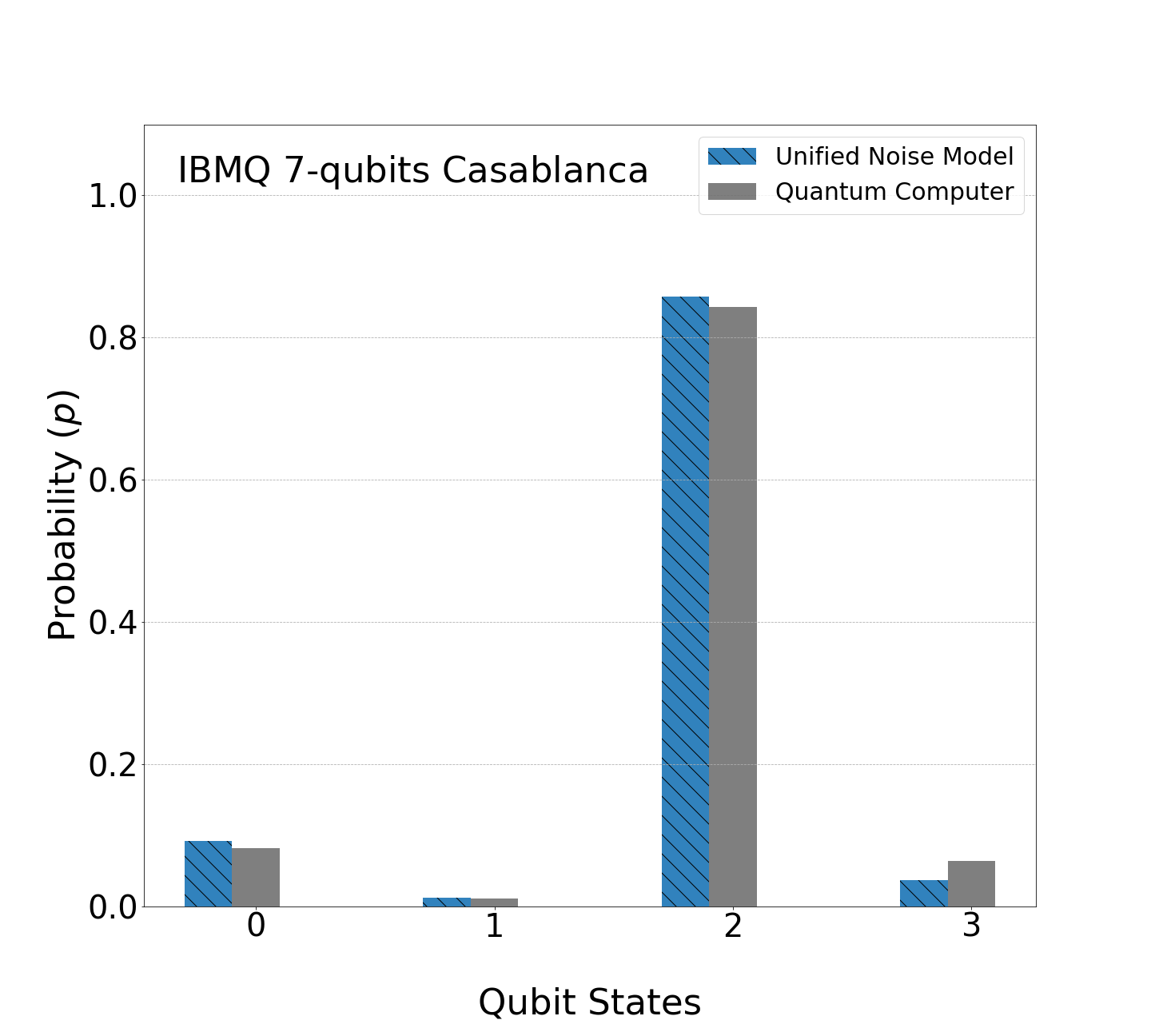} & \hspace{3em}
          \includegraphics[width=6.5cm]{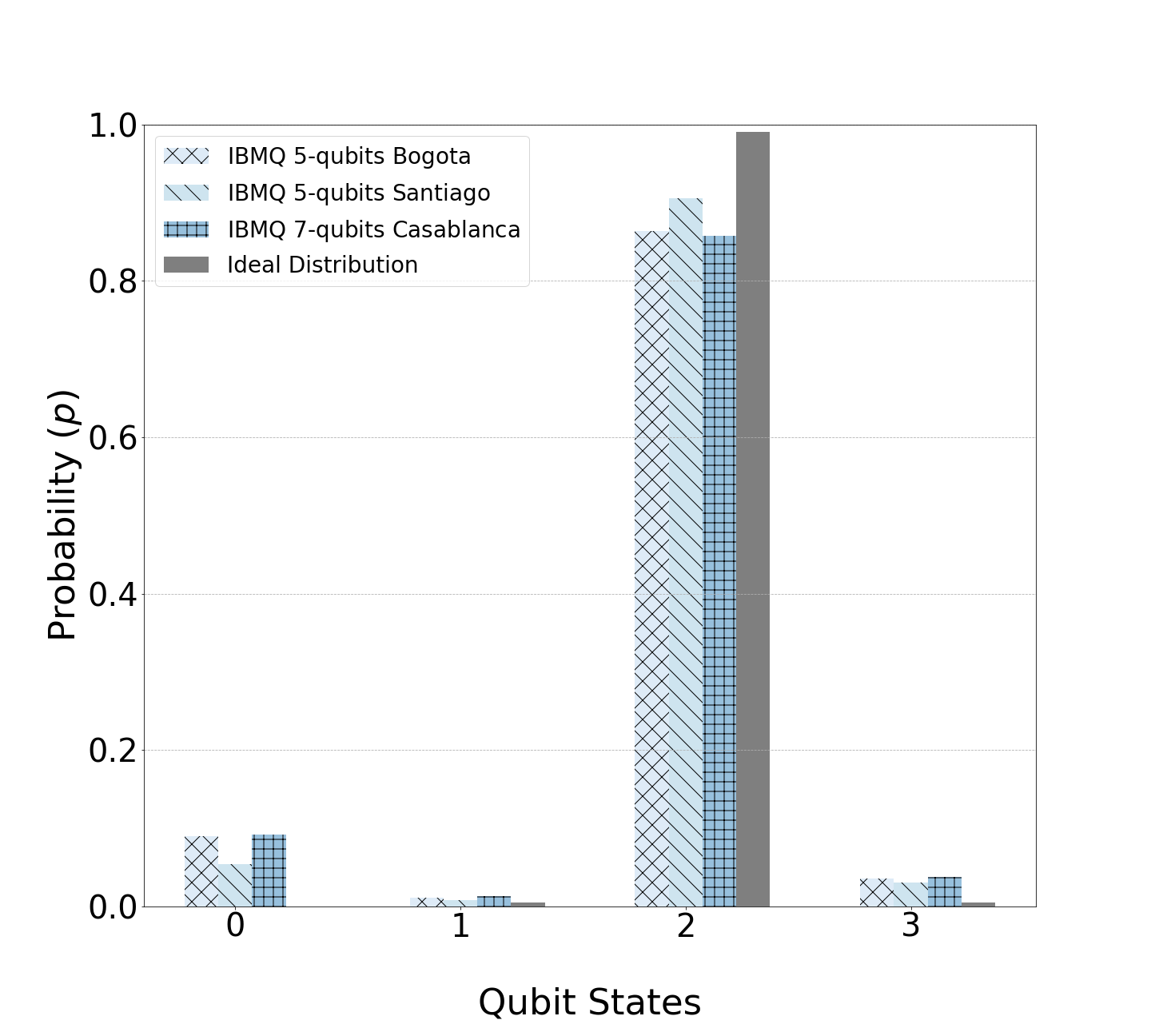} \\
        (c) & \hspace{3em} (d) \\[6pt]
    \end{tabular}
    \caption{Comparison between the probability distributions of the continuous-time quantum walk execution on each quantum machine: (a) the IBMQ $5$-qubit Bogota, (b) IBMQ $5$-qubit Santiago and (c) IBMQ $7$-qubit Casablanca machines; (d) comparison between the UNM simulations for each machine and the ideal distribution.}
    \label{fig:2qCTQW}
\end{figure*}

\begin{figure*}[!tp]
    \begin{tabular}{cc}
          \includegraphics[width=6.5cm]{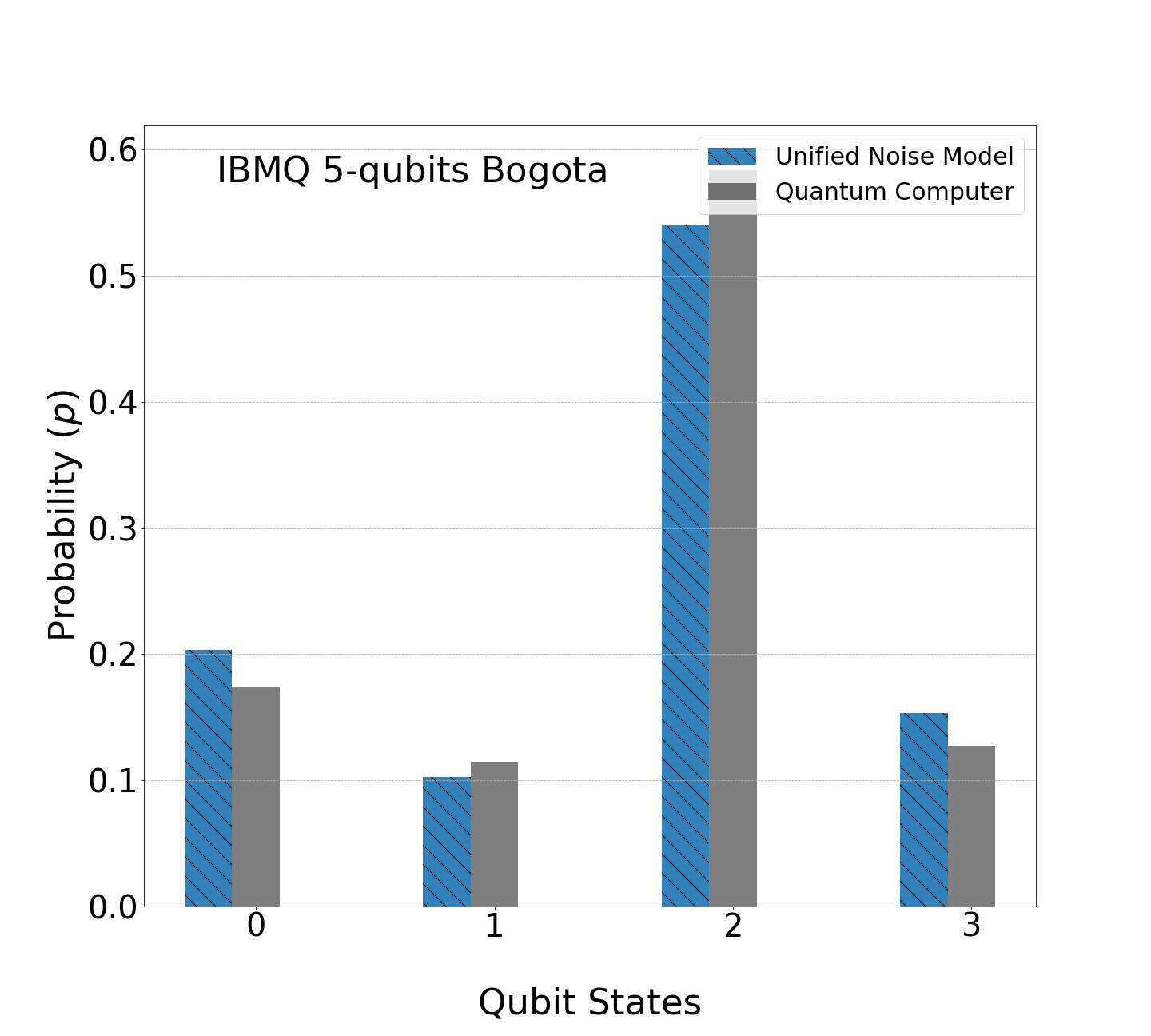} & \hspace{3em} \includegraphics[width=6.5cm]{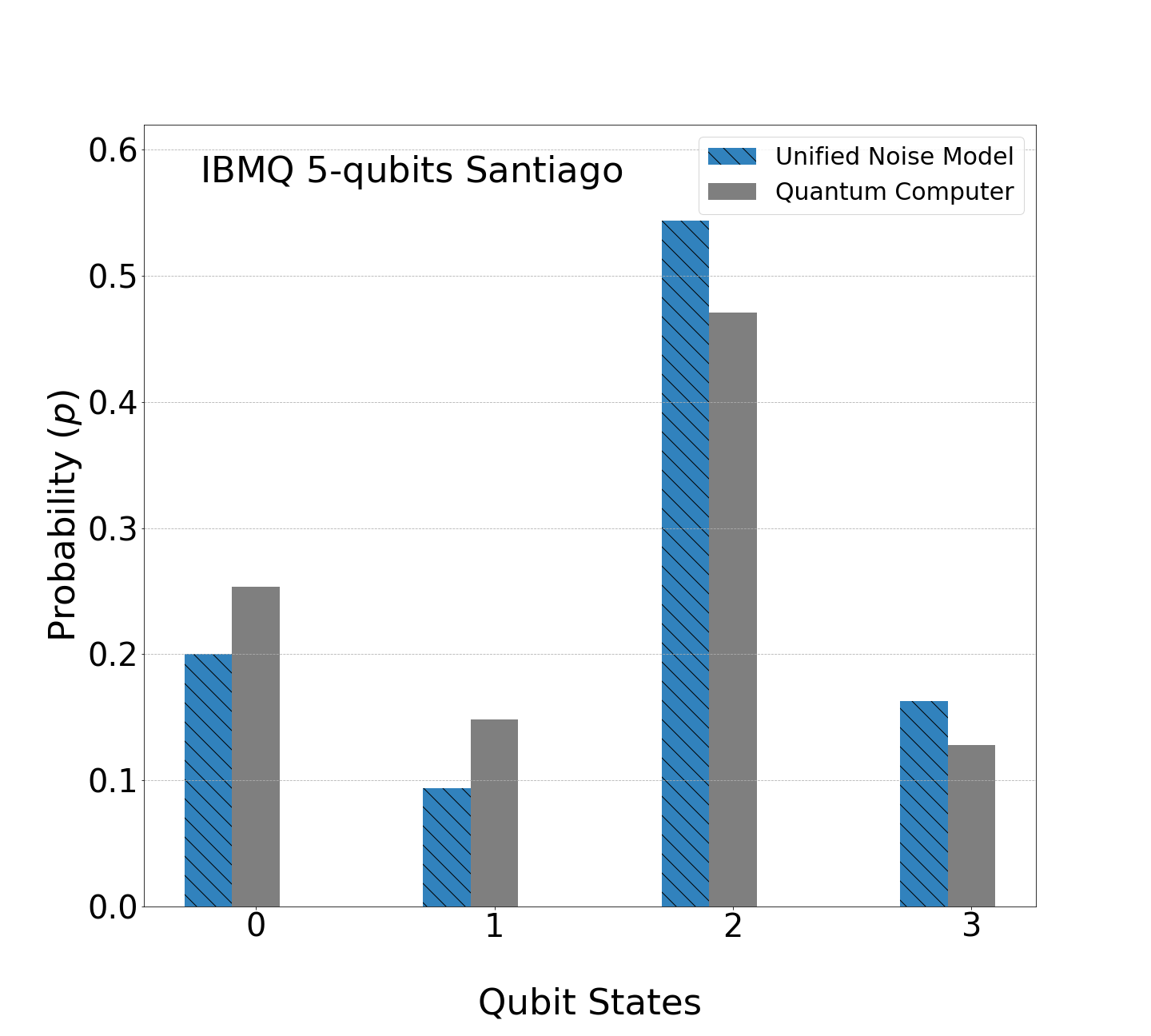} \\
        (a) & \hspace{3em} (b) \\[6pt]
          \includegraphics[width=6.5cm]{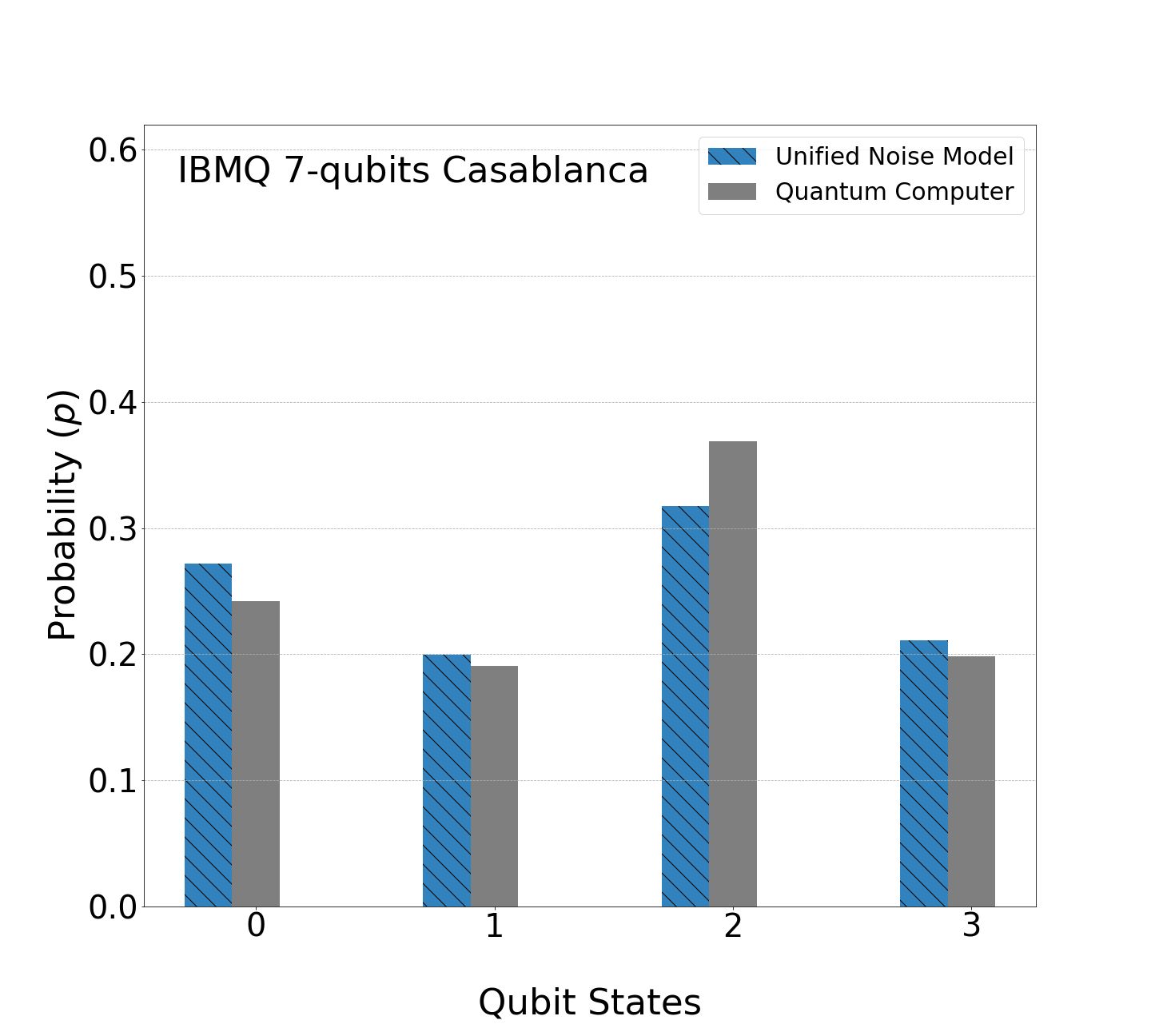} & \hspace{3em}
          \includegraphics[width=6.5cm]{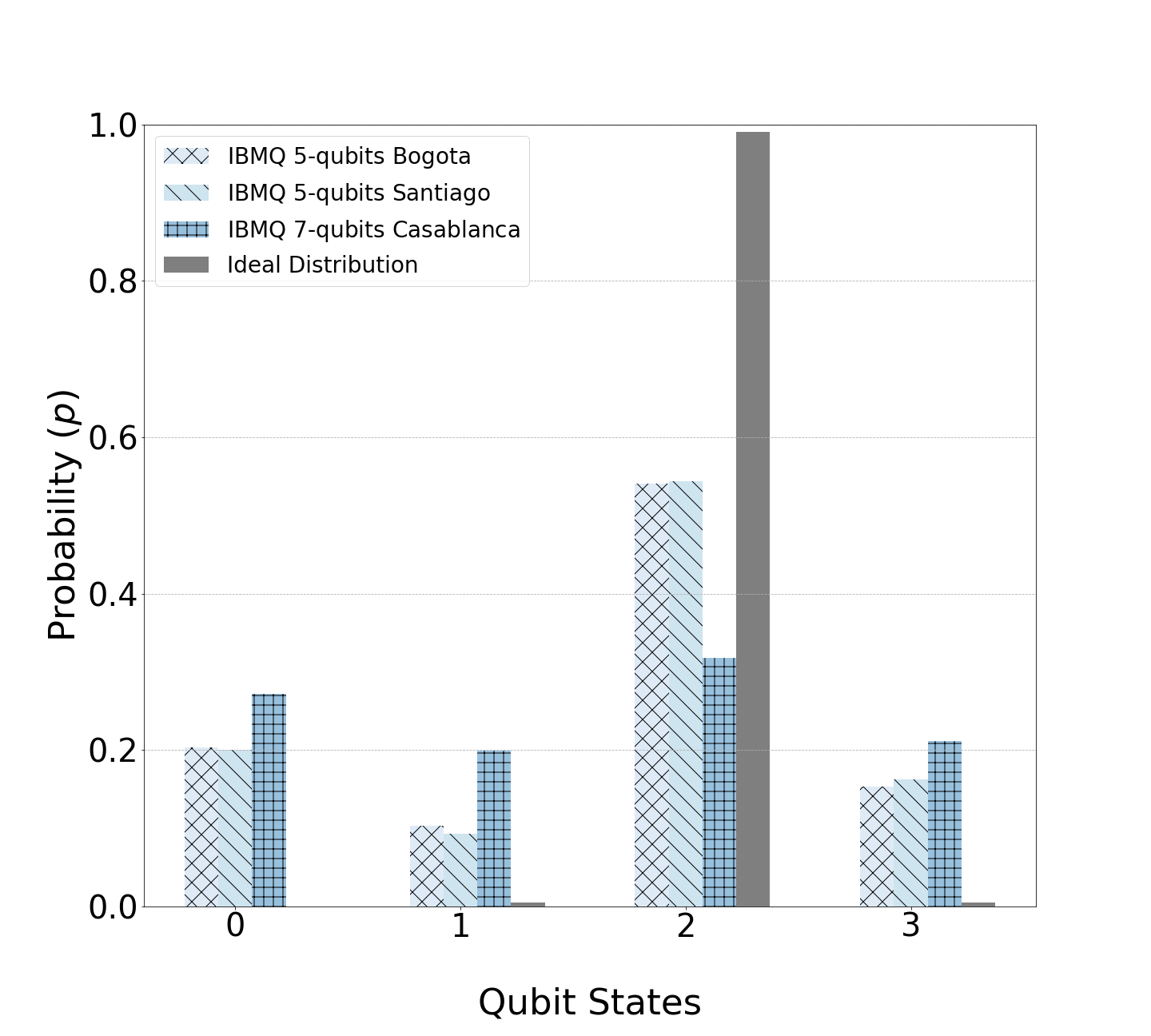} \\
        (c) & \hspace{3em} (d) \\[6pt]
    \end{tabular}
    \caption{Comparison between the probability distributions of the Pauli decomposition of the continuous-time quantum walk execution on each quantum machine: (a) the IBMQ $5$-qubit Bogota, (b) IBMQ $5$-qubit Santiago and (c) IBMQ $7$-qubit Casablanca machines; (d) comparison between the UNM simulations for each machine and the ideal distribution.}
    \label{fig:2qPD}
\end{figure*}

\begin{figure*}[!tp]
    \begin{tabular}{cc}
          \includegraphics[width=6.5cm]{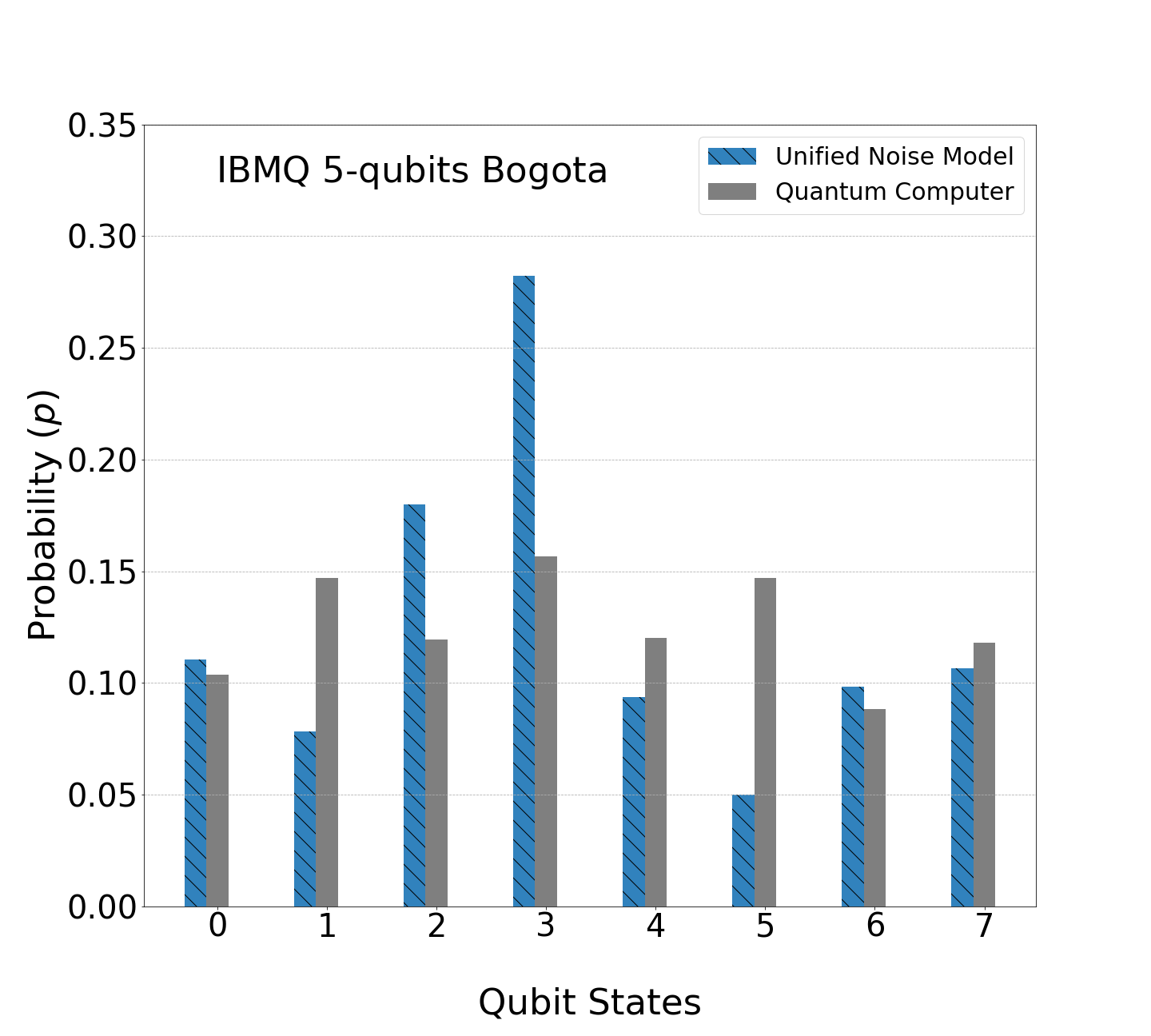} & \hspace{3em} \includegraphics[width=6.5cm]{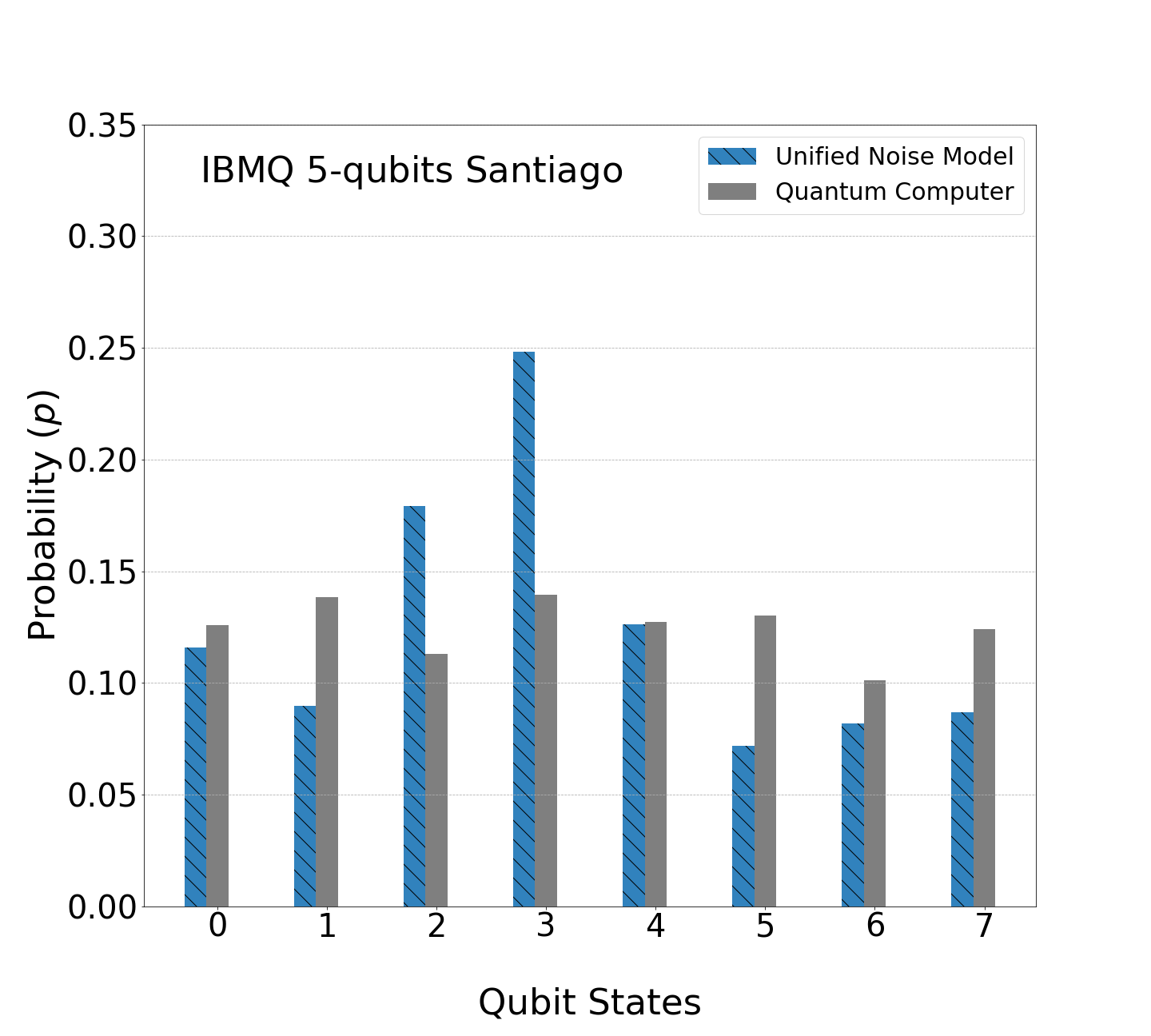} \\
        (a) & \hspace{3em} (b) \\[6pt]
          \includegraphics[width=6.5cm]{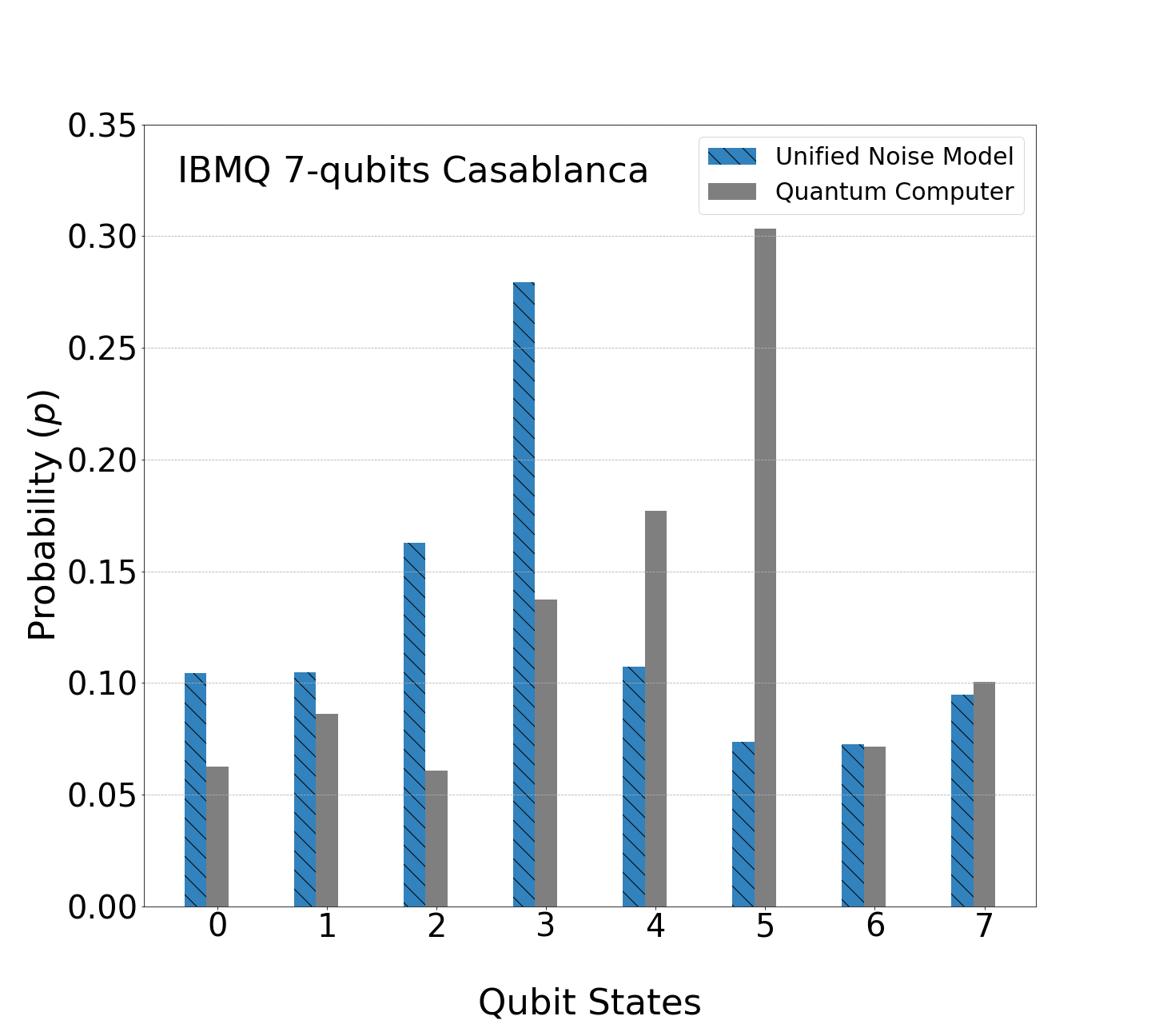} & \hspace{3em}
          \includegraphics[width=6.5cm]{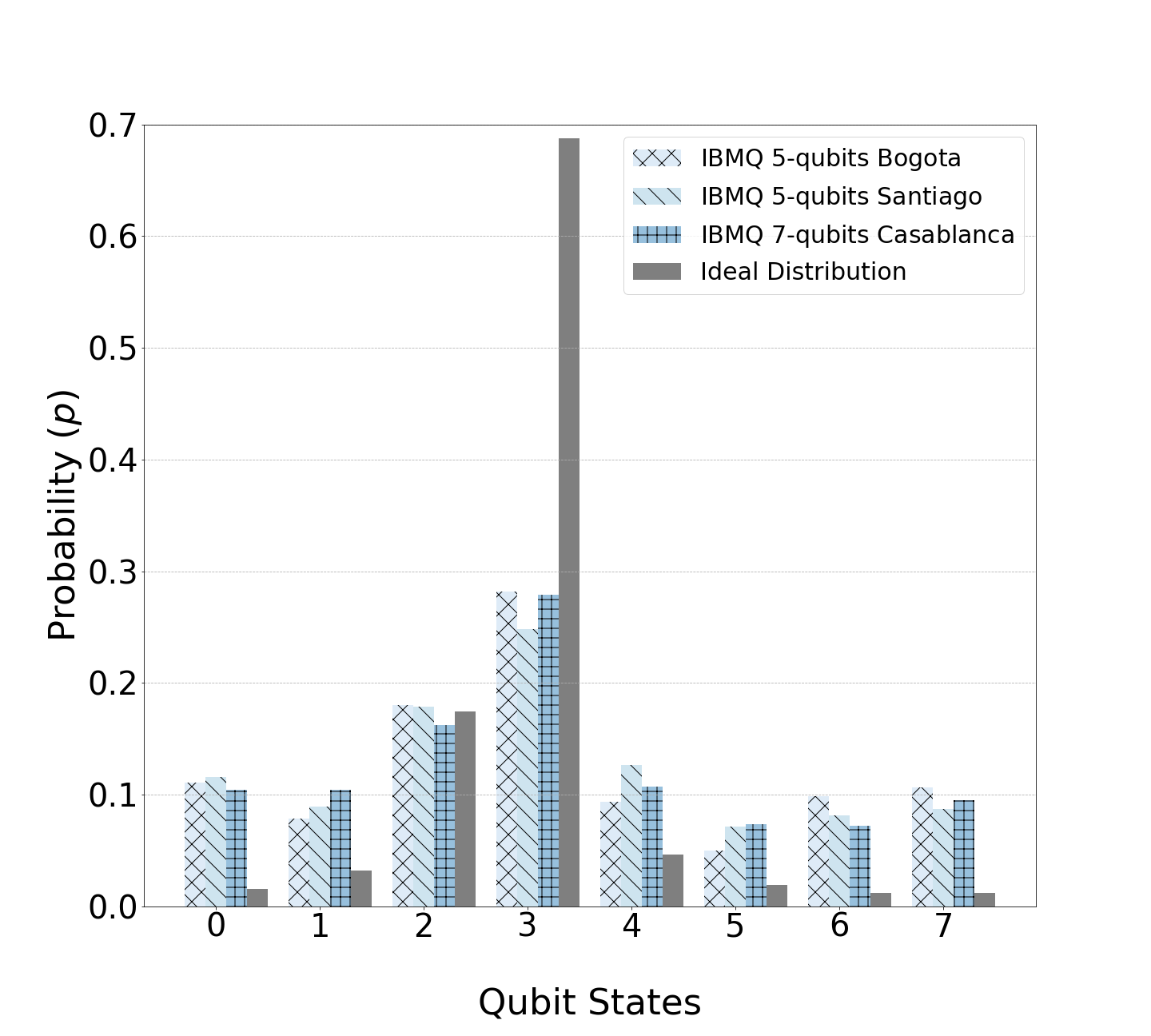} \\
        (c) & \hspace{3em} (d) \\[6pt]
    \end{tabular}
    \caption{Comparison between the probability distributions of the quantum phase estimation execution on each quantum machine: (a) the IBMQ $5$-qubit Bogota, (b) IBMQ $5$-qubit Santiago and (c) IBMQ $7$-qubit Casablanca machines; (d) comparison between the UNM simulations for each machine and the ideal distribution.}
    \label{fig:4qQPE}
\end{figure*}

\begin{figure*}[!tp]
    \begin{tabular}{cc}
          \includegraphics[width=6.5cm]{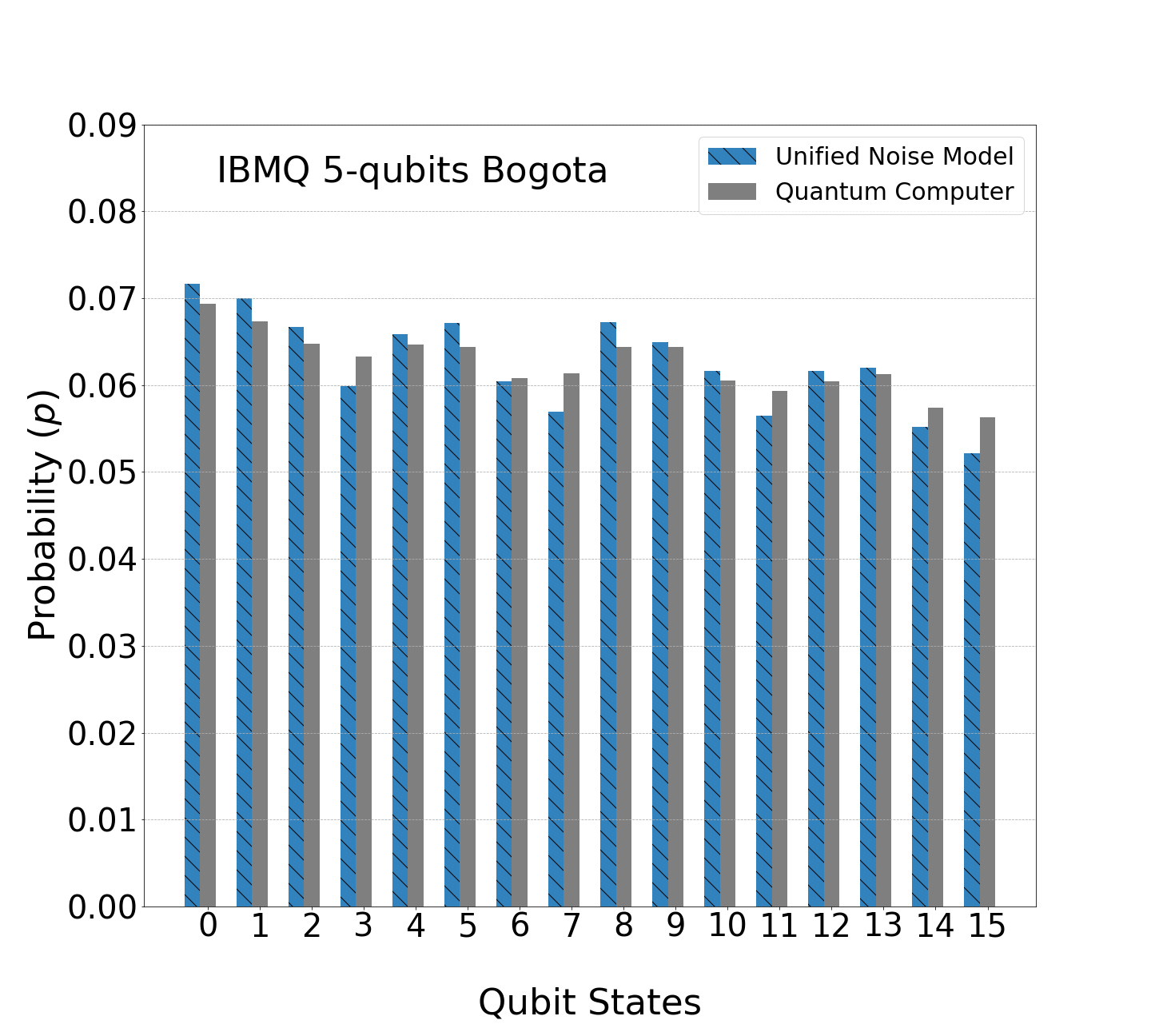} & \hspace{3em} \includegraphics[width=6.5cm]{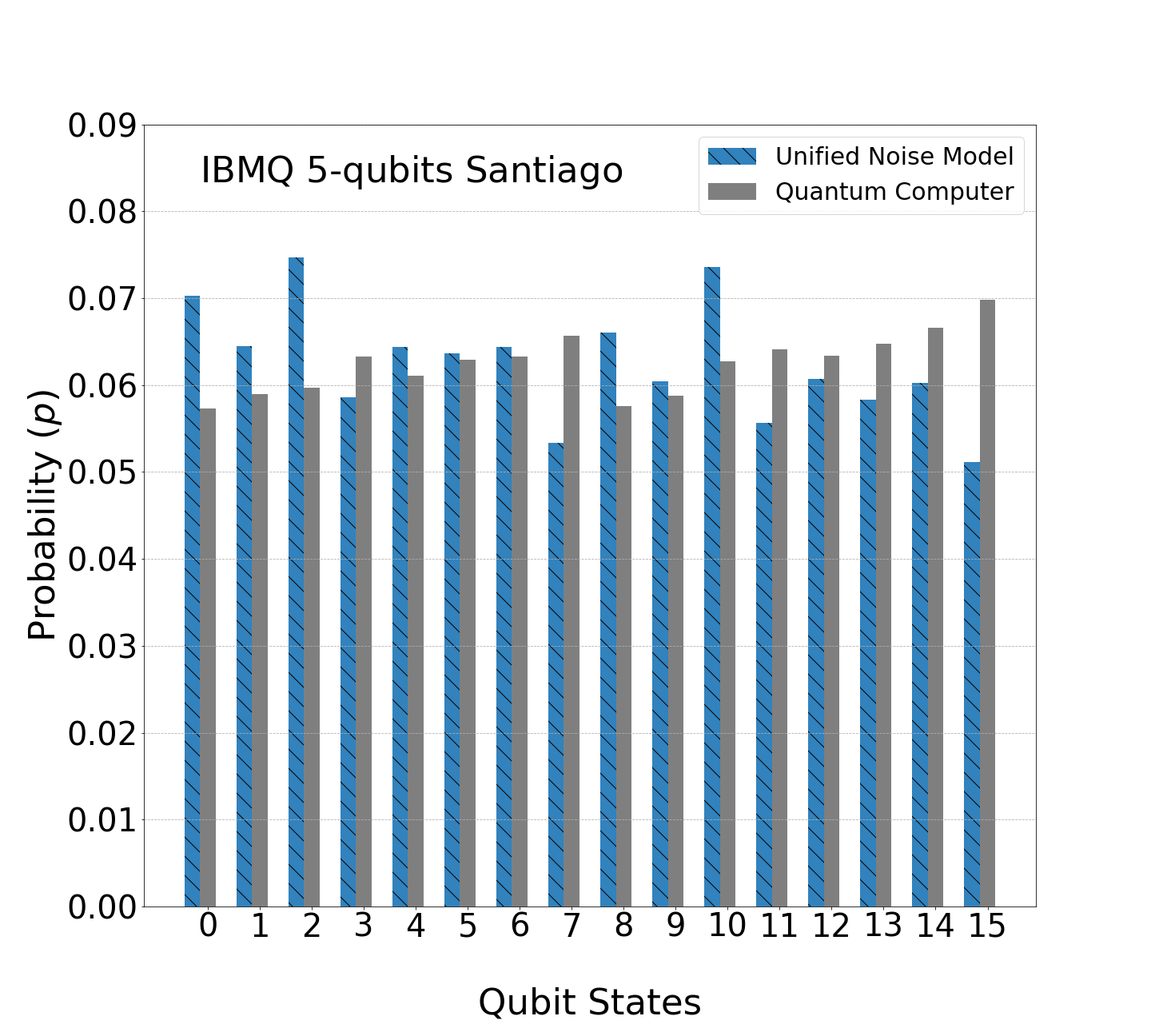} \\
        (a) & \hspace{3em} (b) \\[6pt]
          \includegraphics[width=6.5cm]{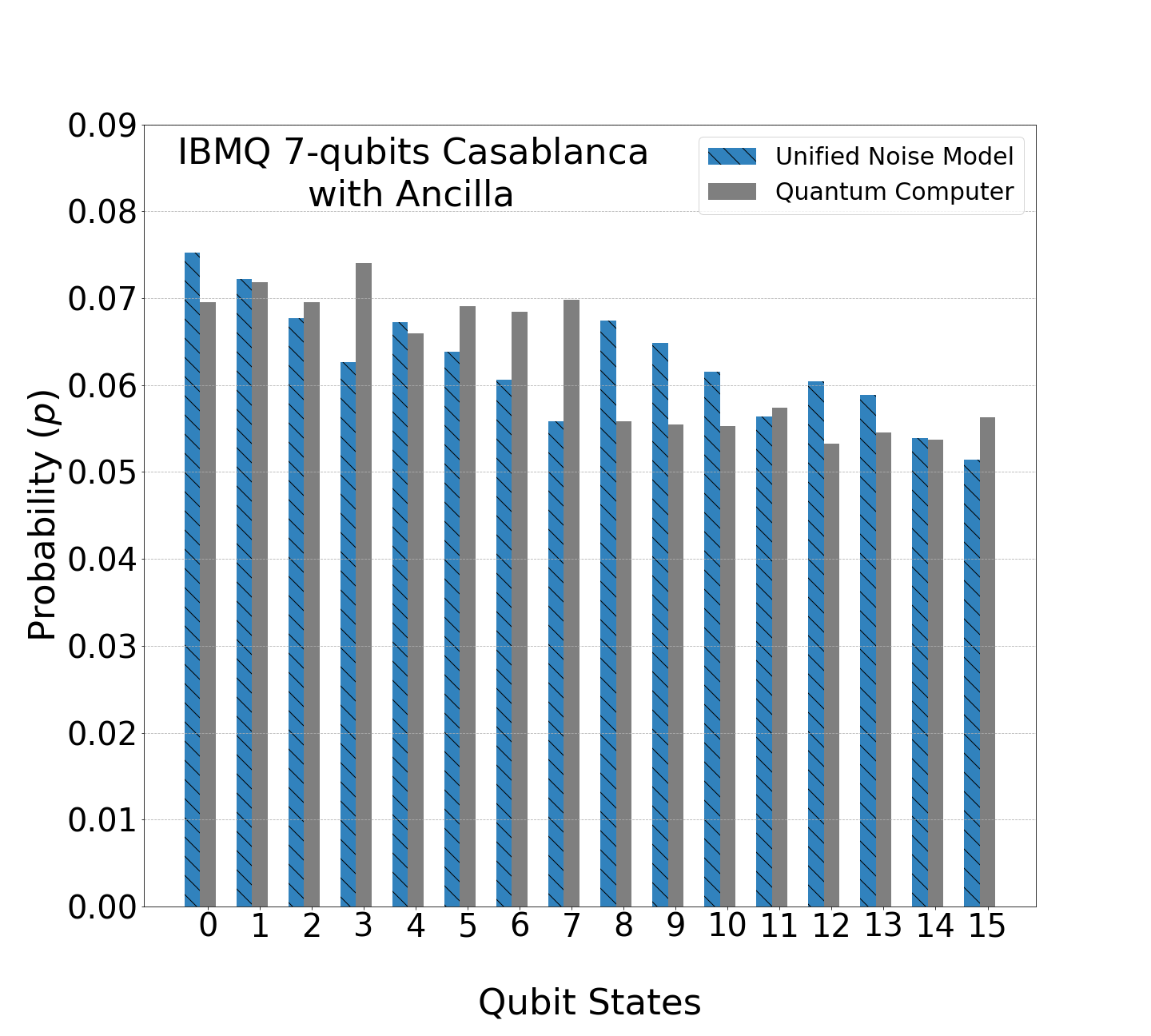} & \hspace{3em}
          \includegraphics[width=6.5cm]{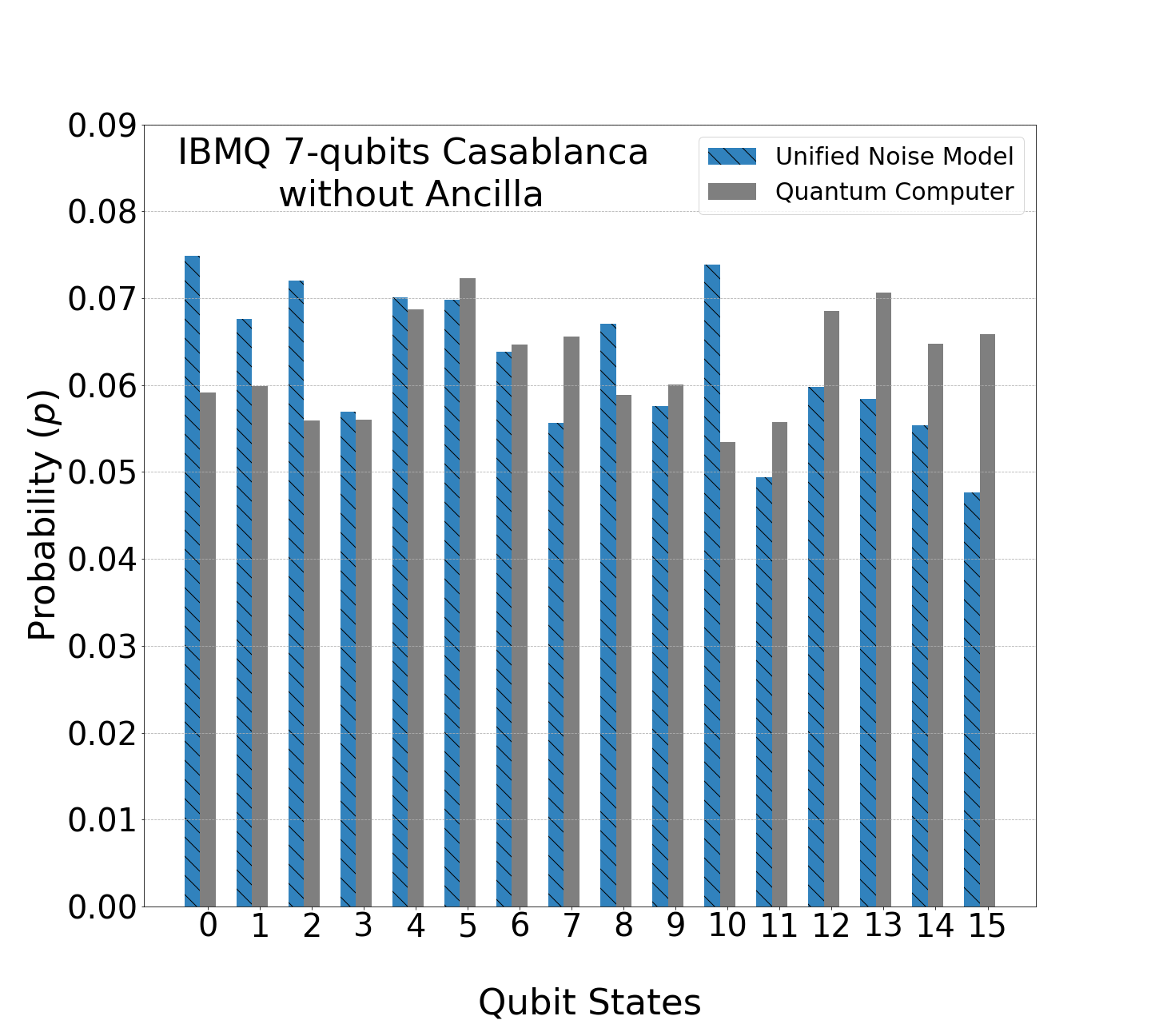} \\
        (c) & \hspace{3em} (d) \\[6pt]
    \end{tabular}
    \centering
    \begin{tabular}{c}
        \includegraphics[width=6.5cm]{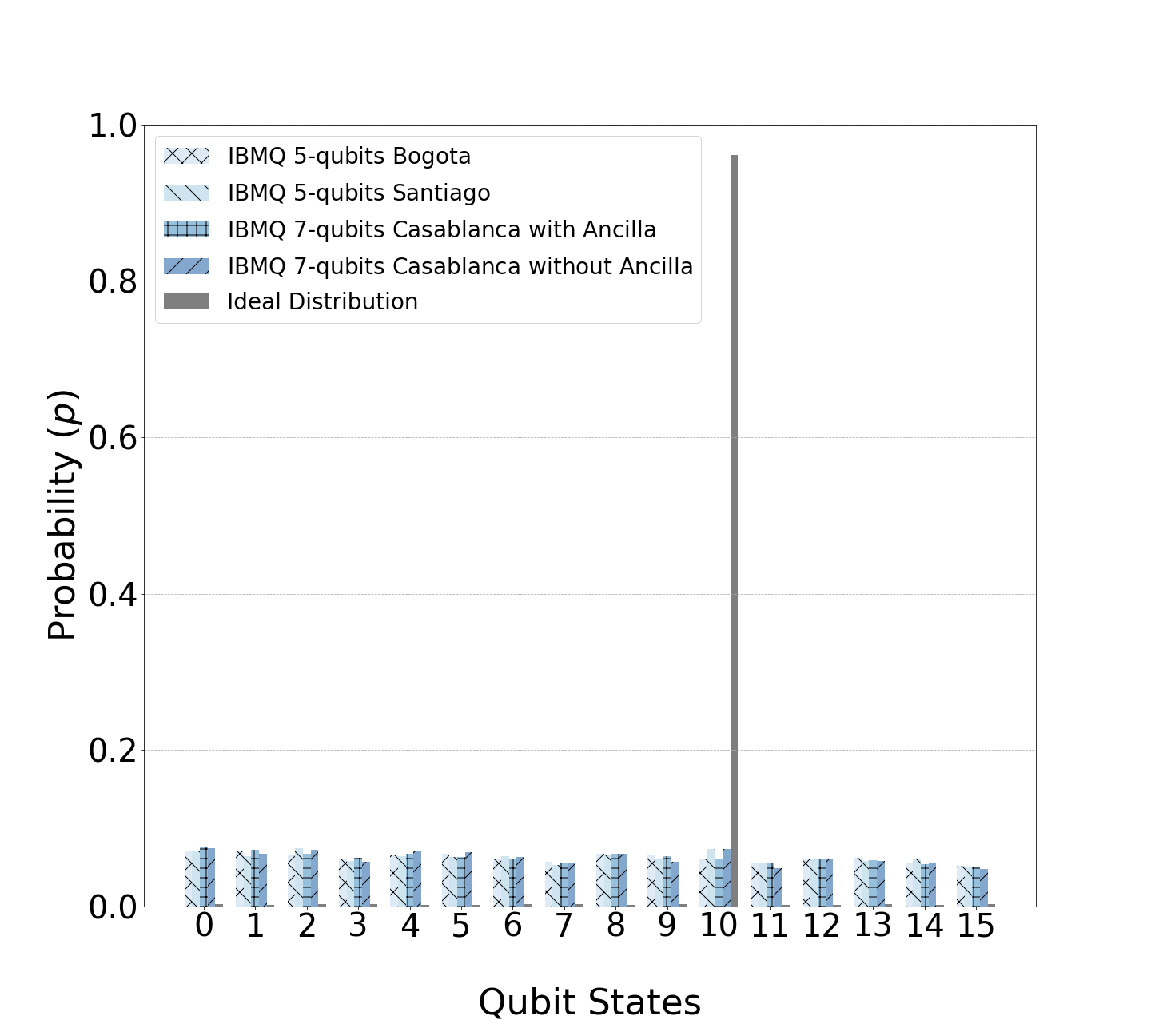} \\
        (e) \\[6pt]
    \end{tabular}
    \caption{Comparison between the probability distributions of the quantum search execution on each quantum machine: (a) the IBMQ $5$-qubit Bogota, (b) IBMQ $5$-qubit Santiago and (c) IBMQ $7$-qubit Casablanca machines; (d) comparison between the UNM simulations for each machine and the ideal distribution.}
    \label{fig:4qQS}
\end{figure*}

In the context of visual comparisons, we present a series of figures that show the difference between the probability distributions of each quantum computer benchmarked and the unified noise model (UNM). Each figure includes a set of plots that represent each algorithm used for benchmarking, i.e., for the discrete-time quantum walk (Figure \ref{fig:2qQW}), the continuous-time quantum walk (Figure \ref{fig:2qCTQW}), the Pauli decomposition of the continuous-time quantum walk Hamiltonian (Figure \ref{fig:2qPD}), the quantum phase estimation algorithm (Figure \ref{fig:4qQPE}) and the quantum search algorithm (Figure \ref{fig:4qQS}). Finally, each figure contains a plot of the probability distributions resulting from the UNM simulations and the ideal evolution for each quantum algorithm, thus including every meaningful comparison that derives from our benchmarks.

\end{document}